\documentclass{aa}  
\usepackage{natbib}
\usepackage{graphicx}
\usepackage{txfonts}
\usepackage{color}
\usepackage[colorlinks=true,linkcolor=blue,citecolor=blue,urlcolor=blue]{hyperref}
\usepackage{hyperref}
\newcommand{\hyperfootnote}[1][]{\def\ArgI\hyperfootnoteRelay}

\newcommand{\mm}{\, \mu \mathrm{m}}

\begin{document} 

\title{Near-infrared observations of star formation and gas flows in the NUGA galaxy NGC 1365\thanks{Based on observations with ESO-VLT, STS-Cologne GTO proposal ID 094.B-0009(A).}}
\titlerunning{Star formation and gas flows in NGC 1365}
\author{Nastaran Fazeli \inst{1}, Gerold Busch \inst{1}, M\'onica Valencia-S. \inst{1}, Andreas Eckart \inst{1,2}, Michal Zaja\v{c}ek \inst{1,2}, Fran\c{c}oise Combes \inst{3}, \and Santiago Garc\'ia-Burillo \inst{4}
}
\authorrunning{Nastaran Fazeli et al.}

\institute{
I. Physikalisches Institut der Universit\"at zu K\"oln, Z\"ulpicher Str. 77, 50937 K\"oln, Germany\\
\email{fazeli@ph1.uni-koeln.de}
\and
Max-Planck-Institut f\"ur Radioastronomie, Auf dem H\"ugel 69, 53121 Bonn, Germany
\and
LERMA, Observatoire de Paris, Coll\`ege de France, PSL, CNRS, Sorbonne Univ., UPMC, 75014 Paris, France
\and
 Observatorio Astron\'omico Nacional (OAN) - Observatorio de Madrid, Alfonso XII 3, 28014 Madrid, Spain}

  \date{Received 25 Sept 2018/ Accepted 19 Dec 2018}

  \abstract{
In the framework of understanding the gas and stellar kinematics and their relations to AGNs and galaxy evolution scenarios, we present spatially resolved distributions and kinematics of the stars and gas in the central $\sim 800$-pc radius of the nearby Seyfert galaxy NGC 1365.
We obtained $H+K$- and $K$-band near-infrared (NIR) integral-field observations from VLT/SINFONI. 
Our results reveal strong broad and narrow emission-line components of ionized gas (hydrogen recombination lines Pa$\alpha $ and Br$\gamma$) in the nuclear region, as well as hot dust with a temperature of $\sim 1300$ K, both typical for type-1 AGNs.
From $M_{\mathrm{BH}}-\sigma_*$ and the broad components of hydrogen recombination lines, we find a black-hole mass of $(5-10)\times 10^6 \, M_{\odot}$. In the central $\sim 800$ pc, we find a hot molecular gas mass of $\sim 615$ M$_{\odot}$, which corresponds to a cold molecular gas reservoir of $(2-8)\times 10^8$ M$_{\odot}$. However, there is a molecular gas deficiency in the nuclear region. 
The gas and stellar-velocity maps both show rotation patterns consistent with the large-scale rotation of the galaxy. However, the gaseous and stellar kinematics show deviations from pure disk rotation, which suggest streaming motions in the central $< 200$ pc and a velocity twist at the location of the ring which indicates deviations in disk and ring rotation velocities in accordance with published CO kinematics. We detect a blueshifted emission line split in Pa$\alpha$, associated with the nuclear region only. 
We investigate the star-formation properties of the hot spots in the circumnuclear ring which have starburst ages of $\lesssim 10\,\mathrm{Myr}$ and find indications for an age gradient on the western side of the ring.
In addition, our high-resolution data reveal further substructure within this ring which also shows enhanced star forming activity close to the nucleus.
}

\keywords{galaxies: active -- galaxies: Seyfert -- galaxies: nuclei --  galaxies: individual: NGC 1365 -- galaxies: kinematics and dynamics -- infrared: galaxies -- galaxies: star formation.}

\maketitle

%________________________________________________________________

\section{Introduction}

In the center of most massive galaxies resides a supermassive black hole \citep[SMBH; e.g., ][]{1969Natur.223..690L, 2013ARA&A..51..511K}. Accretion of mass onto the SMBH is observed as a powerful nonstellar radiation from an unresolved region that can even outshine the host galaxy, known as active galactic nucleus (AGN). However, only $\sim 10\%$ of all galaxies are currently in a strong AGN phase (Seyfert and Quasi-stellar object (QSO)) and $\sim 40\%$ show at least low-luminosity AGN activity \citep{2008ARA&A..46..475H}. 

The question of how gas is funneled from kiloparsec scales to the accretion disk of the SMBH in order to feed the black hole and maintain the AGN activity is an active field of research \citep[e.g., reviews by][]{1990Natur.345..679S,2005Ap&SS.295...85K,2006LNP...693..143J,2012NewAR..56...93A}. While high-luminosity AGNs (at redshift $\gtrsim 2$) are often triggered by external perturbations \citep[i.e., galaxy interactions and major mergers; e.g.,][]{1988ApJ...325...74S,1989Natur.338...45S,2005A&A...437...69B}, AGNs in the nearby universe are mostly dominated by secular evolution \citep[e.g., ][and references therein]{2013seg..book....1K}. Torques due to large-scale bars and spiral arms transport the gas to the central region ($\lesssim 1$ kpc), where secondary/nuclear bars and spirals can take over \citep[e.g., ][]{2004A&A...414..857C}. The bars lead to Lindblad resonances, where the gas is stalled and compressed in rings which can lead to starburst \citep[e.g., ][]{1996FCPh...17...95B}. Stellar winds from these star-forming regions might also play a role in AGN feeding in cases where the wind material is captured at the corresponding Bondi radius \citep[e.g., as seems to be the case for the low-luminous Galactic center;][]{2018MNRAS.479.4778Y}.  

The aim of the NUGA project (NUclei of GAlaxies) is to study the mechanisms that lead to fueling of a galactic nucleus \citep[PIs: Santiago
García-Burillo and Françoise Combes;][]{2003A&A...407..485G}. The project therefore started by mapping the distribution and dynamics of the (cold) molecular gas in the central kiloparsec of nearby galaxies (where the host galaxy can be resolved at scales  of tens parsecs), using the IRAM Plateau de Bure Interferometer (PdBI) and 30-m single-dish. With the Atacama Large Millimeter/sub-millimeter Array (ALMA) being established, NUGA expanded to the southern hemisphere. The NUGA sample comprises several nearby galaxies with variant nuclear activities (Seyfert, LINER, starburst) that are studied from kiloparsec scales down to scales of a few tens of parsecs. 

The observations of these galaxies with integral-field spectroscopy in the near-infrared (NIR) are complementary to the millimeter data \citep[e.g., NGC 1433, 1566, and 1808;][]{2014A&A...567A.119S, 2015A&A...583A.104S, 2017A&A...598A..55B}. The NIR data reveal information about the hot surface layer ($\sim$2000 K) of the molecular gas clouds and the ionized gas phase and are essential to study the distribution and kinematics of the mass-dominating old stellar population. The NIR wavelength region also provides valuable information on star formation history and properties of the central engine of the galaxy, in particular in the presence of obscuring dust \citep[e.g.,][]{2008AJ....135..479B,2009ApJ...698.1852B,2012A&A...544A.105S,2014MNRAS.438..329F,2015A&A...575A.128B}. 

\begin{figure}
\centering
\includegraphics[width=\columnwidth]{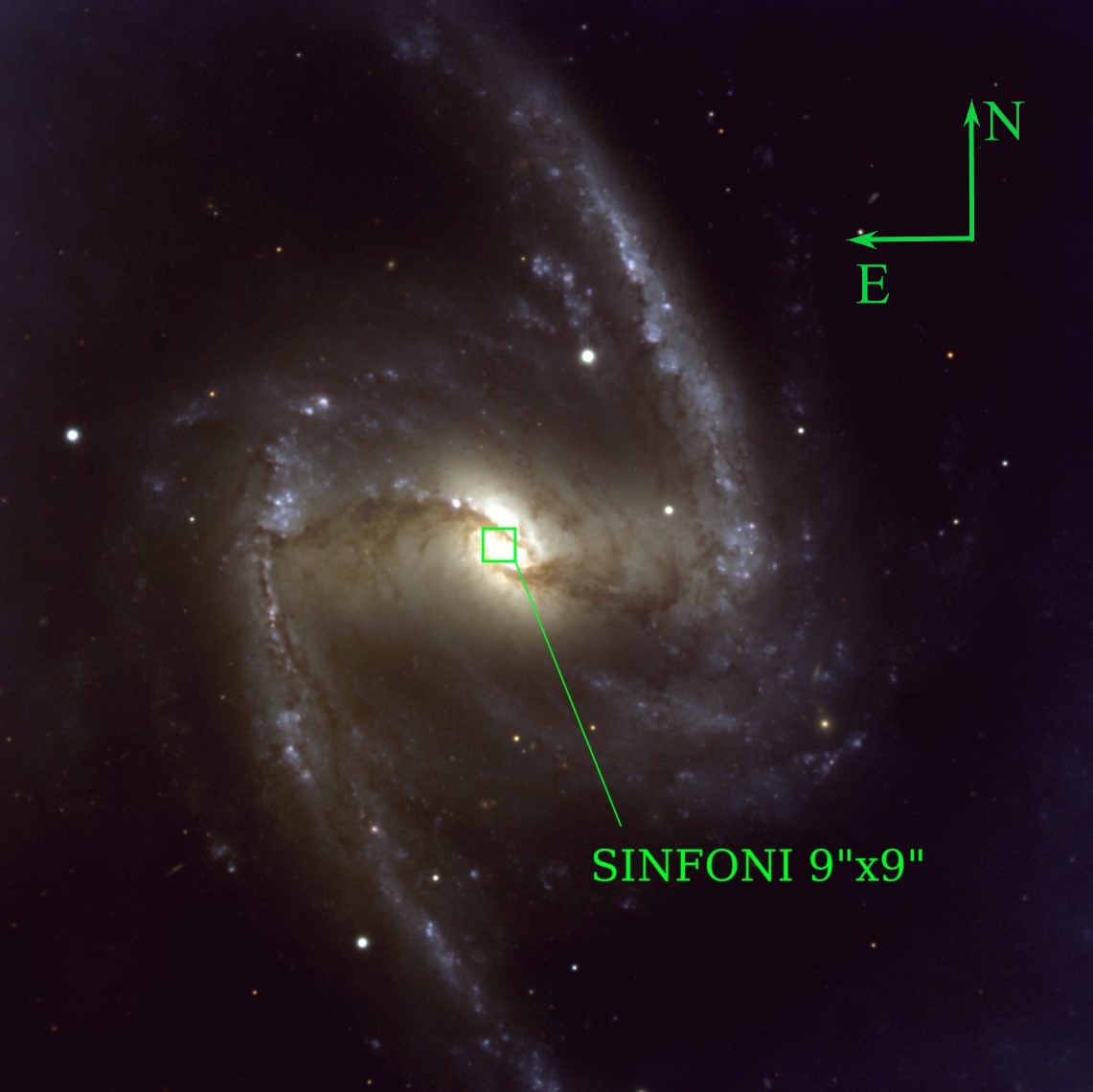}
\caption{A three-color optical image of NGC 1365 combined from three exposures with the FORS1 multi-mode instrument at VLT UT1, in the B (blue), V (green), and R (red) optical bands. The field measures about $7 \times 7$ arcmin$^2$. North is up and east is left. The green box shows the SINFONI field-of-view of $9\times 9$ arcsec$^2$. Credit: ESO.}
\label{fig:NGC1365}
\end{figure}

In this paper, we present NIR integral-field spectroscopy of the NUGA source NGC 1365 (Fig.~\ref{fig:NGC1365}), which is a barred, spiral, and ringed galaxy (SB(s)b) in the Fornax cluster \citep{1980MNRAS.191..685J, 1991S&T....82Q.621D}, located at a redshift of $z\approx 0.005457 $. This grand design spiral galaxy has been intensely investigated \citep[a review of early work can be found in][]{1999A&ARv...9..221L}. The inflow of gas is carried out by the striking long bar spanning about 3\arcmin \citep{1997AJ....114..965R}, which is accompanied by a prominent dust lane to the central 2 kpc. At the transition region between $x_1$- and $x_2$-like gas streamlines \citep[shape of the main families of periodic orbits in a barred galaxy, i.e., $x_1$ orbits are parallel to the bar and $x_2$ orbits are perpendicular and inside the corotation, see e.g.,][]{1980A&A....92...33C, 1989A&ARv...1..261C} in the bar there is an oval-shaped inner Lindblad resonance (ILR) ($r\approx 1$ kpc) \citep{1996A&A...313...65L}. 

This ring was reported early on as strong emitting ``hot spots'' in optical wavelengths \citep{1958PASP...70..364M, 1965PASP...77..287S}. These hot spots trace hot stellar clusters and their corresponding supernova remnants, and have been detected in H$\alpha$ and [\ion{N}{II}] by \cite{1997A&A...328..483K}, in radio by \cite{1995A&A...295..585S}, \cite{1998MNRAS.300..757F}, \cite{1999MNRAS.306..479S}, in mid-IR by \cite{2005A&A...438..803G} and also by \citet[][studied far-IR data]{2012MNRAS.425..311A} who suggest that most of the star formation is obscured by the dust lane crossing the nuclear region. \cite{2007ApJ...654..782S} detect large amounts of molecular gas in the oval shaped ring from their CO observations. The ring is also reported to have strong soft X-ray emissions \citep{2009ApJ...694..718W}. 

The [\ion{O}{III}] emission line observations suggest that this galaxy also has a bi-conical outflow from the nucleus. The cone towards the SE  has stronger emission and was found first \citep[e.g.,][]{1988MNRAS.234..155E,1991MNRAS.250..138S}. The cone towards the NW has fainter emission and is thought to be located under the galactic plane \citep{2003AJ....126.2185V,2010ApJ...711..818S}. As \cite{2018arXiv180901206V} show from high-resolution optical IFS data, the outflow is traced predominantly by high-ionization lines such as [\ion{O}{iii}], while lower ionization lines such as the Balmer lines are tracing the AGN and the star formation regions in the circumnuclear ring.

The driver of this bi-conical outflow is a well-known ``changing-look'' AGN \citep[according to extensive studies of X-ray variations, e.g.,][]{2005ApJ...623L..93R, 2014ApJ...795...87B} and has been classified as Seyfert 1.5 to 2 in the optical \citep[e.g.,][]{1993ApJ...418..653T, 1995ApJ...454...95M, 2006A&A...455..773V}. \cite{1995A&A...295..585S} suggest that the AGN has a weak radio jet towards the SE. However, the presence of this jet has not been confirmed by other authors \citep[e.g.,][]{1999MNRAS.306..479S}. 

This paper concentrates on the NIR emission lines and continuum to study the gas streaming motions and star formation in the circumnuclear region of NGC 1365. It is structured as follows: In Sect.~\ref{sec:obs+datared}, we present the observations and data reduction used for this analysis. In Sect.~\ref{sec:Analysis} we present the SINFONI data cubes, explain details of the emission line fitting procedure, and present the emission line maps and stellar and gaseous kinematics. These maps are then further discussed and compared to previous results from the literature in Sect.~\ref{sec:discussion}. A short summary and conclusions of our results are presented in Sect.~\ref{sec:conclusions}.

Throughout the paper, we adopt a luminosity distance of $D_L = (18.1 \pm 0.5)\,\mathrm{Mpc}$, measured with the Tip of the Red Giant Branch (TRGB) method by \citet{2018ApJ...852...60J}, which corresponds to a scale of $87\,\mathrm{pc}\,\mathrm{arcsec}^{-1}$.

\section{Observations and data reduction}
\label{sec:obs+datared}

NGC 1365 was observed on October 7 2014 with the integral-field spectrograph SINFONI mounted at the Unit Telescope 4 (UT4, Yepun) of the ESO Very Large Telescope in Chile \citep[VLT,][]{2003SPIE.4841.1548E,2004Msngr.117...17B}.

The observations were taken in seeing-limited mode, the field-of-view (FOV) of single exposures in this mode is 8$\arcsec \times 8\arcsec$ and the spatial sampling is $0\farcs 125 \, \mathrm{pixel}^{-1}$. Using a jitter pattern with offsets $\pm 1\arcsec$ we minimize the effect of bad pixels and increase the FOV of the combined cube to $9\arcsec \times 9\arcsec$, which corresponds to a linear scale of 785 pc. The gratings used are in the $H+K$-band ($1.45-2.45 \,\mm$) with a spectral resolution of $R\approx 1500$ and $K$-band with a resolution of $R\approx 4000$. We spent an integration time of $150\,\mathrm{s}$ per exposure in a TST nodding sequence (T: target, S: sky). This leads to an overall on-source integration time of $1500\,\mathrm{s}$ ($10\times 150\,\mathrm{s}$) for the $H+K$-grating and $3000\,\mathrm{s}$ ($20\times 150\,\mathrm{s}$) for the $K$-band grating, as well as an additional $750\,\mathrm{s}$ in $H+K$-band and $1500\,\mathrm{s}$ in $K$-band on-sky.

We used the pipeline delivered by ESO to reduce the data up to single-exposure-cube reconstructions. First we corrected for some detector pattern features imprinted in the SINFONI data using the method developed by \citet{2014A&A...567A.119S}. For alignment, final coaddition, and telluric correction, we use our own \textsc{Python} and \textsc{Idl} routines.  

The G2V star HIP 039102 was observed in $H+K$ and $K$-bands in between the science target observations. Two object sky pairs for each band were taken, with an integration time of $1\,\mathrm{s}$. The  spectrum of this star was used to correct for atmospheric telluric absorption and for flux calibration. Prior to using the spectrum we eliminated the intrinsic spectral features and blackbody shape of the standard star. This procedure was done using a high-resolution solar spectrum \citep{1996AJ....111..537M}, which was convolved with a Gaussian to match the resolution of SINFONI. A blackbody function with temperature $T=5800\,\mathrm{K}$ was used for the spectral region between the two bands and at the spectral edges to extrapolate the solar spectrum. 

To perform flux calibration, we extracted a spectrum from the telluric standard star with an aperture of size $3 \times \mathrm{FWHM}_\mathrm{PSF}$ \citep{2000hccd.book.....H}. The FWHM of the point-spread-function (PSF), measured by fitting a 2D Gaussian to the telluric star, is $0\farcs 5$. The calibration for standard star counts was done at $\lambda 2.1 \mm$ for both $H+K$ and $K$-bands, to the flux taken from the 2MASS All-sky Point Source Catalog \citep{2006AJ....131.1163S}. 

We show an optical image of NGC1365 in Fig.~\ref{fig:NGC1365}, where the large-scale structure, in particular the large-scale bar and spirals, become apparent. The green box denotes the $9\arcsec \times 9\arcsec$ FOV of the SINFONI observations that we analyze in the following.

\section{Results and analysis}
\label{sec:Analysis}

\begin{figure}
\centering
\includegraphics[width=\columnwidth]{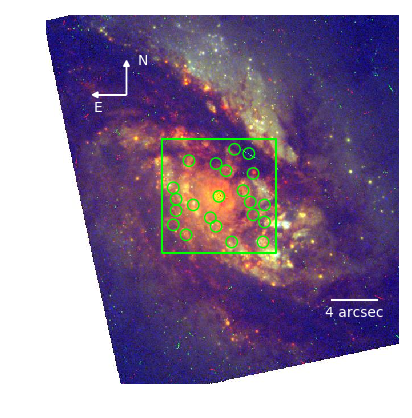}
\caption{HST composite image of the central region of NGC 1365 (red: F814W, green: F555W, blue: F336W). The box indicates the FOV of SINFONI, the circles the apertures that we analyze in detail.}
\label{fig:HST-comp}
\end{figure}

The SINFONI FOV and apertures that we analyze in this paper are marked in green over the optical image (Fig. \ref{fig:HST-comp}). Here the circumnuclear star formation regions and the prominent dust lanes at the inside of the large-scale spiral arms are illustrated. The image is taken with the \emph{Wide Field and Planetary Camera 2} of the \emph{Hubble Space Telescope}, where red corresponds to the F814W, green to the F555W, and blue to the F336W filter.\footnote{Retrieved from the Hubble Legacy Archive (\url{http://hla.stsci.edu/}). The observations took place in Jan 1995 (proposal ID 5222, PI: John Trauger).}.

\begin{figure}
\centering
\includegraphics[width=\columnwidth]{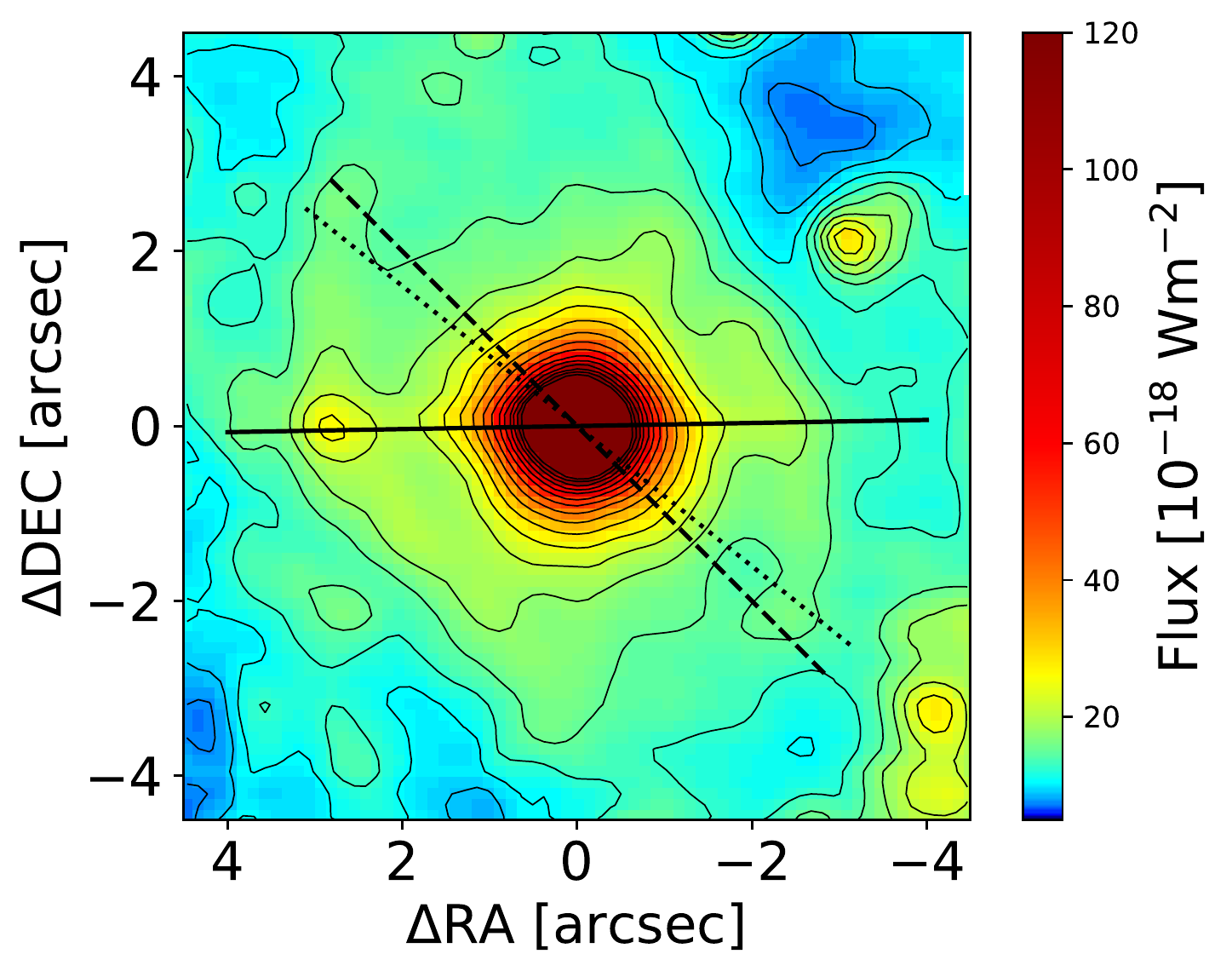}
\caption{$K$-band continuum image extracted from the SINFONI data cube by taking the median in a line-free region between $2.15\mm$ and $2.16\mm$. The solid line indicates the position angle of the large-scale bar ($91^\circ$), the dashed line the PA of the secondary bar/nuclear disk \citep[$45^\circ$,][]{2001A&A...368...52E}, and the dotted line the PA of the line-of-nodes of the stellar kinematics (Sect.~\ref{sec:stellarkinematics}).}
\label{fig:kbandcont}
\end{figure}

We obtained a $K$-band continuum map of the inner $9\arcsec$ radius region of NGC1365 from the SINFONI cube by taking the median of the spectrum between $2.15\mm$ and $2.16\mm$. In this spectral region there are no strong emission or absorption lines (Fig.~\ref{fig:kbandcont}). The strongest feature of the map is the bright nuclear point source component, which is emitted from hot dust heated by the type-1 active nucleus (see Sect.~\ref{sec:AGN}). The extended emission shows an elongation from NE to SW in the map. This structure was first referred to as a secondary bar with an extent of $\sim 10\arcsec$ by \cite{1997A&AS..125..479J}. However, \citet{2001A&A...368...52E}, studying the circumnuclear kinematics, interpret this feature to be a nuclear disk whose ellipticity is solely due to inclination (the dashed line in Fig. \ref{fig:kbandcont} shows the PA $\sim 45^{\circ}$ of the disk). This disk is surrounded by a starburst ring. Some of the off-nuclear emission flux peaks in the map are from clusters which belong to this ring.

\subsection{Emission line fitting}
\label{sec:line_fitting}
\begin{figure*}
\includegraphics[width=\linewidth]{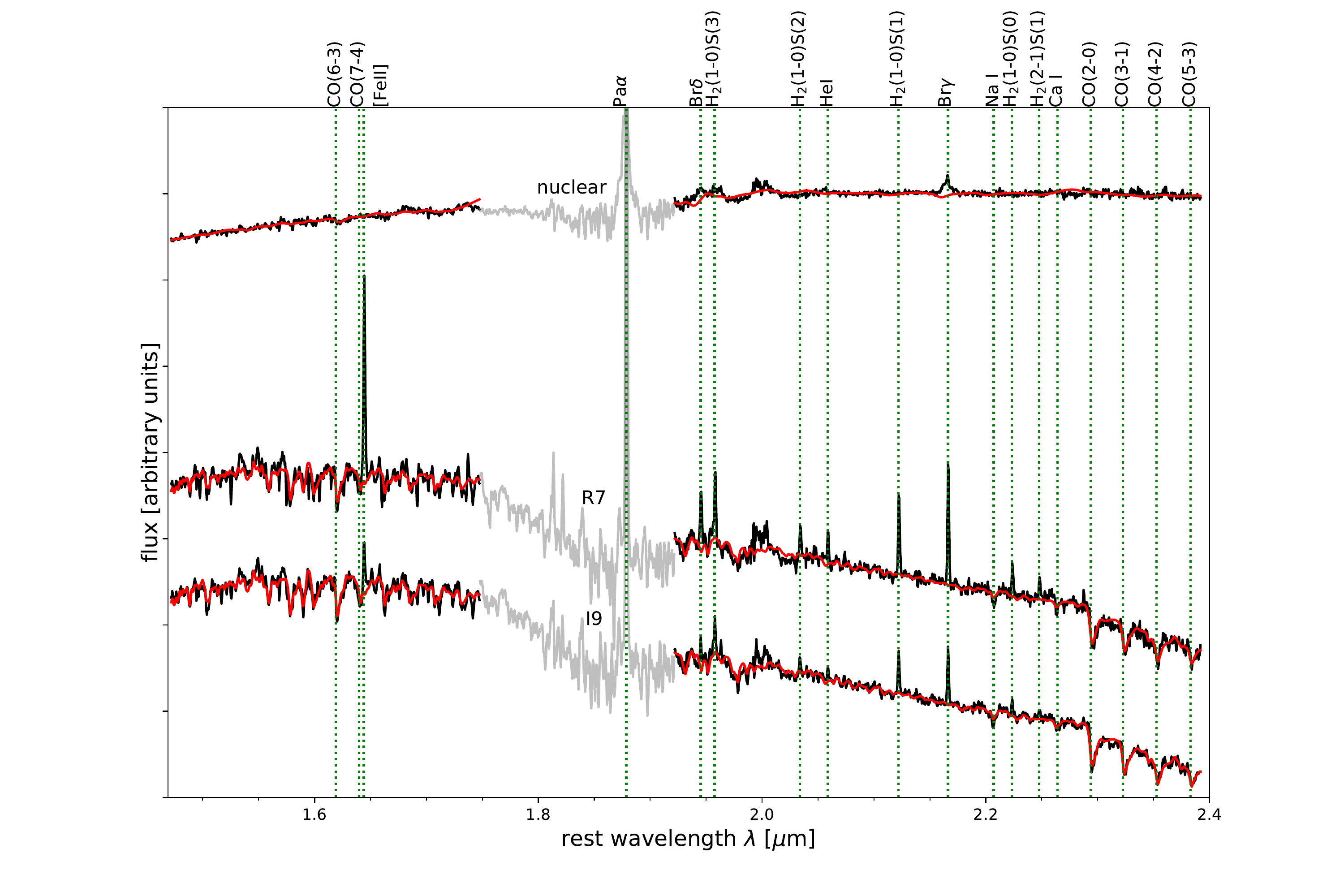}
\caption{Representative spectra ($H$- and $K$-bands in black and between the two bands in gray) from the nuclear, R7, and I9 apertures, as indicated in the flux maps (Fig. \ref{fig:line_fluxes}). Several prominent emission and absorption lines are indicated with green dotted lines, the stellar continuum fit is plotted in red. The region between the bands has higher uncertainties in telluric correction (Sect. \ref{sec:obs+datared}).}
\label{fig:regions_specs}
\end{figure*}

We defined several apertures within the FOV that we analyze in more detail, including the nuclear aperture ("N"), eleven apertures showing parts of the ILR region ("R1" - "R11"; $r\approx 1$ kpc), and ten apertures of star forming regions between the ILR and the nucleus ("I1" - "I10"; $r\approx 0.3$ kpc). All the apertures have a radius of $0\farcs5$ (matching the PSF size). The apertures are marked in Figs.~\ref{fig:HST-comp} and \ref{fig:line_fluxes}. The $H+K$ spectra of the three example apertures (nuclear, R7 and I9) are presented in Fig. \ref{fig:regions_specs}. The spectra of all the other apertures are shown in Figs.~\ref{fig:spectra_subt_R} and \ref{fig:spectra_subt_I} in the Appendix.

We identified several strong emission and absorption lines in the $H+K$-band spectra of NGC 1365. The most prominent emission lines are: the hydrogen recombination lines Pa$\alpha$ and Br$\gamma$, which trace fully ionized gas regions, several molecular hydrogen (H$_{2}$) rotational-vibrational lines, and forbidden [\ion{Fe}{ii}] lines, which originate in partially ionized regions. Furthermore, stellar absorption features, for example  CO(2-0), CO(3-1), CO(2-4), are detected at the red end of the $K$-band spectra.

The emission line fluxes are measured after subtraction of a stellar continuum from each of the selected apertures. For this purpose we used the \textsc{Python} implementation of the Penalized Pixel-Fitting method\footnote{\url{www-astro.physics.ox.ac.uk/~mxc/software/}\linebreak {\small\texttt{PPXF}} version (V6.7.8).} \citep[{\small\texttt{PPXF}},][]{2004PASP..116..138C,2017MNRAS.466..798C} with default parameters. We employed a set of synthetic model spectra from \citet{2007A&A...468..205L}, which are broadened by convolving with a Gaussian kernel to match the spectral resolution of SINFONI in $H+K$-band. The stellar model spectra have assumed solar abundances, and cover an effective temperature range $T_\mathrm{eff} = 2900 - 5900\,\mathrm{K}$, gravities of $\log(g/\mathrm{cm}\,\mathrm{s}^{-2}) = [-1,-0.5,0,+1]$ and masses $1\,M_\odot$ and $15\,M_\odot$. The stellar continuum fits in the analyzed apertures are shown in red in Fig.~\ref{fig:regions_specs}. 

After stellar continuum subtraction, all lines, except in the nuclear aperture, are modeled with single Gaussian functions. In order to increase the reliability of the fits, we decided to fix the emission line width of the Pa$\alpha$, Br$\delta,$ and \ion{He}{i} to that found in the Br$\gamma$ line, assuming that they originate in the same region. Similarly, for the H$_2$ emission lines, we chose to fix the width to that found in the strongest line, H$_2$(1-0)S(1)$\lambda 2.122\mm$. For all fits, unless stated otherwise, we use the Levenberg-Marquardt algorithm included in the \textsc{Python} implementation of \textsc{LMFIT} \citep{2009ASPC..411..251M,newville_2014_11813}. The results of the line fits in the apertures, corrected for extinction (see Sect.~\ref{sec:extinction}) together with their $1 \sigma$ standard errors, which are derived from the covariance matrix, are presented in Tables \ref{tab:emissionlines1} and \ref{tab:emissionlines2}, as well as in Tables \ref{tab:h2lines1} and \ref{tab:h2lines2} for the H$_2$ lines. 

\begin{table*}
\caption{Emission line fluxes in the apertures in the circumnuclear ring and derived quantities, visual extinction and ionized gas mass.}
\label{tab:emissionlines1}
\centering
\begin{tabular}{cccccccc}
\hline
\hline 
Aperture & [\ion{Fe}{ii}] & Pa$\alpha$ & Br$\delta$ & \ion{He}{i} & Br$\gamma$ & A$_V$ & \ion{H}{ii} mass \\
 & $\lambda 1.644\mm$ & $\lambda 1.876\mm$ & $\lambda 1.945\mm$ & $\lambda 2.059\mm$ & $\lambda 2.166\mm$ & [mag] & [$10^4\,M_\odot$] \\
\hline
R1 & $1.46 \pm 0.23$ & $13.67 \pm 1.44$ & $0.79 \pm 0.09$ & $0.39 \pm 0.05$ & $1.12 \pm 0.09$ & $6.9 \pm 0.8$ & $10.6$ \\  
R2 & $1.31 \pm 0.14$ & $17.07 \pm 1.10$ & $0.95 \pm 0.06$ & $0.50 \pm 0.04$ & $1.40 \pm 0.06$ & $8.6 \pm 0.5$ & $13.3$ \\                      
R3 & $0.40 \pm 0.04$ & $6.51 \pm 0.11$ & $0.28 \pm 0.03$ & $0.18 \pm 0.02$ & $0.51 \pm 0.02$ & $- $ & $4.8$ \\                        
R4 & $0.51 \pm 0.03$ & $8.69 \pm 0.12$ & $0.41 \pm 0.02$ & $0.17 \pm 0.02$ & $0.67 \pm 0.02$ & $-$ & $6.4$ \\                        
R5 & $1.87 \pm 0.26$ & $18.77 \pm 1.75$ & $1.03 \pm 0.11$ & $0.46 \pm 0.05$ & $1.54 \pm 0.10$ & $6.0 \pm 0.7$ & $14.6$ \\                      
R6 & $5.65 \pm 0.98$ & $24.18 \pm 2.87$ & $1.26 \pm 0.18$ & $0.71 \pm 0.09$ & $1.98 \pm 0.17$ & $9.0 \pm 0.9$ & $18.8$ \\                      
R7 & $4.32 \pm 0.54$ & $14.70 \pm 1.36$ & $0.70 \pm 0.07$ & $0.33 \pm 0.04$ & $1.20 \pm 0.08$ & $7.0 \pm 0.7$ & $11.4$ \\                      
R8 & $1.21 \pm 0.15$ & $13.45 \pm 1.11$ & $0.64 \pm 0.05$ & $0.51 \pm 0.04$ & $1.10 \pm 0.06$ & $4.0 \pm 0.6$ & $10.5$ \\                      
R9 & $3.04 \pm 0.37$ & $18.37 \pm 1.61$ & $1.06 \pm 0.10$ & $0.58 \pm 0.06$ & $1.50 \pm 0.09$ & $5.8 \pm 0.6$ & $14.3$ \\                      
R10 & $1.90 \pm 0.25$ & $15.50 \pm 1.46$ & $0.83 \pm 0.08$ & $0.44 \pm 0.05$ & $1.27 \pm 0.08$ & $4.7 \pm 0.7$ & $12.1$ \\                     
R11 & $0.87 \pm 0.05$ & $11.15 \pm 0.14$ & $0.53 \pm 0.03$ & $0.30 \pm 0.03$ & $0.90 \pm 0.03$ & $-$ & $8.5$ \\  
\hline
\end{tabular}
\tablefoot{Emission lines are corrected for extinction and given in units of $10^{-18}$ Wm$^{-2}$. The apertures all have a radius of 0\farcs5.}
\end{table*}

\begin{table*}
\caption{As in Table \ref{tab:emissionlines1}, but for apertures within the circumnuclear ring.}
\label{tab:emissionlines2}
\centering
\begin{tabular}{cccccccc}
\hline
\hline
Aperture & [\ion{Fe}{ii}] & Pa$\alpha$ & Br$\delta$ & \ion{He}{i} & Br$\gamma$ & A$_V$ & \ion{H}{ii} mass \\
 & $\lambda 1.644\mm$ & $\lambda 1.876\mm$ & $\lambda 1.945\mm$ & $\lambda 2.059\mm$ & $\lambda 2.166\mm$ & [mag] & [$10^4\,M_\odot$] \\
\hline
I1 & $0.38 \pm 0.11$ & $3.95 \pm 0.74$ & $0.15 \pm 0.04$ & $0.09 \pm 0.02$ & $0.32 \pm 0.04$ & $2.9 \pm 1.4$ & $3.1$ \\                                     
I2 & $0.99 \pm 0.30$ & $6.17 \pm 1.17$ & $0.28 \pm 0.07$ & $0.12 \pm 0.04$ & $0.51 \pm 0.07$ & $7.2 \pm 1.4$ & $4.8$ \\     
I3 & $0.24 \pm 0.05$ & $3.88 \pm 0.10$ & $0.13 \pm 0.03$ & $0.09 \pm 0.02$ & $0.29 \pm 0.02$ & $-$ & $2.8$ \\
I4 & $0.24 \pm 0.04$ & $4.44 \pm 0.08$ & $0.17 \pm 0.02$ & $0.08 \pm 0.02$ & $0.31 \pm 0.02$ & $-$ & $2.9$ \\ 
I5 & $0.34 \pm 0.03$ & $5.00 \pm 0.09$ & $0.22 \pm 0.02$ & $0.11 \pm 0.02$ & $0.39 \pm 0.02$ & $-$ & $3.7$ \\ 
I6 & $0.24 \pm 0.05$ & $2.50 \pm 0.10$ & $0.04 \pm 0.02$ & $0.07 \pm 0.02$ & $0.20 \pm 0.02$ & $-$ & $1.9$ \\
I7 & $0.50 \pm 0.04$ & $3.45 \pm 0.09$ & $0.12 \pm 0.02$ & $0.06 \pm 0.02$ & $0.26 \pm 0.02$ & $-$ & $2.4$ \\
I8 & $0.34 \pm 0.09$ & $3.88 \pm 0.63$ & $0.16 \pm 0.04$ & $0.08 \pm 0.02$ & $0.32 \pm 0.04$ & $0.6 \pm 1.2$ & $3.0$ \\  
I9 & $1.98 \pm 0.41$ & $11.23 \pm 1.56$ & $0.56 \pm 0.10$ & $0.25 \pm 0.05$ & $0.92 \pm 0.09$ & $7.5 \pm 1.0$ & $8.7$ \\
I10 & $0.22 \pm 0.03$ & $3.27 \pm 0.08$ & $0.16 \pm 0.02$ & $0.07 \pm 0.02$ & $0.27 \pm 0.02$ & $-$ & $2.5$ \\  
\hline
\end{tabular}
\tablefoot{Emission lines are corrected for extinction and given in units of $10^{-18}$ Wm$^{-2}$. The apertures all have a radius of 0\farcs5.}
\end{table*}

\begin{table*}
\caption{Molecular hydrogen emission lines in the apertures in the circumnuclear ring, and derived quantities hot and cold molecular gas mass.}
\label{tab:h2lines1}
\centering
\begin{tabular}{cccccccc}
\hline
\hline 
Aperture & H$_2$(1-0)S(3) & H$_2$(1-0)S(2) & H$_2$(1-0)S(1) & H$_2$(1-0)S(0) & H$_2$(2-1)S(1) & hot H$_2$ mass & cold gas mass \\
 & $\lambda 1.958\mm$ & $\lambda 2.034\mm$ & $\lambda 2.122\mm$ & $\lambda 2.223\mm$ & $\lambda 2.248\mm$ & [$M_\odot$] & [$10^6\,M_\odot$] \\
\hline
R1 & $53.0 \pm 7.7$ & $24.9 \pm 6.6$ & $56.6 \pm 5.1$ & $19.7 \pm 3.1$ & $13.3 \pm 2.6$ & $9$ & $3 - 15$ \\
R2 & $52.4 \pm 4.4$ & $25.9 \pm 3.5$ & $66.1 \pm 4.2$ & $27.6 \pm 2.6$ & $14.1 \pm 2.2$ & $11$ & $3 - 18$ \\
R3 & $24.2 \pm 2.3$ & $10.1 \pm 1.5$ & $28.1 \pm 2.0$ & $11.2 \pm 1.8$ & $5.7 \pm 1.5$ & $5$ & $1 - 7$ \\
R4 & $22.4 \pm 2.0$ & $10.2 \pm 2.1$ & $27.1 \pm 1.6$ & $9.9 \pm 1.7$ & $5.6 \pm 1.8$ & $4$ & $1 - 7$ \\
R5 & $63.2 \pm 9.8$ & $21.7 \pm 5.2$ & $63.8 \pm 6.9$ & $18.3 \pm 4.3$ & $11.2 \pm 4.3$ & $11$ & $3 - 17$ \\
R6 & $119.4 \pm 16.5$ & $42.6 \pm 7.0$ & $102.2 \pm 11.2$ & $31.5 \pm 6.0$ & $22.7 \pm 4.8$ & $17$ & $5 - 27$ \\
R7 & $77.1 \pm 7.7$ & $38.5 \pm 4.3$ & $86.9 \pm 6.3$ & $30.8 \pm 2.9$ & $18.5 \pm 2.5$ & $14$ & $4 - 23$ \\
R8 & $52.8 \pm 5.2$ & $23.3 \pm 3.8$ & $57.2 \pm 3.7$ & $19.6 \pm 2.3$ & $9.4 \pm 2.0$ & $10$ & $3 - 15$ \\
R9 & $97.4 \pm 8.5$ & $47.5 \pm 5.1$ & $109.3 \pm 7.3$ & $36.5 \pm 3.0$ & $21.3 \pm 3.0$ & $18$ & $5 - 29$ \\
R10 & $78.5 \pm 7.4$ & $32.8 \pm 4.5$ & $90.2 \pm 6.7$ & $27.4 \pm 2.7$ & $15.3 \pm 2.3$ & $15$ & $5 - 24$ \\
R11 & $30.3 \pm 2.6$ & $17.4 \pm 3.3$ & $37.6 \pm 2.1$ & $12.0 \pm 1.5$ & $8.8 \pm 1.6$ & $6$ & $2 - 10$ \\
\hline
\end{tabular}
\tablefoot{Emission lines are corrected for extinction and given in units of $10^{-20}$ Wm$^{-2}$. The apertures all have a radius of 0\farcs5.}
\end{table*}

\begin{table*}
\caption{As in Table \ref{tab:h2lines1},  but for apertures within the circumnuclear ring.}
\label{tab:h2lines2}
\centering
\begin{tabular}{cccccccc}
\hline
\hline 
Aperture & H$_2$(1-0)S(3) & H$_2$(1-0)S(2) & H$_2$(1-0)S(1) & H$_2$(1-0)S(0) & H$_2$(2-1)S(1) & hot H$_2$ mass & cold gas mass \\
 & $\lambda 1.958\mm$ & $\lambda 2.034\mm$ & $\lambda 2.122\mm$ & $\lambda 2.223\mm$ & $\lambda 2.248\mm$ & [$M_\odot$] & [$10^6\,M_\odot$] \\
\hline
I1 & $36.9 \pm 6.6$ & $16.2 \pm 3.2$ & $37.1 \pm 5.2$ & $14.5 \pm 2.1$ & $8.8 \pm 1.6$ & $6$ & $2 - 10$ \\
I2 & $40.9 \pm 8.1$ & $22.4 \pm 4.8$ & $44.7 \pm 6.9$ & $19.3 \pm 3.3$ & $7.6 \pm 2.4$ & $7$ & $2 - 12$ \\
I3 & $24.4 \pm 2.4$ & $9.4 \pm 2.0$ & $22.9 \pm 2.1$ & $10.9 \pm 1.7$ & $5.4 \pm 1.5$ & $4$ & $1 - 6$ \\
I4 & $20.9 \pm 1.7$ & $10.7 \pm 1.7$ & $25.4 \pm 1.7$ & $12.1 \pm 1.5$ & $4.1 \pm 1.5$ & $4$ & $1 - 7$ \\
I5 & $28.0 \pm 2.3$ & $9.7 \pm 1.8$ & $32.3 \pm 1.7$ & $12.8 \pm 1.7$ & $6.5 \pm 1.4$ & $5$ & $2 - 9$ \\
I6 & $19.0 \pm 2.5$ & $11.3 \pm 2.6$ & $24.4 \pm 2.4$ & $11.3 \pm 1.7$ & $5.7 \pm 1.8$ & $4$ & $1 - 7$ \\
I7 & $24.3 \pm 2.0$ & $10.6 \pm 1.4$ & $27.0 \pm 1.9$ & $12.9 \pm 1.5$ & $5.0 \pm 1.3$ & $4$ & $1 - 7$ \\
I8 & $26.2 \pm 5.5$ & $11.4 \pm 2.4$ & $29.0 \pm 3.9$ & $14.1 \pm 2.3$ & $5.9 \pm 1.7$ & $5$ & $1 - 8$ \\
I9 & $54.8 \pm 10.8$ & $31.0 \pm 4.9$ & $70.6 \pm 7.9$ & $23.5 \pm 3.1$ & $12.1 \pm 2.6$ & $12$ & $4 - 19$ \\
I10 & $17.6 \pm 1.8$ & $10.2 \pm 1.4$ & $25.4 \pm 1.8$ & $10.8 \pm 1.2$ & $6.2 \pm 1.3$ & $4$ & $1 - 7$ \\
\hline
\end{tabular}
\tablefoot{Emission lines are corrected for extinction and given in units of $10^{-20}$ Wm$^{-2}$. The apertures all have a radius of 0\farcs5.}
\end{table*}

\subsection{Emission line maps}
\begin{figure*}
\centering
\includegraphics[width=0.33\linewidth]{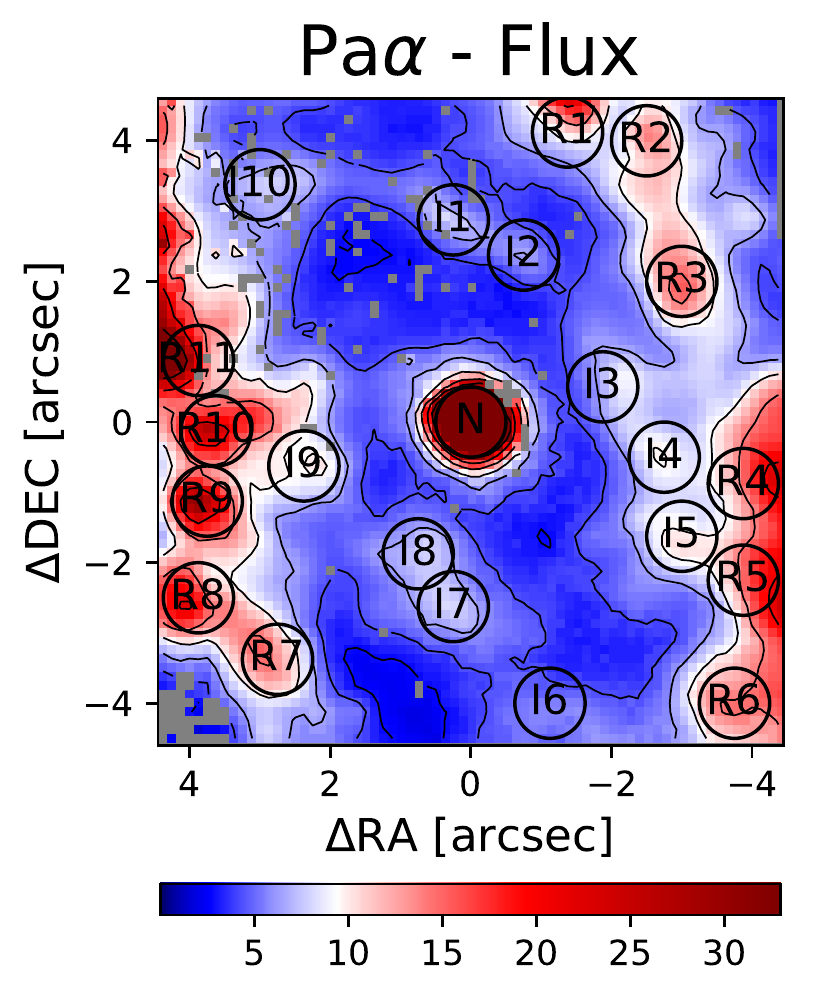}
\includegraphics[width=0.33\linewidth]{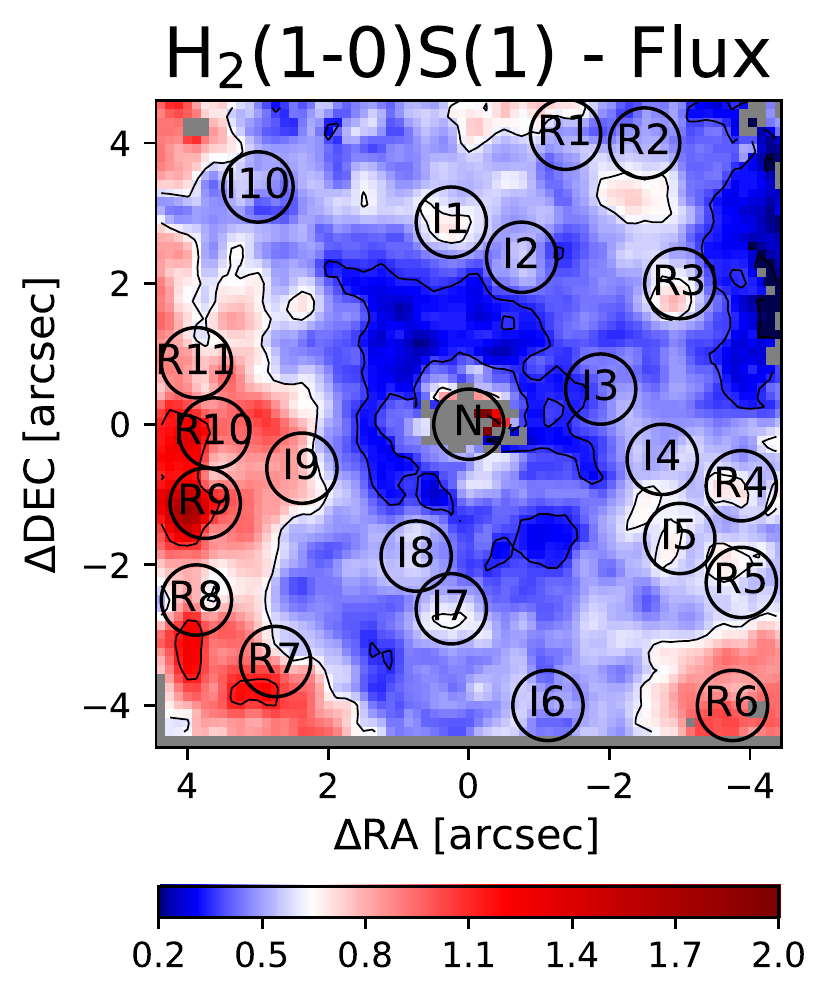}
\includegraphics[width=0.33\linewidth]{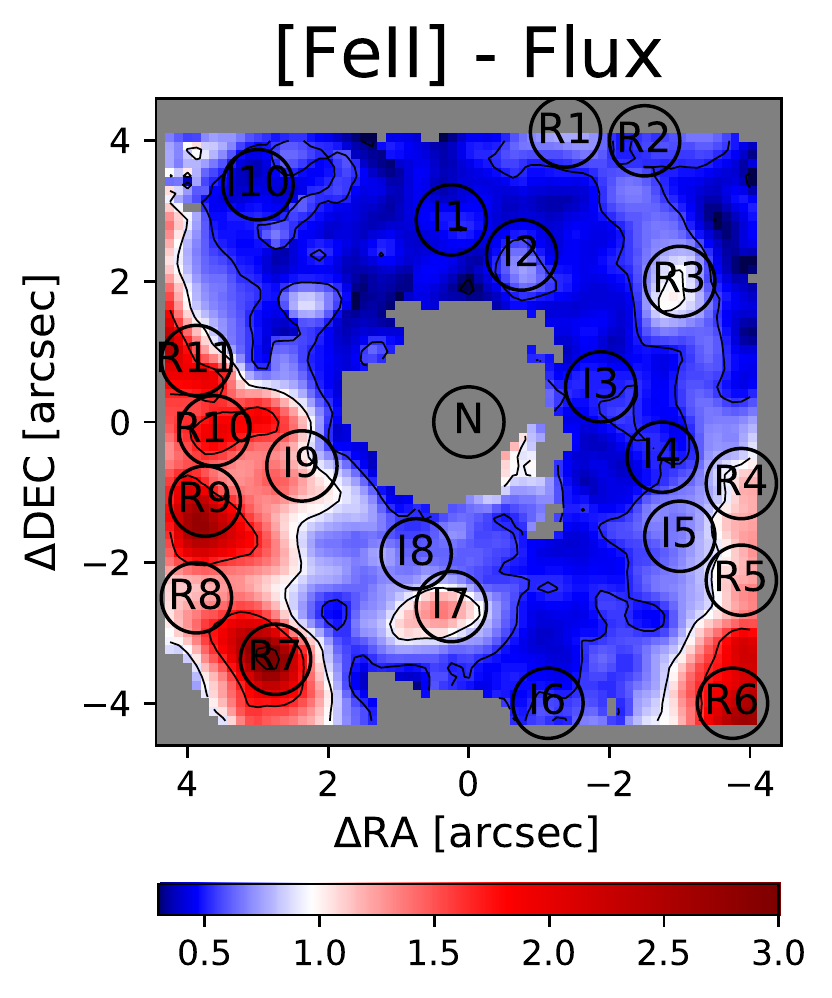}
\caption{Flux maps of the Pa$\alpha$ $\lambda$1.875, H$_{2}$ $\lambda$2.12 $\mm$, [\ion{Fe}{ii}] $\lambda$1.644 $\mm$ emission lines. In all of the maps the flux units are $10^{-20}$ Wm$^{-2}$. The gray pixels show the clipped regions with low S/N, where line fit uncertainties are above 50$\%$.}
\label{fig:line_fluxes}
\end{figure*}

We use the algorithm explained in the previous section to fit the emission lines in the spatial pixels of the IFU cube to get the flux distribution maps of emission lines. In particular the maps for Pa$\alpha$ $\lambda$1.876 $\mm$, [\ion{Fe}{ii}] $\lambda$1.644 $\mm$, and H$_{2}$ $\lambda$2.12 $\mm$ are shown in Fig. \ref{fig:line_fluxes}. Here the gray pixels show where we clipped the data to obtain uncertainties of less than 50 $\%$. In the maps, the nuclear region where the continuum flux peaks is marked as N .

The hydrogen recombination lines show a complex profile in the nuclear region. Since the broad line region is a point source (unresolved), we fixed the width and position of the broad component to that fitted in a central aperture with $1\farcs5$ radius and fit the line using a three-component Gaussian (Fig. \ref{fig:nuclearPaafit}). These additional components are only extended to a PSF size region in the center (with $r\sim 0\farcs.5$), which means out of the central region the line can be fitted with only one component.

The flux distribution maps reveal several off-nuclear hot spots with clumps of enhanced line emission. These hot spots are also bright in the optical (Fig. \ref{fig:HST-comp}) and X-ray \citep[e.g., ][]{2009ApJ...694..718W} images and are mostly referred to as a ring. In the $9\arcsec \times 9\arcsec$ aperture of SINFONI we see part of this ring (Fig. \ref{fig:line_fluxes}).

The observed emission line maps show several variations in distribution. The ionized gas, traced by the hydrogen recombination line Pa$\alpha$, is detected in the whole FOV. However the flux of this line is strongest in the center and in the ILR ($r\sim 1$ kpc), and the emission is symmetric in the east and west side of the ring. We also detect emitting patches from this line (I apertures in Fig. \ref{fig:line_fluxes}) in regions closer to the nucleus. Due to spatial projection effects and limited resolution it is difficult to recognize the underlying structure of these patches, however the form suggests a partial ring or spiral morphology within the ILR (at distances of $r\sim 0.3 \, \mathrm{kpc}$ from the nucleus).

The hot molecular hydrogen emission on the other hand is not detected in the nuclear region. We calculate an upper limit for the H$_2$ line in the center to be $30 \times 10^{-20}$ W s$^{-2}$ , which is less than the measured fluxes in the off-nuclear star forming regions (\cite{2003MNRAS.343..192R} also finds no H$_{2}$ emission in the nuclear region using NIR slit spectroscopy of NGC 1365). The emission line shows less symmetric distribution compared to Pa$\alpha$ maps and is stronger in the E, SE, and SW parts of the ring. From a radius of about 1$\arcsec$ from the center it is observed more or less over the entire FOV. 

We eliminate the $^{12}$CO(7-4) absorption line at $\lambda 1.6397$, which is located very close to the [\ion{Fe}{ii}] prior to producing the flux distribution map of this line. Here the image is smoothed using a boxcar algorithm (to get a higher signal-to-noise ratio (S/N)) and then the stellar continuum is fitted and subtracted (with the procedure explained in Sect. \ref{sec:line_fitting}).
This line emits strongly in the ILR patches, however similar to H$_2$ map it has an asymmetric distribution and shows low emission in the NW of the FOV. There is a prominent dust lane in this region that can extinct the emission lines. However, \citet{2007ApJ...654..782S} also report a CO emission deficiency in this region, meaning the emission can be intrinsically weaker. Similar to Pa$\alpha$ and Br$\gamma$ maps, the [\ion{Fe}{ii}] emission in the E, SE, and SW parts of the ring is stronger. However there are some small differences in the flux level of [\ion{Fe}{ii}] and Pa$\alpha,$ especially in the S and SE of the nucleus, where there are regions with higher emission flux. This can be used to estimate starburst ages and also a possibility of an outflow/inflow. Even with the effort spent to subtract the stellar continuum, the [\ion{Fe}{ii}] line is not detected in the nuclear region.

\subsection{Stellar kinematics}
\label{sec:stellarkinematics}

To study the stellar kinematics, in particular the stellar line-of-sight-velocity (LOSV), we fit the region around the CO band heads between $2.28\mm$ and $2.42\mm$. Since these stellar absorption features show generally a lower S/N than strong emission lines, we need to smooth the data. A boxcar-filter with a width of 3 pixels ensures the necessary S/N to perform reliable fits.

We use the penalized Pixel-Fitting {\small\texttt{PPXF}} method \citep{2004PASP..116..138C,2017MNRAS.466..798C} with default parameters, to fit stellar spectral templates to the CO absorption features. The template library contains 29 late-type stars with spectral classes of F7 to M5 which were observed with the IFU of the Gemini Near-Infrared Spectrograph \citep[GNIRS;][]{2009ApJS..185..186W}. The stars spectra are convolved with a Gaussian kernel to match the spectral resolution of SINFONI in $K$-band. 

\begin{figure}
\centering
\includegraphics[width=\linewidth]{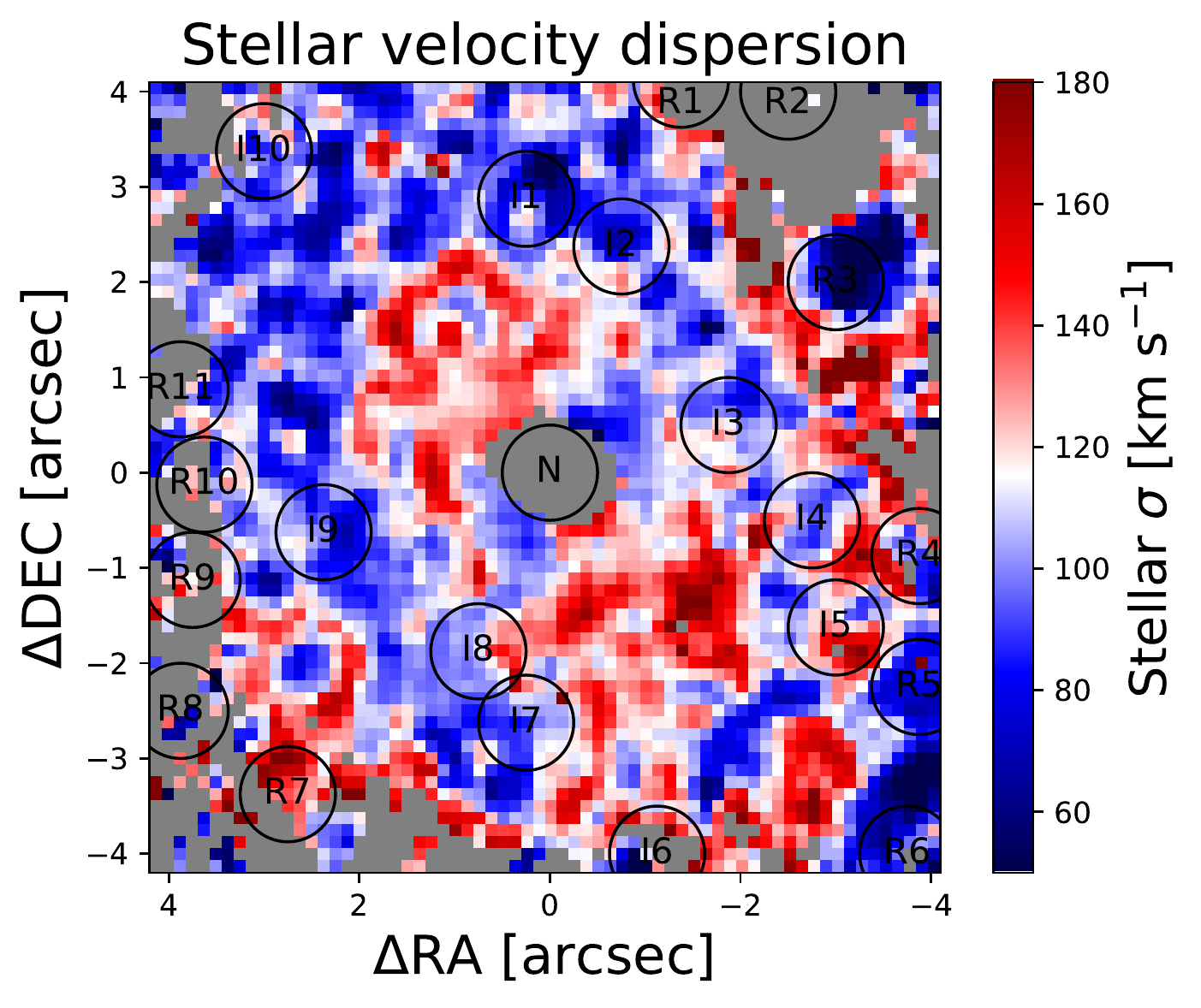}
\caption{Stellar velocity dispersion map derived from CO band heads (around 2.3 $\mm$) fit results. The gray pixels show where we clipped the spectrum due to lower S/N.}
\label{fig:sigmastarmap}
\end{figure}

The stellar velocity dispersion map (Fig. \ref{fig:sigmastarmap}) displays velocity dispersions in the range of 40 to 170 km s$^{-1}$. The map shows a ring-like structure with lower velocity dispersion with values around 80 km s$^{-1}$. These spots mostly belong to regions that we named ``I apertures''. The lower-velocity dispersion could be due to young- to intermediate-age stellar population in these regions \citep[e.g.,][]{2003A&A...409..469W, 2011MNRAS.416..493R,2014MNRAS.438.2036M,2017A&A...598A..55B}.

\begin{figure*}
\centering
\includegraphics[width=0.33\linewidth]{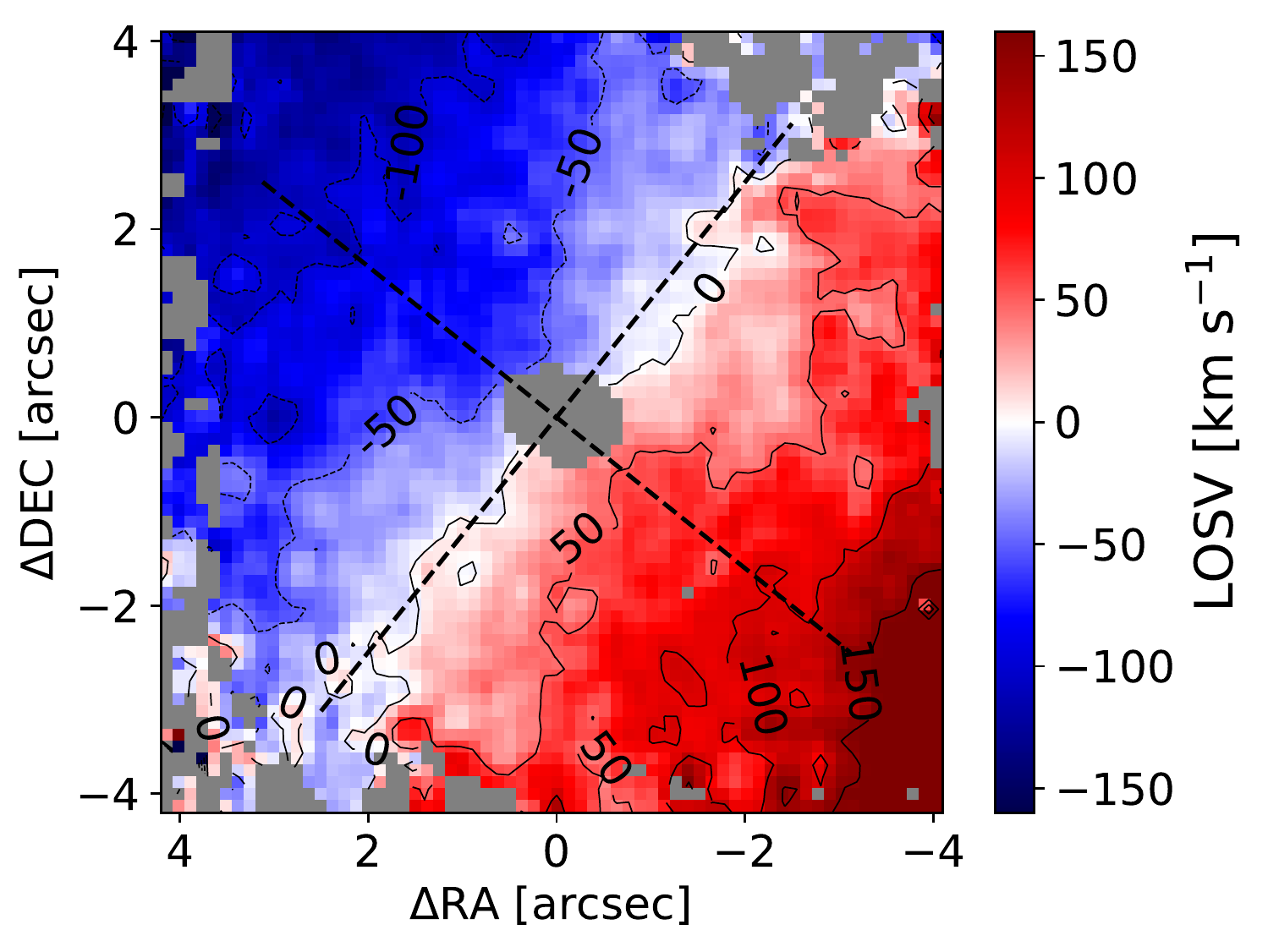}
\includegraphics[width=0.33\linewidth]{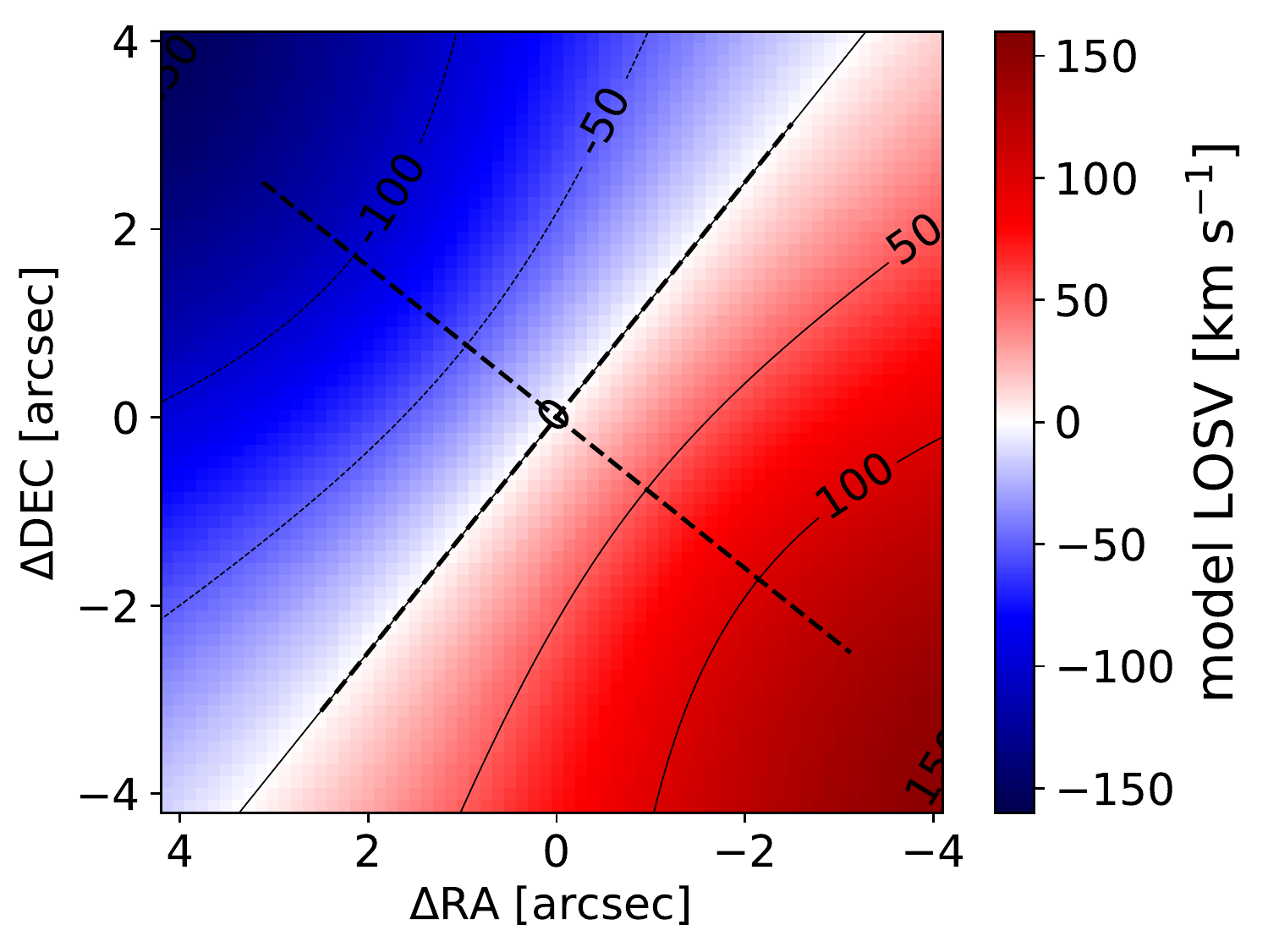}
\includegraphics[width=0.33\linewidth]{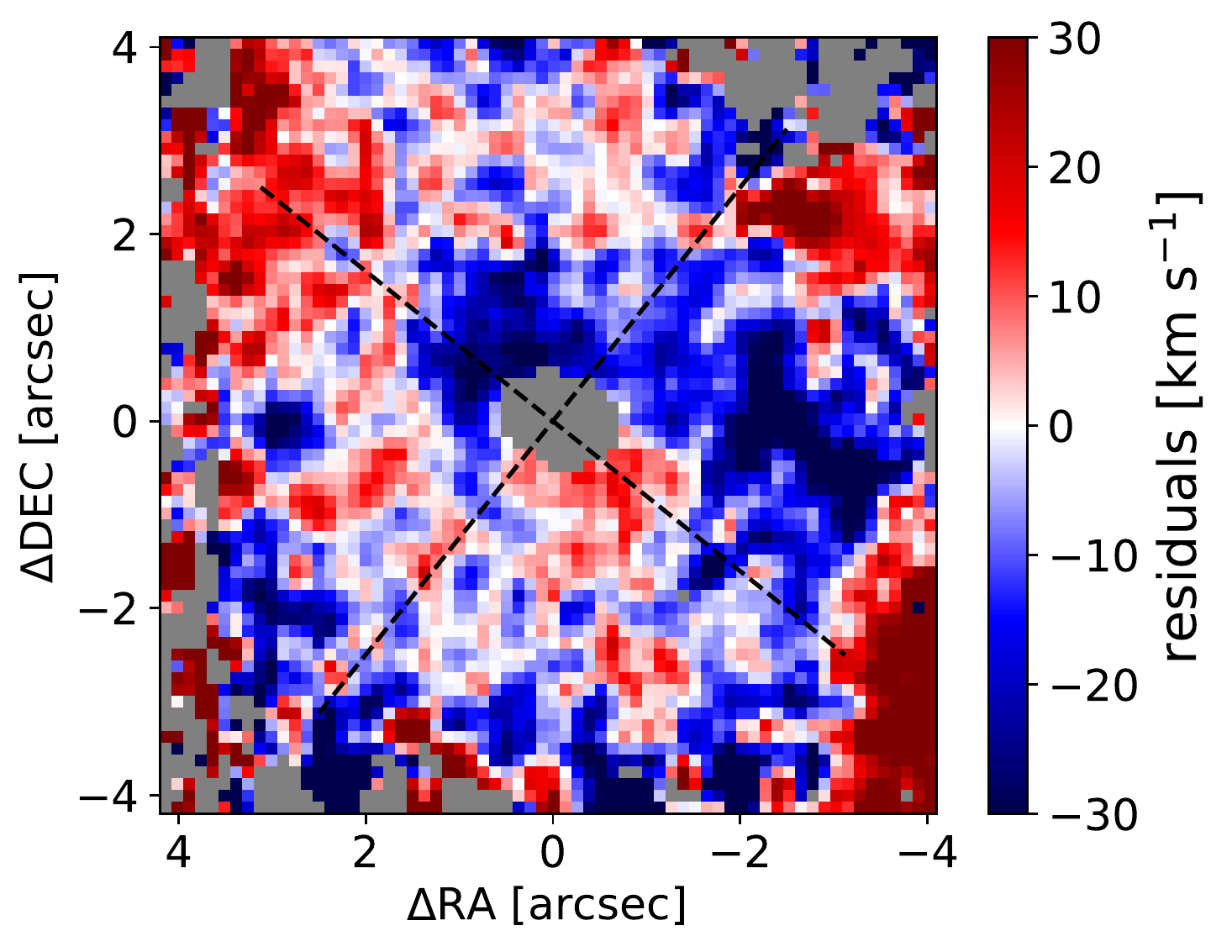}
\caption{\emph{Left:} Stellar line-of-sight velocity obtained by fitting the spectral region around the CO band heads at $2.3\mm$. \emph{Middle:} Rotating disk model fitted to the stellar LOSV. \emph{Right:} Residuals between observation and model. The dashed lines show the zero-velocity line and the line-of-nodes of the model and where they cross indicates the center of the continuum emission. Clipped regions are indicated in gray.}
\label{fig:stell_losv}
\end{figure*}

The stellar LOSV map is shown in the left panel of Fig. \ref{fig:stell_losv}. The gray pixels show areas with lower S/N, that is, at the corners of the FOV, where we have less data (due to jittering) and at the center, where the CO absorption lines are diluted by dust and AGN continuum emission \citep[e.g., ][]{2000AJ....120.2089F}. This map shows a clear rotation structure with redshift to the SW and blueshift to the NE. Slightly irregular isovelocity contours are seen which can be indicative of disturbances in the velocity field, for example by nuclear bars/spiral or ovals. 

In order to obtain kinematical and geometrical information about this rotation field, we fit a rotating disk model, which assumes circular orbits in the plane of the galaxy \citep{1991ApJ...373..369B}:
\begin{equation}
\begin{split}
v_r &= v_s +\\ &\frac{A R \cos(\Psi - \Psi_0) \sin(i) \cos^p(i)}{\left( R^2 \left[\sin^2(\Psi - \Psi_0) + \cos^2(i) \cos^2(\Psi - \Psi_0) \right] + c_0^2  \cos^2(i) \right)^{p/2}},
\end{split}
\label{eq:rotdisc}
\end{equation}
where $v_r$ is the rotation velocity as a function of the polar coordinates $R$ and $\Psi$, $v_s$ is the systemic velocity, $A$ is the amplitude of the rotation field, $\Psi_0$ is the position angle of the line-of-nodes, $i$ is the inclination of the disk, $p$ is a measure for the slope of the rotation field at large radii, and $c_0$ is a concentration parameter. The inclination from the large-scale kinematics has been measured as $i=41^\circ$ and the maximum rotation amplitude as $A\approx 300\,\mathrm{km}\,\mathrm{m}^{-1}$ \citep{1995AJ....110.2037J, 2008ApJ...674..797Z}. Furthermore, we assume an asymptotically flat disk ($p=1$). We fix these values in the fit to have a lower number of free parameters. In addition, we assume that the center of the rotation coincides with the peak of the continuum emission, which we assume to be the location of the galactic nucleus.

The fit yields a systemic velocity of $v_s = 1645 \,\mathrm{km}\,\mathrm{s}^{-1}$, applying the heliocentric corrections we derive {$v_s = 1660 \,\mathrm{km}\,\mathrm{s}^{-1}$} which is higher than 1632 km s$^{-1}$ found by \citet[][]{2008ApJ...674..797Z} and slightly lower than 1671 km s$^{-1}$ reported by \citet{2016MNRAS.459.4485L}. However, our result is consistent with 1657 km s$^{-1}$, which is the average systemic velocity derived using the optical measurements listed on the Hyperleda extragalactic database\footnote{\url{http://leda.univ-lyon1.fr/}}. For the position angle of the line-of-nodes, we derive $(51\pm 1)^\circ$ which is higher than the $40^\circ$ that \cite{2008ApJ...674..797Z} derive from the large-scale H$\alpha$ distribution and lower than the $70^\circ$ that \cite{2016MNRAS.459.4485L} derive from the central [\ion{N}{ii}] distribution. However, this is not unexpected since gas and stellar kinematics can show significant deviations and the small- and large-scale kinematics can differ, especially in strongly barred galaxies such as NGC 1365. For the concentration parameter, we get $(398\pm 5)\,\mathrm{pc}$ which is similar to the value derived by \cite{2016MNRAS.459.4485L} from the [\ion{N}{ii}] emission lines, $3\farcs 6$ ($315\,\mathrm{pc}$, together with an amplitude of $218\,\mathrm{km}\,\mathrm{s}^{-1}$).

The model disk is presented in Fig.~\ref{fig:stell_losv} (middle), together with the residuals in the right panel. The residuals show a patchy structure with amplitudes that have absolute values below $30\,\mathrm{km}\,\mathrm{s}^{-1}$ with acceptable uncertainties in a range of 20$\%$. 

\subsection{Gas kinematics}

Studying the gas kinematics in comparison to the stellar kinematics is essential to get an overall view on the dynamics of the host galaxy.

\begin{figure*}
\centering
\includegraphics[width=0.33\linewidth]{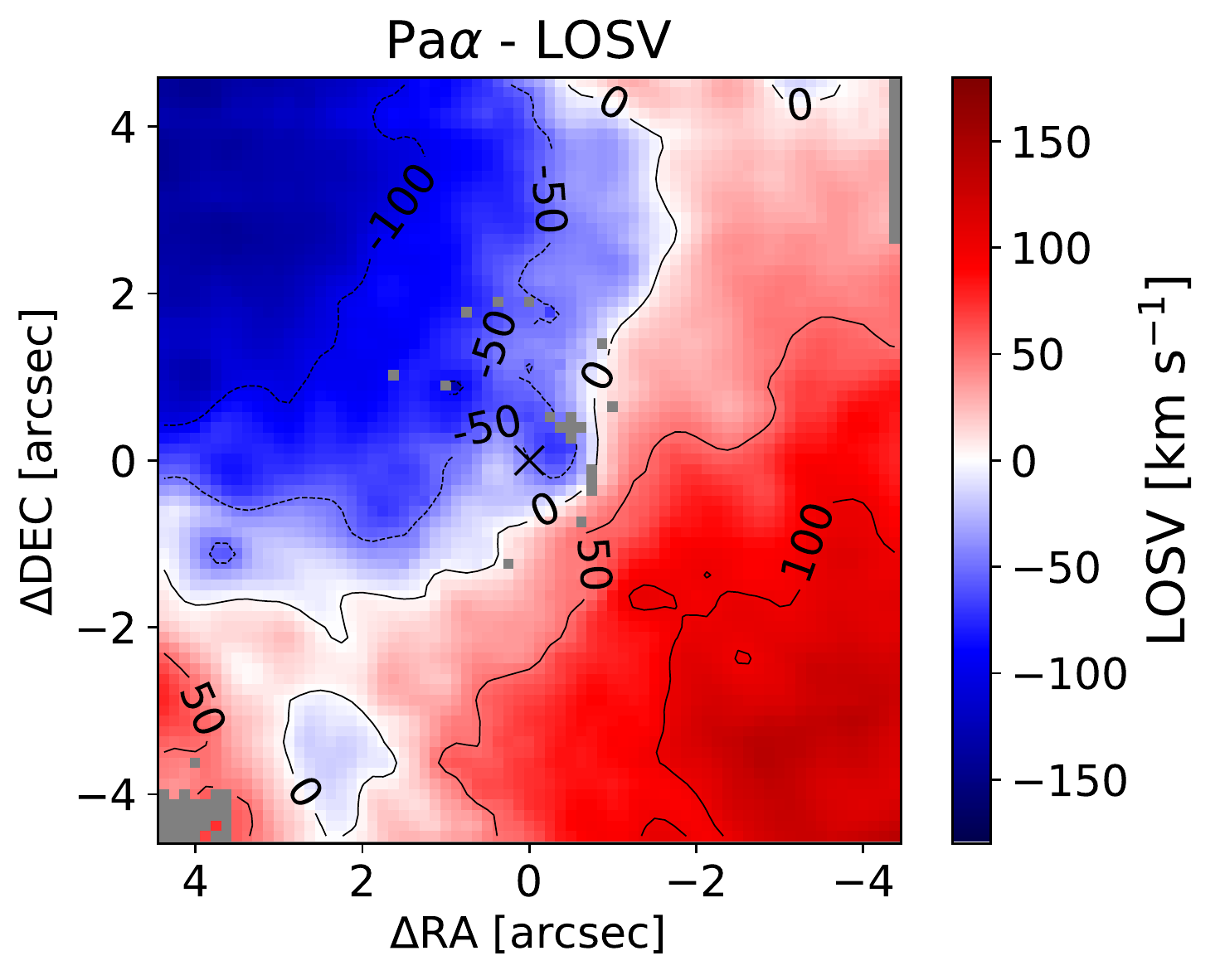}
\includegraphics[width=0.33\linewidth]{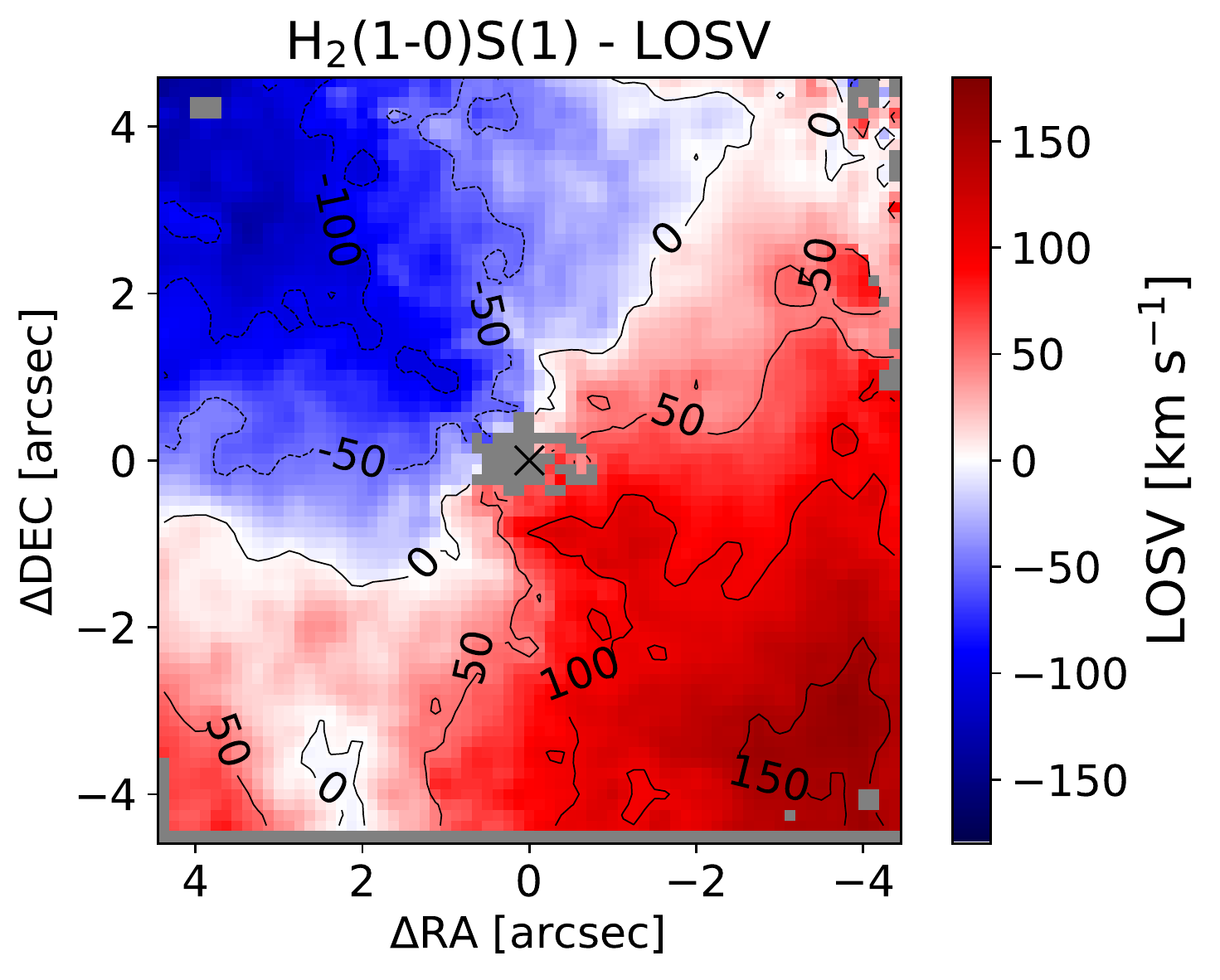}
\includegraphics[width=0.33\linewidth]{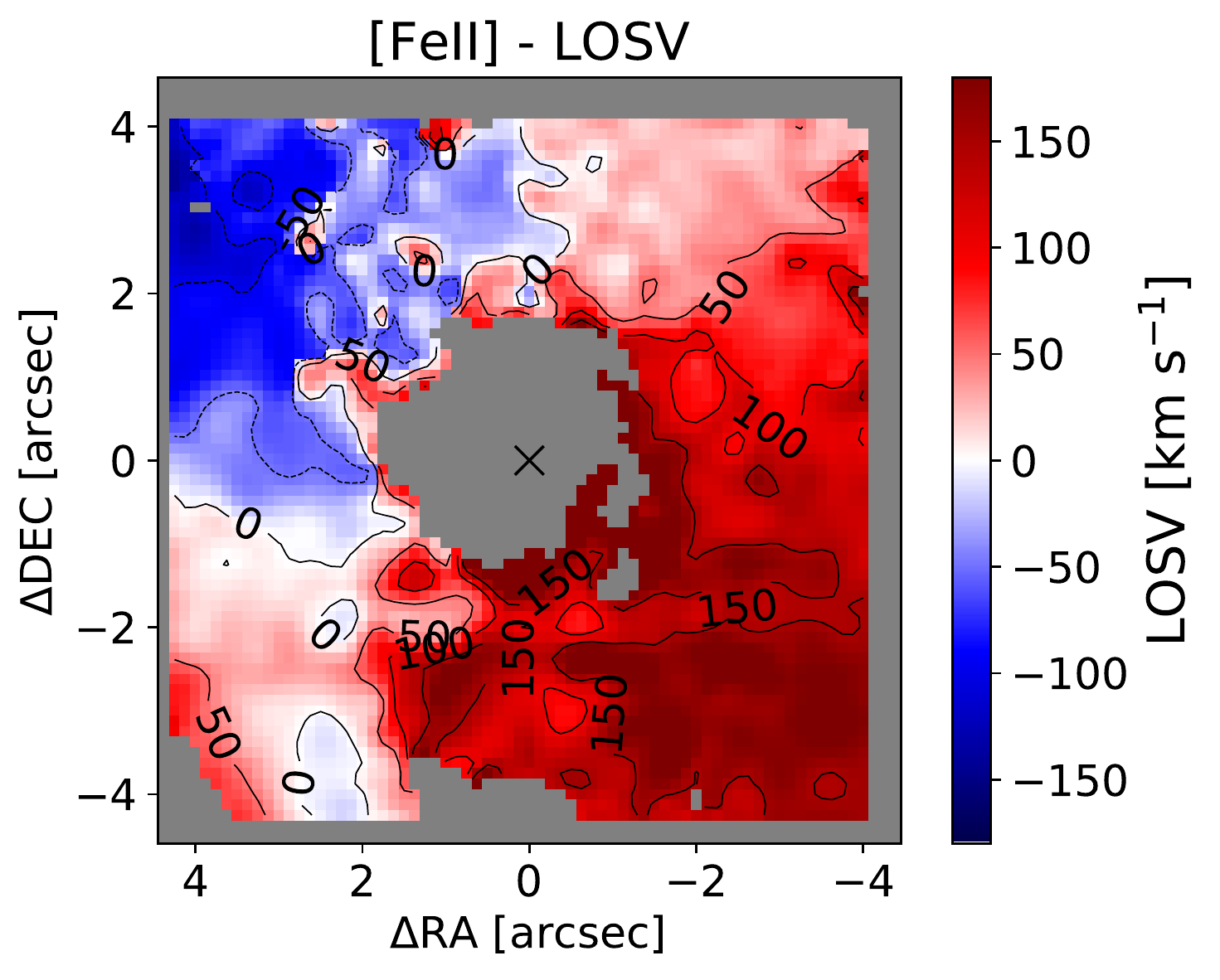}
\caption{\emph{From left to right:} Gas line-of-sight velocities for Pa$\alpha$, H$_2 \lambda 2.122\mm$ and [\ion{Fe}{ii}]$\lambda 1.644\mm$. Clipped regions are indicated in gray. In the maps the nuclear region, where the continuum flux peaks, is marked by a black cross.}
\label{fig:gas_losv}
\end{figure*}

From the fit routine that we use in Sect.~\ref{sec:line_fitting} we get the central positions of the emission-line fits from which we can derive the LOSV. Figure \ref{fig:gas_losv} shows the LOSV fields of Pa$\alpha$, H$_2$, and [\ion{Fe}{ii}]. Here we subtract the systemic velocity 1645 km s$^{-1}$ which was derived from the stellar model fit in Sect.~\ref{sec:stellarkinematics}. The maps for Pa$\alpha$ and [\ion{Fe}{ii}] have been extracted from the $H+K$-band spectra and the H$_2$ has been extracted from the $K$-band spectra since they have a higher spectral resolution (but we compare the map with the one extracted from the $H+K$-band spectra to cross-check). The gray pixels indicate where we clipped the data due to high line-fit errors. Similar to the stellar velocity field maps, the gas shows emission redshifted in the SW and blueshifted in the NE in all maps. However, in contrast to the stellar fields the zero-velocity contours show a disturbed and twisted pattern. Moreover, there is a striking velocity gradient in the SE corner of the FOV. The center is unfortunately absent in the H$_2$ and [\ion{Fe}{ii}] maps, but in the Pa$\alpha$ it shows a blueshift emission of 50 km s$^{-1}$ and has an elongated structure in the NW of the nucleus with higher redshifted emission. Also, in the [\ion{Fe}{ii}] velocity field map, 3$\arcsec$ towards the south of the nucleus the velocity gradient shows higher redshift values compared to ionized and molecular gas maps.

\subsection{Extinction correction}
\label{sec:extinction}

The dust has a remarkable reddening effect on the emission line fluxes which we receive from AGN and starburst host galaxies. Even though this effect is significantly lower in the NIR in comparison to the optical (lower extinction by about a factor of 10 in magnitude), it is still high enough not to be neglected.   

We correct for the extinction using the \cite{2000ApJ...533..682C} extinction law. To get the extinction at a certain wavelength $\lambda$ (in $\mm$) the equation is:
\begin{equation}
A(\lambda) = \frac{A_V}{R'_V} \times \left[ 2.659 \left( -1.857 + \frac{1.04}{\lambda} \right) + R'_V \right],
\end{equation}
with $R'_V=4.05$ and the visual extinction $A_V$.

We can now derive the visual extinction in the gas by comparing observed line ratios to typical intrinsic line ratios (assuming that the deviation is solely caused by intrinsic extinction). For a case B recombination scenario\footnote{In case B scenario the gas is optically thick for the Lyman emission lines \citep[][for more information on this topic]{1960ApJ...131..202P,2006agna.book.....O}.} with a typical electron density of $n_e = 10^2\,\mathrm{cm}^{-3}$ and temperature of $10^4\,\mathrm{K}$, the line ratio is $\mathrm{Pa}\alpha/\mathrm{Br}\gamma = 12.19$ \citep{2006agna.book.....O}. The visual extinction is then
\begin{equation}
A_V = 51.4 \times \log \left(\frac{12.19}{F_{\mathrm{obs},\mathrm{Pa}\alpha}/F_{\mathrm{obs},\mathrm{Br}\gamma}} \right).
\label{eq:extinction}
\end{equation}
While producing the $A_V$ map, we use a boxcar smoothing algorithm to get a higher S/N. 

\begin{figure*}
\centering
\includegraphics[width=0.34\linewidth]{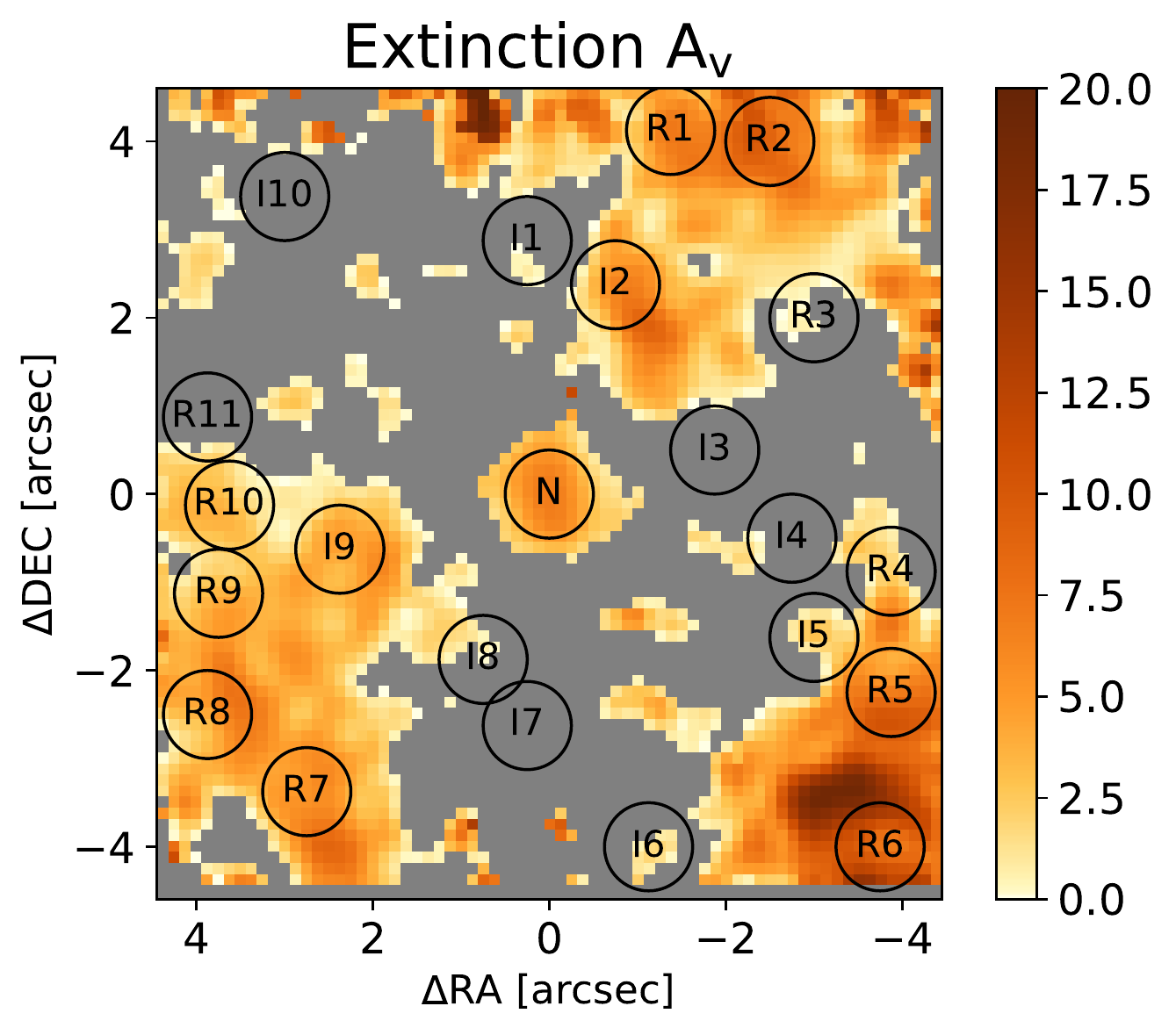}
\includegraphics[width=0.34\linewidth]{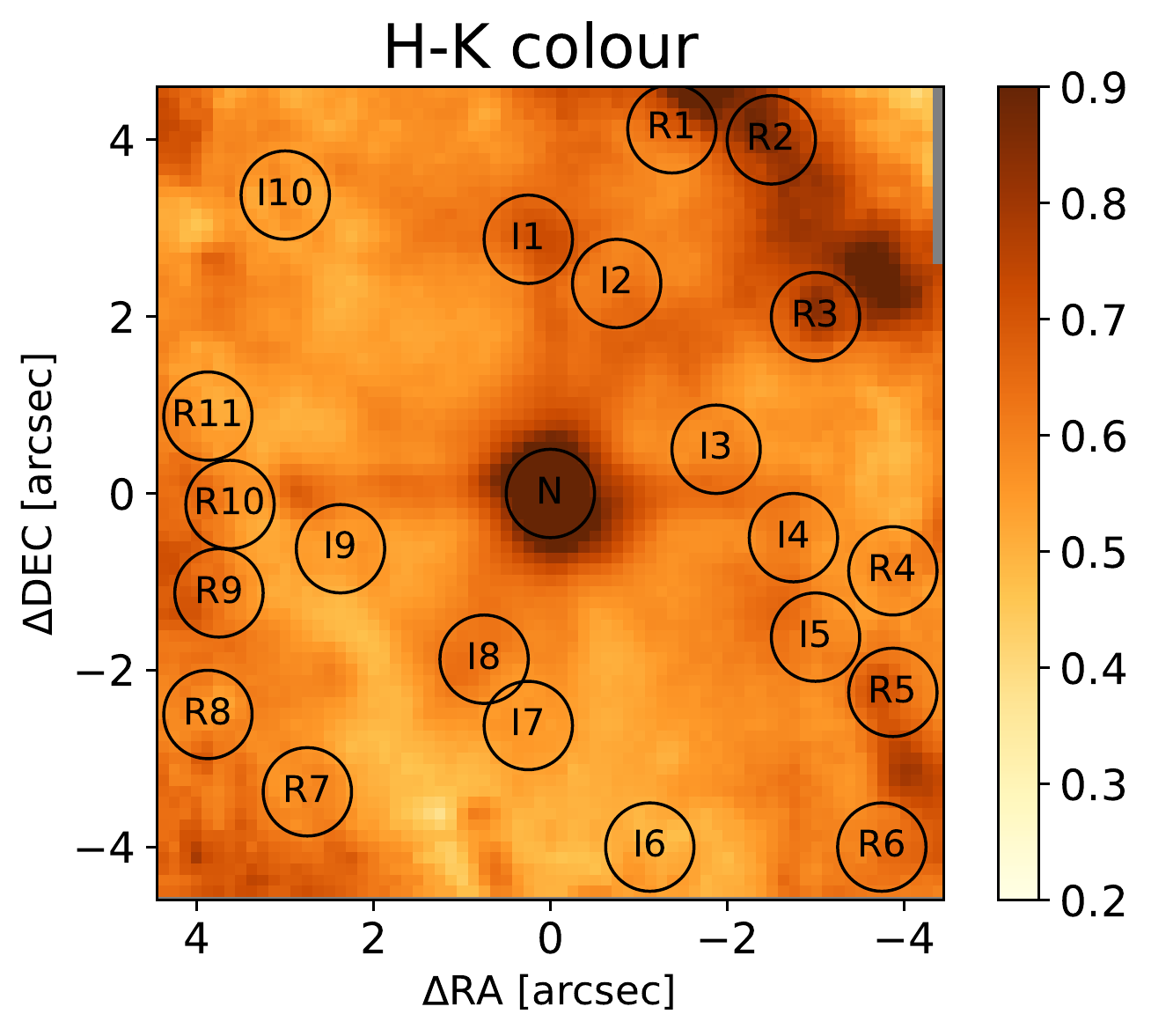}
\includegraphics[width=0.29\linewidth]{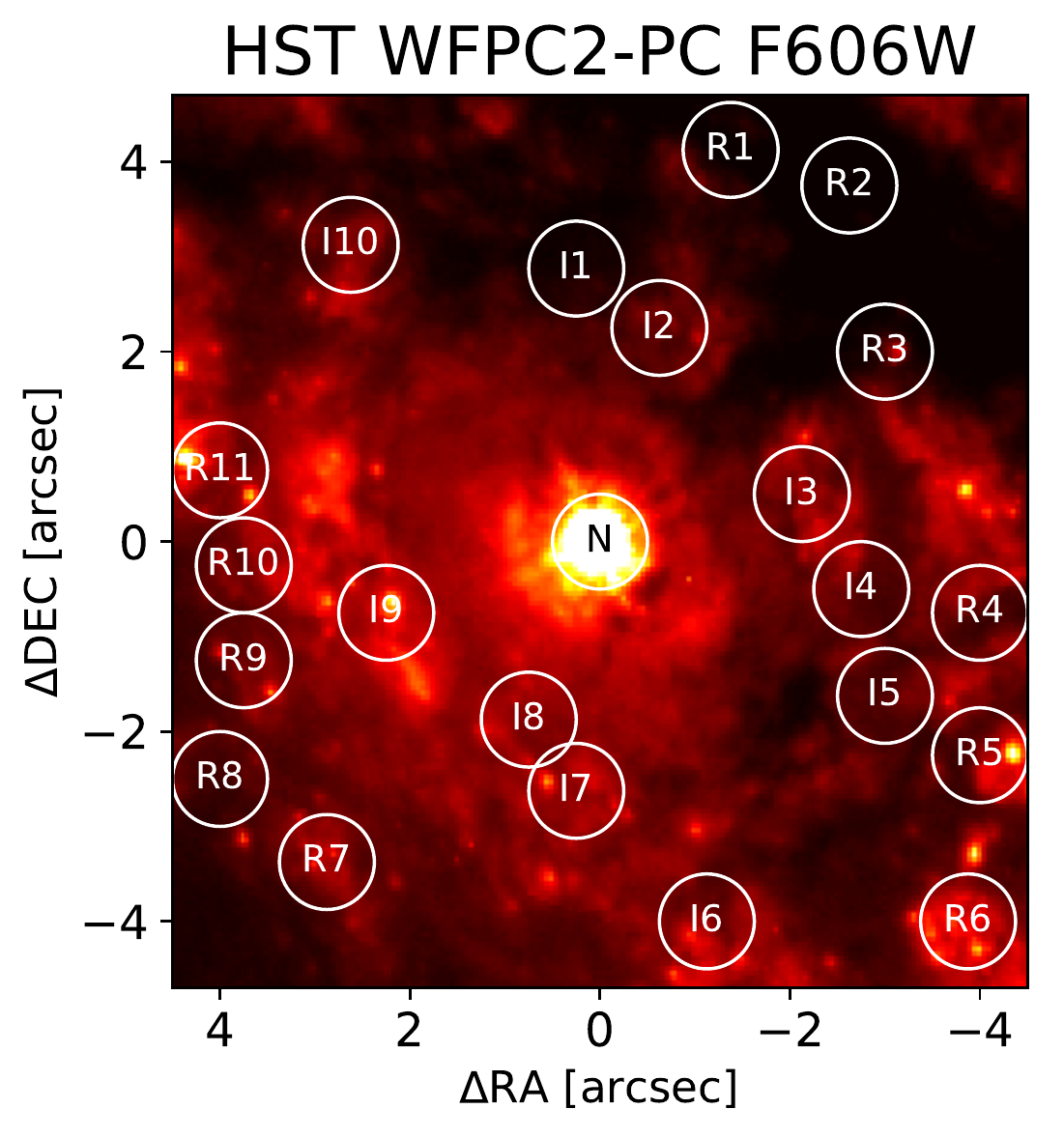}
\caption{\emph{Left:} map of the visual extinction $A_V$ derived from the line ratio Pa$\alpha$/Br$\gamma$. \emph{Middle:} $H-K$ color map derived from the $H+K$ spectra. \emph{Right:} HST F606W image. The center is marked with ``N'' and the apertures are shown to guide the  reader. The gray pixels show where we clipped the spectrum due to lower S/N.}
\label{fig:ext}
\end{figure*}

In Fig.~\ref{fig:ext} we present maps of the extinction $A_V$, the $H-K$ color which we derive from the $H+K$ spectra, and a HST F606W image in which dust lanes become apparent. In the $H-K$ color map there are some artifacts produced by the diffraction spikes made by the support vanes of the secondary mirror.

The HST F606W image and $H-K$ color map show consistency, meaning where there are dark dust lanes in the optical image, the color map has higher extinction. The $A_V$ map on the other hand is slightly different from the other two maps, which may be due to the fact that the emission lines and continuum emission originate from different layers along the line of sight. The results show that there is a highly extincted region in the NW corner, and more regions with high extinction between apertures I2 and I3 in the SW and in the SE. 

The extinction value in the central aperture where the AGN is located has higher inaccuracy in the $A_V$ map due to multiple emission line components. In the $H-K$ map we obtain values of  approximately one, which is consistent with the results found by \citet{1982MNRAS.199..943H}, \citet{1984MNRAS.211..461G}, \citet{1998ApJ...495..196A}, \citet{2006A&A...452..827F}. They find that a normal galaxy (not active) has an $H-K$ value of approximately $ 0.2$ and a zero-redshift quasar of approximately
$1.2$. The two maps can be compared with each other using $A_V/E(H-K)=14.6$ \citep{1999A&A...351..303T}. 

Extinction values in the analyzed apertures are estimated using Eq. \ref{eq:extinction} and are listed in Tables \ref{tab:emissionlines1} and \ref{tab:emissionlines2}. Emission line fluxes in those tables are already corrected for extinction.

\section{Discussion}
\label{sec:discussion}

\subsection{The active nucleus}
\label{sec:AGN}

\begin{figure}
\centering
\includegraphics[width=1.\linewidth]{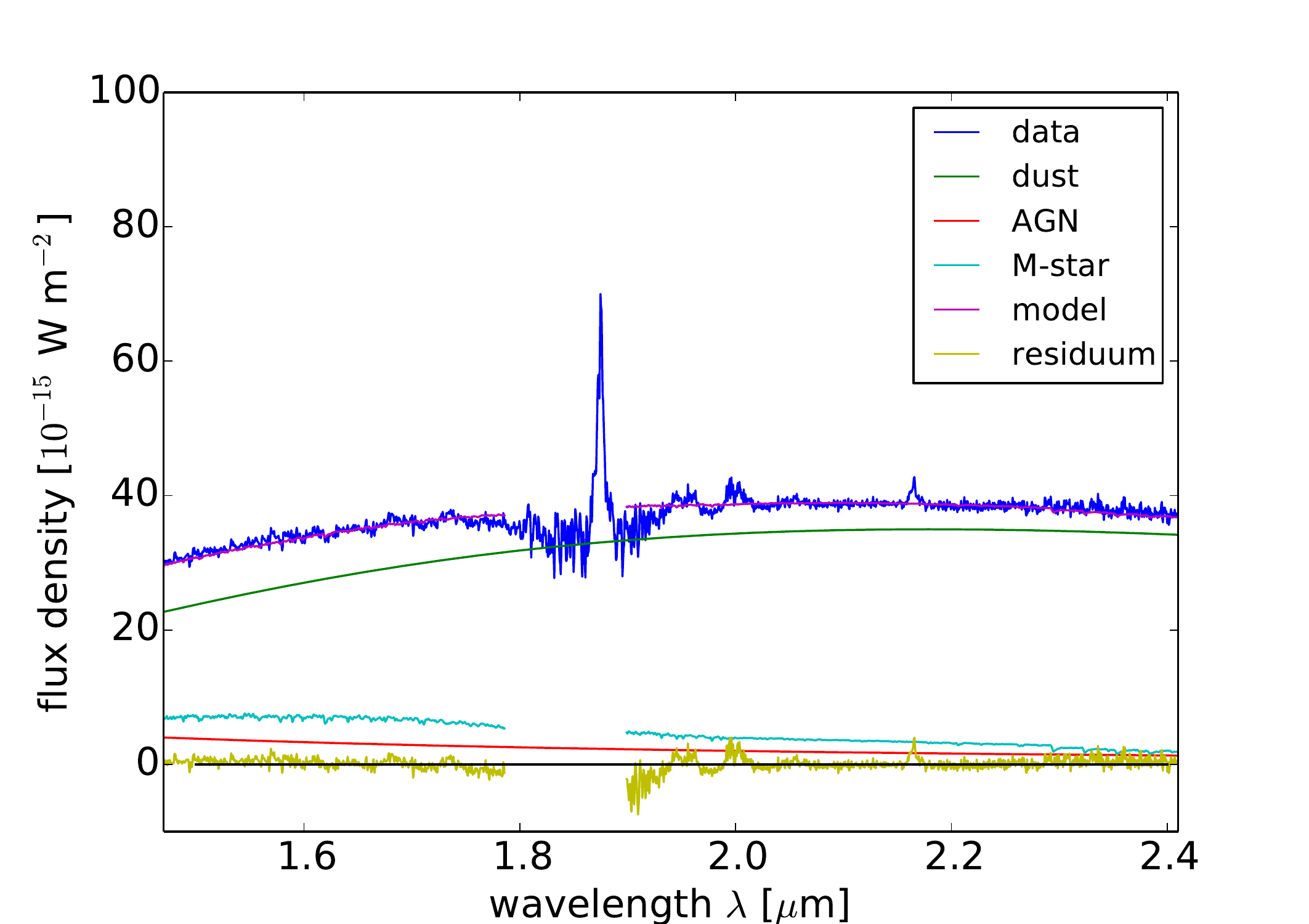}
\caption{Continuum fit of NGC 1365. The spectrum is extracted from an aperture with a radius $r\sim 45\, \mathrm{pc}$.}
\label{fig:irtf_fit}
\end{figure}

NGC 1365 is well known for hosting an AGN, and has been classified as a Seyfert 1, 1.5, 1.8, and 2 by different authors in the past \citep[e.g.,][]{1980A&A....87..245V,1993ApJ...418..653T,1995ApJ...454...95M,2007ApJ...659L.111R, 2017ApJS..232...11T}. 
Due to its strong variations in X-ray flux and spectral shape, most probably due to variations in the line-of-sight absorber rather than intrinsic changes in the accretion process, it has been classified as a changing-look AGN \citep[e.g.,][]{2005ApJ...623L..93R,2009MNRAS.393L...1R,2013MNRAS.429.2662B}.
%Broad components of hydrogen emission lines (see Fig.~\ref{fig:regions_specs}) support this classification. 
Also we find a typical reddening of the spectrum peaking towards the $K$-band which can be associated with dust close to its sublimation temperature \citep{1987ApJ...320..537B} which is an essential ingredient of the unified model of AGNs \citep{1993ARA&A..31..473A,1995PASP..107..803U}. A spectral decomposition of the central $0\farcs5$ into a black body representing the thermal emission of hot dust, a stellar template for the star light of the host galaxy \citep[M-star from the NASA Infrared Telescope Facility (IRTF) spectral library of cool stars;][]{2009ApJS..185..289R}, a power-law function for the AGN accretion disk emission, and an extinction contribution reveals that the emission from the dust torus is dominating in the center and contributes to more than 90$\%$ of the emission (Fig.\ref{fig:irtf_fit}). The temperature of the black body is $\approx 1300\,\mathrm{K}$ which is typical for a type-1 AGN \citep[e.g.,][]{2011MNRAS.414..218L,2016A&A...587A.138B}

\subsubsection{Black hole mass}
\label{sec:BHmass}

The mass of the supermassive black hole can be derived from the emission stemming from the broad-line region; in particular the broad components of hydrogen recombination lines. In the NIR, \cite{2010ApJ...724..386K} found the relation
\begin{equation}
M_\mathrm{BH} = 10^{7.16\pm 0.04} \left(\frac{L_{\mathrm{Pa}\alpha}}{10^{35}\,\mathrm{W}} \right)^{0.49\pm 0.06} \left(\frac{\mathrm{FWHM}_{\mathrm{Pa}\alpha}}{10^3\,\mathrm{km}\,\mathrm{s}^{-1}} \right)^2\,M_\odot
,\end{equation}
used to estimate the black hole mass $M_\mathrm{BH}$ from the luminosity of the broad component of Pa$\alpha$, $L_{\mathrm{Pa}\alpha}$, and its FWHM$_{\mathrm{Pa}\alpha}$.

\begin{figure*}
\centering
\includegraphics[width=0.48\linewidth]{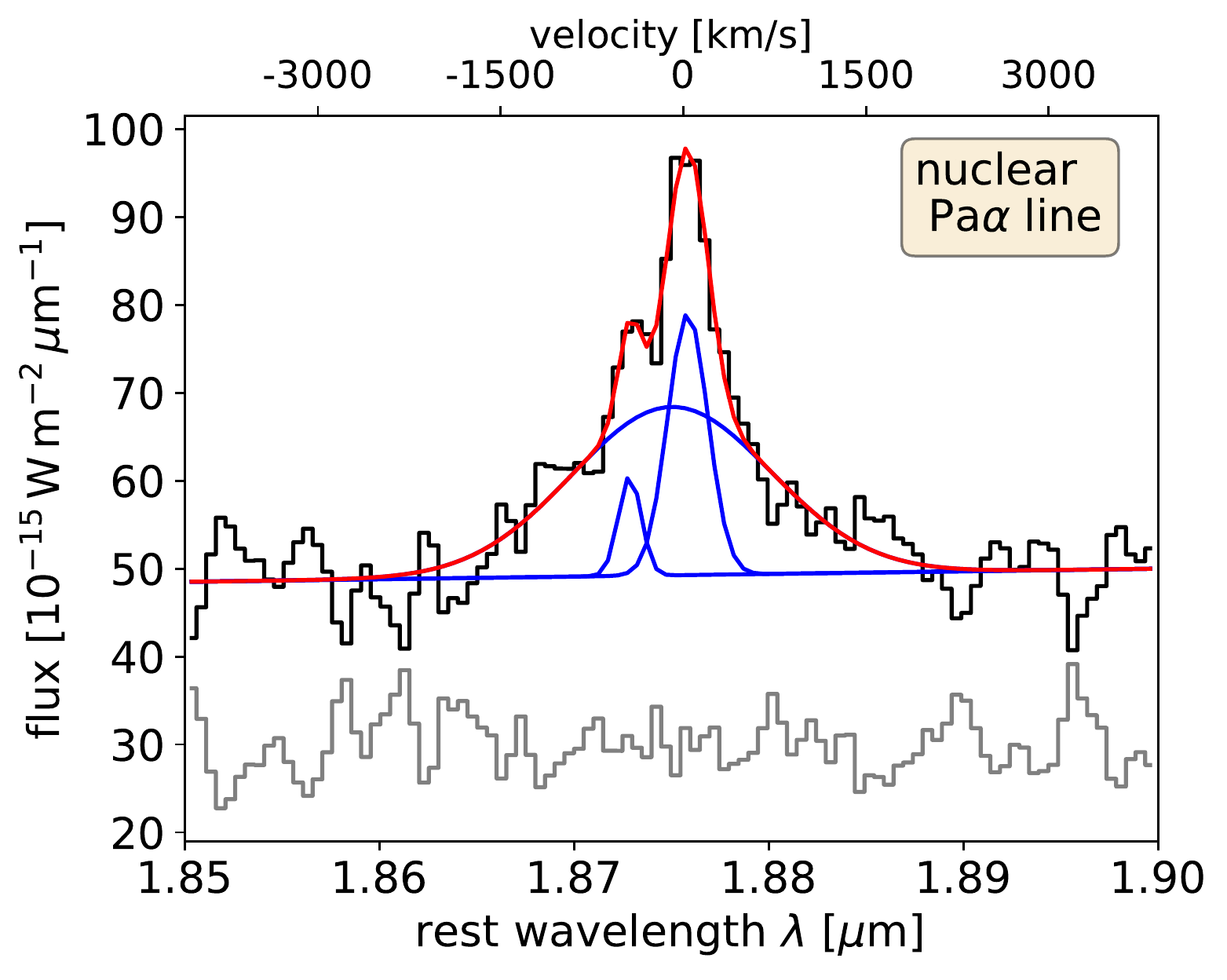}
\includegraphics[width=0.48\linewidth]{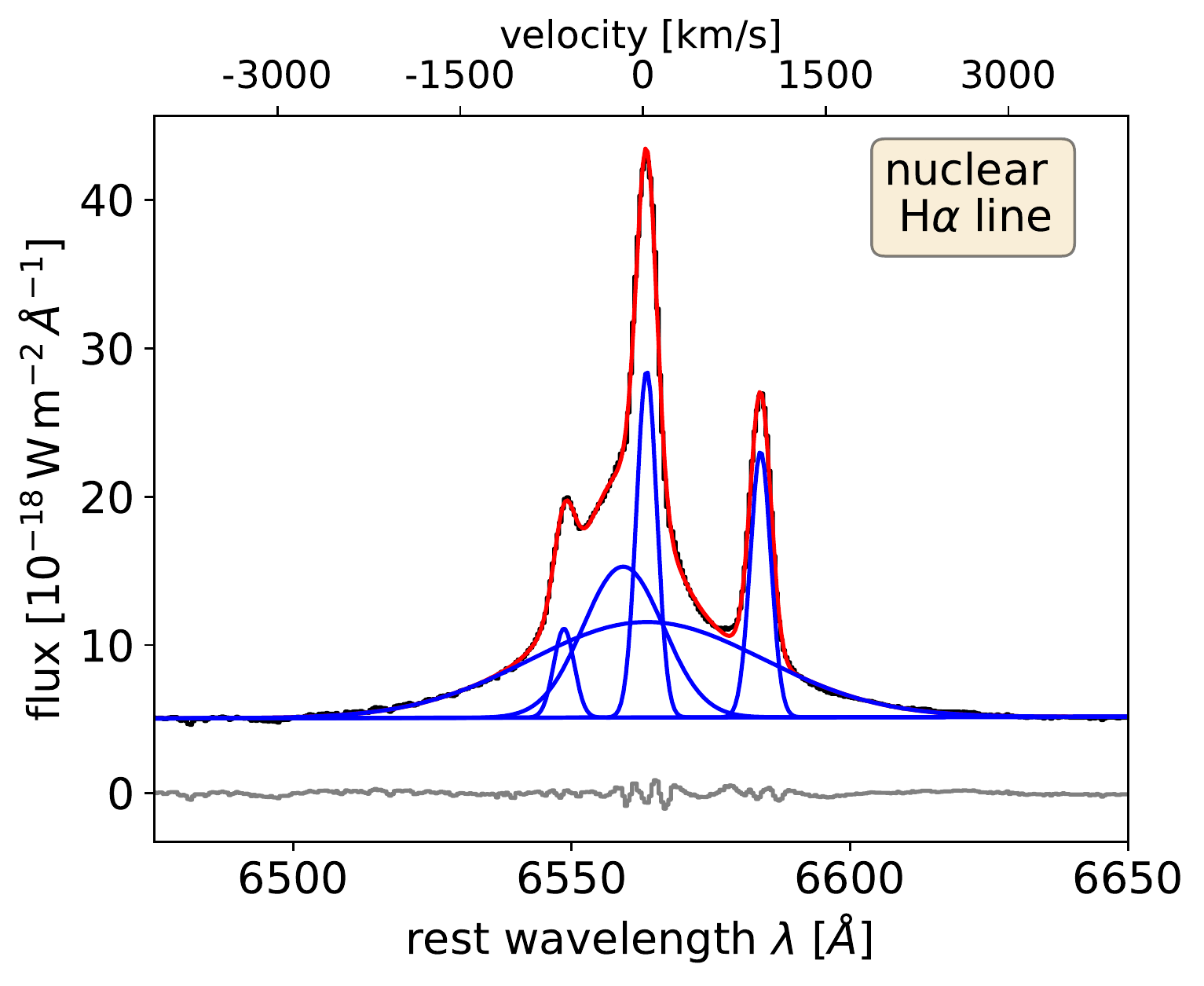}
\caption{\emph{Left}: Pa$\alpha$ emission line in a central aperture with radius $1\farcs 5$ which corresponds to $3\times \mathrm{PSF}$. Here we use three components to fit the line. \emph{Right}: H$\alpha$ $+$ [\ion{N}{ii}] emission lines complex in a central aperture with diameter $4\arcsec$ from the S7 survey \citep{2017ApJS..232...11T}. Here we use five components to fit these lines. The observed data is in black, the fitted components in blue and the full model in red.}
\label{fig:nuclearPaafit}\label{fig:nuclearHafit}
\end{figure*}

We measure the broad Pa$\alpha$ line in a spectrum that we extract from an aperture with radius $1\farcs5$ which corresponds to $3\times \mathrm{FWHM}_\mathrm{PSF}$. The fit consists of three components to account for the broad-line region, the narrow-line region and a third component which is clearly seen as line split in Fig.~\ref{fig:nuclearPaafit}. The luminosity is $L_{\mathrm{Pa}\alpha} = (9.6\pm 0.9)\times 10^{32}\,\mathrm{W}$ and the width, corrected for instrumental broadening, is $\mathrm{FWHM}_{\mathrm{Pa}\alpha} = (1925\pm 214)\,\mathrm{km}\,\mathrm{s}^{-1}$. This results in a BH mass of $\log(M_\mathrm{BH}) = 6.7$ which is consistent with the mass of $\log(M_\mathrm{BH}) = 6.65$ ($\mathrm{FWHM} = 1971\,\mathrm{km}\,\mathrm{s}^{-1}$) that \cite{2017MNRAS.468L..97O} derive from the broad Pa$\beta$ line, observed with the slit-spectrograph ISAAC.

An optical spectrum of the central $4\arcsec$ is available from the Siding Spring Southern Seyfert Spectroscopic Snapshot Survey \citep[S7,][]{2015ApJS..217...12D,2017ApJS..232...11T}. We fit the H$\alpha$ complex with five components, accounting for the two [\ion{N}{ii}] lines, the broad-line region and the narrow-line region of H$\alpha,$ and one blue component of H$\alpha$ for the potential outflow (Fig.~\ref{fig:nuclearHafit}).
We derive a luminosity of the broad H$\alpha$ component of $L_{\mathrm{H}\alpha} = (1.55\pm 0.02)\times 10^{33}\,\mathrm{W}$ and a width of FWHM$_{\mathrm{H}\alpha} = (1930\pm 20)\,\mathrm{km}\,\mathrm{s}^{-1}$. This is of the same order of magnitude as the luminosity of $L_{\mathrm{H}\alpha} = 2.4\times 10^{33}\,\mathrm{W}$ that \cite{1994A&A...288..425S} derived from an aperture of $2\farcs6\times 6\farcs5$.
From the luminosity and the width of the broad H$\alpha$, we can then estimate the black hole mass following \cite{2012MNRAS.423..600S},
\begin{equation}
M_\mathrm{BH} = 10^{7.4} \left( \frac{L_{\mathrm{H}\alpha}}{10^{37}\,\mathrm{W}} \right)^{0.545} \left( \frac{\mathrm{FWHM}_{\mathrm{H}\alpha}}{10^3\,\mathrm{km}\,\mathrm{s}^{-1}} \right)^{2.06}\,M_\odot,
\end{equation}
and get $\log(M_\mathrm{BH}) = 5.9$. The lower estimated value compared to the NIR might hint at the extinction that the optical is more prone to.

\begin{figure*}
\sidecaption
\centering
\includegraphics[width=0.6\linewidth]{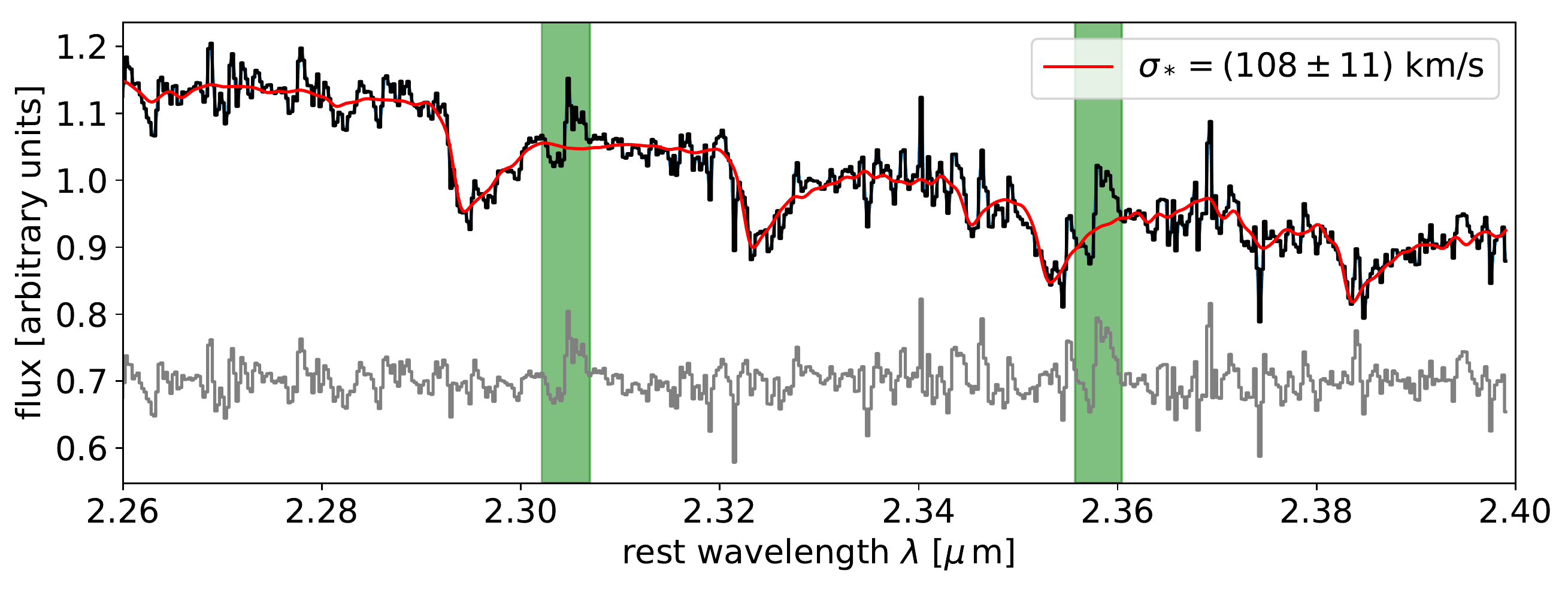}
\caption{Fit of the stellar $K$-band continuum around the CO-band heads at $2.3\mm$. The spectrum was extracted from an annulus with inner radius $0\farcs5$ and outer radius $2\arcsec$ to avoid the nucleus in which the stellar continuum is completely diluted by the AGN continuum and the star forming ring. The observed spectrum is shown in black, the ppxf fit in red, and the (upwards shifted) residuum in gray.}
\label{fig:centralsigma}
\end{figure*}

Several studies show a tight correlation between the central BH mass and the stellar bulge velocity dispersion $\sigma_*$. In order to estimate the BH mass we run \textsc{Ppxf} on the integrated spectrum of a ring-shaped aperture with an inner radius of $r =0\farcs 5$ and an outer radius of $r= 2\arcsec$. The inner radius is chosen in a way to exclude the continuum emission from the AGN, which dilutes the CO absorption lines. The stellar velocity dispersion in star forming regions is often lowered which reflects the dynamically cold gas from which the young stars formed \citep{2011MNRAS.416..493R,2014MNRAS.438.2036M}. Since we are interested in a stellar velocity dispersion which is representative for the mass-dominating old stellar populations, we select the outer radius such that the star-forming ring is excluded. The spectrum and its fit are shown in Fig.~\ref{fig:centralsigma}. The resulting stellar velocity dispersion is $\sigma_*=(108\pm 11)$ km s$^{-1}$. 
Inserting this value into different relations, we get the following $M_\mathrm{BH}$ estimates: $\log(M_\mathrm{BH}/M_{\odot}) = 7.0$ \citep{2009ApJ...698..198G}, $\log(M_\mathrm{BH}/M_{\odot}) = 6.7$ \citep{2013ARA&A..51..511K}, and $\log(M_\mathrm{BH}/M_{\odot}) = 6.5 $ \citep[][using the relation for barred galaxies]{2013ApJ...764..151G}.

As \cite{2015MNRAS.446.2823R} find, the $\sigma_*$ seems to be systematically lower when measured at the CO band heads in the $K$-band compared to optical measurements, which are usually used to calibrate the $M_\mathrm{BH} - \sigma_*$ relation. Applying their correction for spiral galaxies, we find that our measurement corresponds to $\sigma_\mathrm{opt} \approx 138\,\mathrm{km}\,\mathrm{s}^{-1}$. Using this value increases the $M_\mathrm{BH}$ estimate by roughly $0.5\,\mathrm{dex}$. In this case, the BH mass estimate from the $M_\mathrm{BH}-\sigma_*$ would be half an order of magnitude higher than from the broad Pa$\alpha$ line. On the one hand, the broad Pa$\alpha$ luminosity might be underestimated due to extinction in the broad-line region. On the other hand, it is well known that galaxies with strong bars such as NGC 1365 are often outliers in the $M_\mathrm{BH}-\sigma_*$ relation \citep[e.g.,][]{2009ApJ...698..812G,2009MNRAS.399..621G,2013ApJ...764..151G,2014MNRAS.441.1243H}. We can conclude that the BH mass is in the range $(5 - 10)\times 10^6\,M_\odot$. Due to the intrinsic scatter of the relations, we refrain from stating formal errors on the estimates.

\subsection{Gas excitation}
\label{sec:excitation}

Diagnostic diagrams use ratios between emission lines that are representative for different modes or degrees of excitation. They determine the dominating source (e.g., star formation, AGN, shocks) that excites the line-emitting gas. Originally they were established in the optical \citep[so-called BPT or seagull-diagrams,][]{1981PASP...93....5B,2001ApJ...556..121K,2003MNRAS.346.1055K}, and later similar tools were established in the NIR. Typically, one considers ratios between star formation tracers like the hydrogen recombination lines (Br$\gamma$, Pa$\beta$, Pa$\alpha$) and shock tracers like the ro-vibrational emission lines of molecular hydrogen (most prominently H$_2$(1-0)S(1) $\lambda 2.12\mm$) or forbidden iron lines (like [\ion{Fe}{ii}] $\lambda 1.64\mm$ in the $H$-band) to find the dominating excitation mechanism \citep{1998ApJS..114...59L,2004A&A...425..457R,2005MNRAS.364.1041R,2013MNRAS.430.2002R,2015A&A...578A..48C}.

The lowest values in the line ratios, typically H$_2 \lambda 2.122\mm$/Br$\gamma \lesssim 0.4$, are found in star formation regions where young OB stars can photoionize large amounts of hydrogen which results in strong hydrogen recombination line emission. Higher line ratios occur in older stellar populations, where the number of bright OB stars is decreasing and supernova explosions take place. Shocks produced by these supernovae can be traced by H$_2$ or [\ion{Fe}{ii}] lines leading to higher H$_2$/Br$\gamma$ line ratios. Low-ionization nuclear emission-line regions (LINERs) can therefore show line ratios $\gtrsim 0.9$, while Seyfert galaxies usually show values in between \citep[e.g.,][]{1993ApJ...406...52M, 1994ApJ...422..521G, 1997ApJ...482..747A}.

\begin{figure}
\centering
\includegraphics[width=\columnwidth]{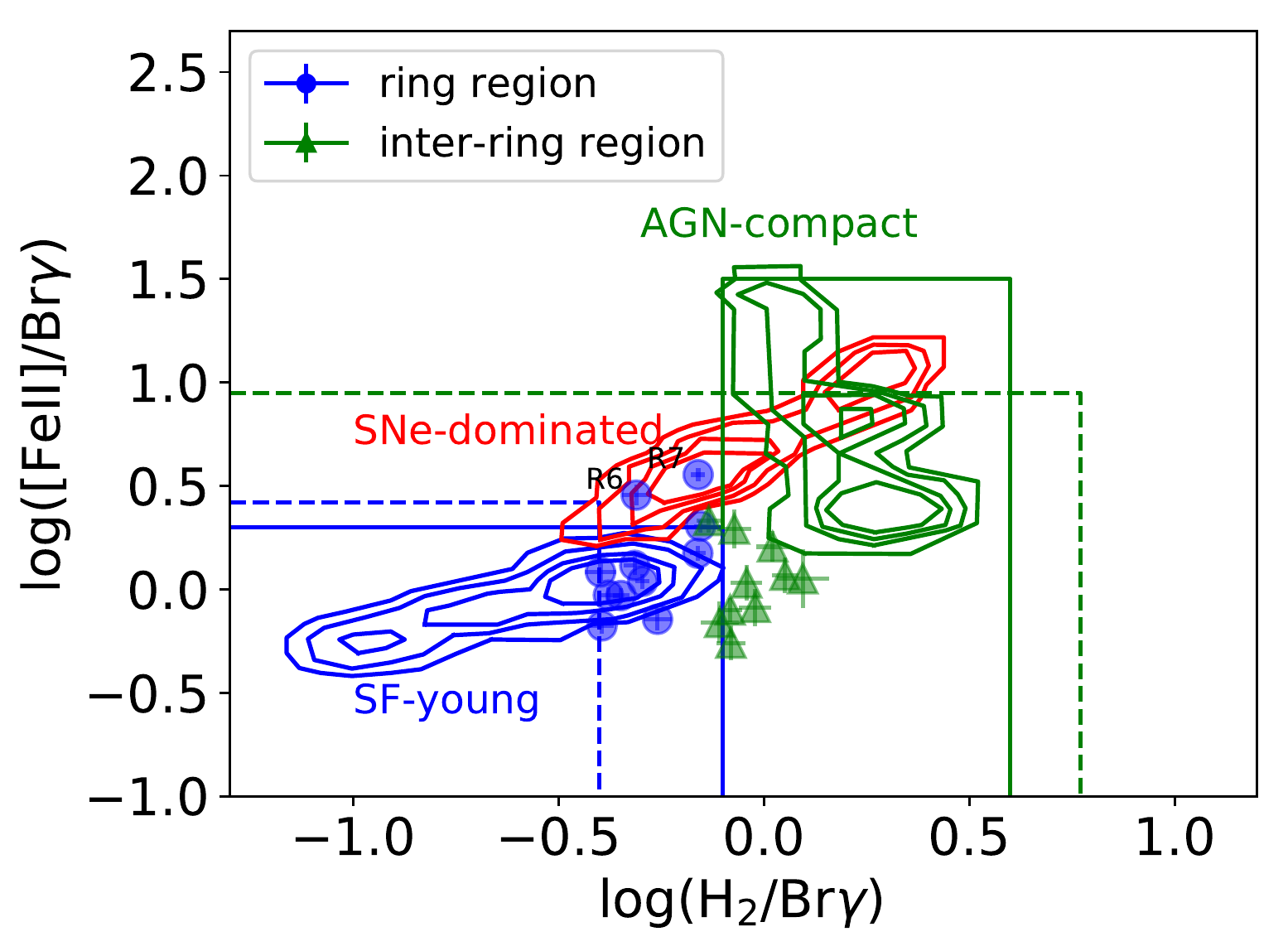}
\caption{Diagnostic diagram with the NIR emission line ratios H$_2$(1-0)S(1)$\lambda 2.122\mm$/Br$\gamma$ and [\ion{Fe}{ii}]$\lambda 1.644\mm$/Br$\gamma$. Apertures in the circumnuclear ring are shown in blue circles and apertures from the region within the ring in green triangles. Contours denote regions for young star formation, supernovae, and compact AGNs, while solid lines denote upper limits for young star formation and AGNs, both derived from integral-field spectroscopy data \citep{2015A&A...578A..48C}. Dashed lines denote upper limits for star formation and AGN derived from slit-spectroscopy \citep{2013MNRAS.430.2002R}.}
\label{fig:diagndiagram}
\end{figure}

We place the line ratios for our apertures in a NIR diagnostic diagram in Fig.~\ref{fig:diagndiagram}. The ratio [\ion{Fe}{ii}]/Br$\gamma$ was corrected for extinction. Due to the proximity of the H$_2 \lambda 2.122\mm$ and the Br$\gamma$ line in wavelength, the extinction correction here would only be of the order of a few percent which is much smaller than the error introduced by the uncertainties in the extinction estimate. Therefore we decide not to correct this line ratio \citep[see discussion in][]{2017A&A...598A..55B}. 

Most ``R'' apertures show line ratios typical for young star formation, indicating that it resembles a star forming ring. Notable exceptions are the apertures R6 and R7 that show increased [\ion{Fe}{ii}] emission (see also flux maps in Fig.~\ref{fig:line_fluxes}) which shifts their position in the diagnostic diagram into the location populated by supernova-dominated regions. This could indicate that the regions are already aged starbursts in which massive stars are exploding in supernovae.

All apertures in ``I'' apertures show increased H$_2$/Br$\gamma$ ratios indicating shock contributions, possibly from the AGN. However, the [\ion{Fe}{ii}]/Br$\gamma$ ratio is lower than expected from the linear relation between the two line ratios seen in  \cite{2013MNRAS.430.2002R} for example. We cannot exclude that this is due to an underestimation of the extinction which is heavily affecting the [\ion{Fe}{ii}]/Br$\gamma$ ratio since these two lines are far apart in wavelength.

Cold molecular hydrogen is supposed to feed AGNs and star formation and is therefore of particular interest. Due to its nature as a diatomic molecule with identical nuclei, H$_2$ has no permanent dipole moment. The lowest vibrational transition is in the NIR at $2.22\mm$ and has an energy of $\sim$$6500\,\mathrm{K}$ which means that in the NIR, we observe hot molecular gas with temperatures $\gtrsim 1000\,\mathrm{K}$ \citep[e.g., see the introduction of][]{2013ARA&A..51..207B}. The excitation temperature of vibrational transitions can be derived by comparing two transitions with equal rotation but consecutive vibrational level. For this, we assume that the population densities in thermal equilibrium follow the Boltzmann distribution $\frac{N''}{N'} = \frac{g''}{g'} e^{-\Delta E/k_B T}$. The population densities are represented by the observed column densities which are derived from the observed line fluxes $f$ via $N_\mathrm{col} \propto \frac{f}{A_{ul}} \frac{\lambda}{hc}$ where $A_{ul}$ is the Einstein coefficient and $hc/\lambda$ the photon energy of the transition. With these relations, we get
\begin{equation}
T_{\mathrm{vib}, J=3} = \frac{5594\,\mathrm{K}}{0.304 + \ln \left( \frac{f_{1-0\mathrm{S}(1)}}{f_{2-1\mathrm{S}(1)}} \right)},
\end{equation}
while the excitation temperature of rotational transitions can be derived by comparing two transitions with equal vibrational but consecutive rotational level, for example
\begin{equation}
T_{\mathrm{rot}, v=1} = \frac{1113\,\mathrm{K}}{1.130 + \ln \left( \frac{f_{1-0\mathrm{S}(0)}}{f_{1-0\mathrm{S}(2)}} \right)}.
\end{equation}
In cases of thermal excitation, these temperatures should be equal, $T_\mathrm{rot} \approx T_\mathrm{vib}$. Examples for this are shock and X-ray excitation with a typical temperature of $T\approx 2000\,\mathrm{K}$ \citep{1989MNRAS.236..929B,1989ApJ...338..197S,1990ApJ...363..464D,1996ApJ...466..561M,2002MNRAS.331..154R,2003MNRAS.343..192R}.

\begin{figure}
\centering
\includegraphics[width=\columnwidth]{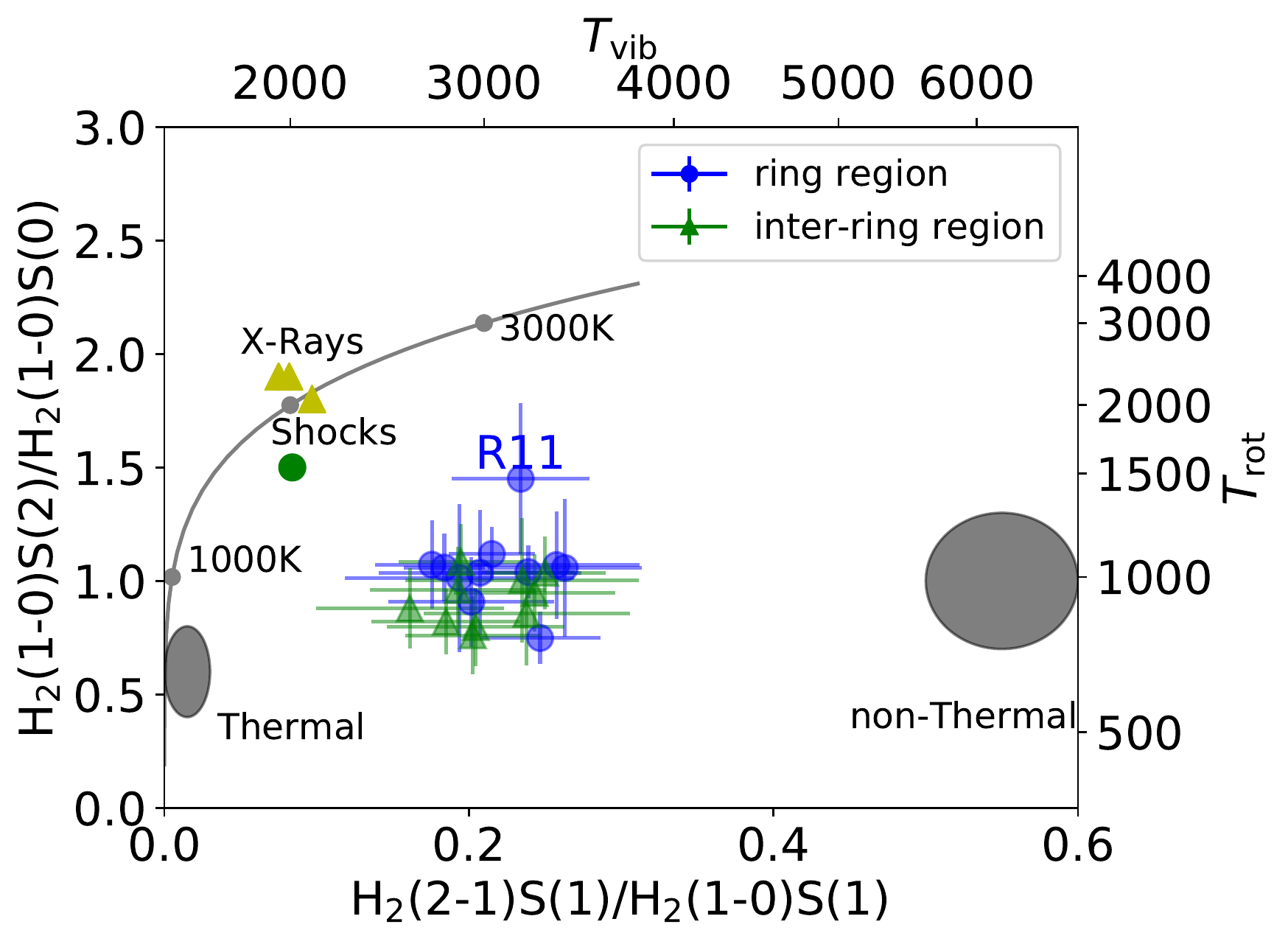}
\caption{Diagnostic diagram with the NIR molecular hydrogen line ratios H$_2$(2-1)S(1)$\lambda 2.248\mm$/H$_2$(1-0)S(1)$\lambda 2.122\mm$ and H$_2$(1-0)S(2)$\lambda 2.034\mm$/H$_2$(1-0)S(0)$\lambda 2.223\mm$ \citep{1994ApJ...427..777M}. Apertures in the circumnuclear ring are shown in blue circles and apertures from the region within the rings in green triangles. The thermal emission from 1000 to 3000 K (solid gray curve), models for thermal UV excitation (gray circle right) \citep{1989ApJ...338..197S}, nonthermal excitation (gray ellipse left) \citep{1987ApJ...322..412B}, X-ray heating (yellow triangles) \citep{1990ApJ...363..464D}, and shock heating (green circle) \citep{1989MNRAS.236..929B} are indicated for comparison.}
\label{fig:H2diagn}
\end{figure}

Strong star formation or AGN continuum can also produce nonthermal excitation, for example by UV fluorescence \citep{1987ApJ...322..412B} which seems to play a major role in H$_2$ excitation in ultra luminous infrared galaxys (ULIRGs) and AGNs \citep{2003ApJ...597..907D,2005ApJ...633..105D}. To distinguish between these mechanisms, one can use diagnostic diagrams such as the one in Fig.~\ref{fig:H2diagn} which shows the H$_2$ line ratios in comparison to those produced by models. We observe line ratios that indicate a mixing of different excitation mechanisms, consistent with many previous studies \citep[e.g.,][]{2007A&A...466..451Z,2013MNRAS.428.2389M,2015A&A...575A.128B,2015A&A...583A.104S}. Analyzed apertures do not lie in the regions with pure nonthermal or thermal excitations and all show rotational temperatures of $T_\mathrm{rot} \approx 1000\,\mathrm{K}$ and vibrational temperatures of $2500\,\mathrm{K} \lesssim T_\mathrm{vib} \lesssim 3500\,\mathrm{K}$. Therefore these spots are not in local thermal equilibrium (LTE) and the significant contribution of nonthermal excitation, which is mirrored in rather low rotational and higher vibrational temperatures, is typical for star formation regions \citep[e.g.,][]{2005MNRAS.364.1041R,2013MNRAS.430.2002R,2017A&A...598A..55B}.

\subsection{Gas masses}

The mass of hot molecular gas can be estimated from the extinction-corrected flux of the H$_2$ 1-0 S(1) $\lambda 2.122\mm$ emission line, $f_{\mathrm{1-0S(1)}}$, as
\begin{equation}
M_{\mathrm{H}_2} = 5.0776\times 10^{16} \left(\frac{D_L}{\mathrm{Mpc}}\right)^2 \left(\frac{f_\mathrm{1-0S(1)}}{\mathrm{W}\,\mathrm{m}^{-2}} \right) \,M_\odot
\label{eq:h2mass}
,\end{equation}
where $D_L$ is the luminosity distance to the galaxy in megaparsec \citep[e.g.,][]{1982ApJ...253..136S,1998ApJS..115..293W,2010MNRAS.404..166R}. What is rarely mentioned in the literature are the two important assumptions that go into this formula, namely that the gas is in LTE and that it has a typical excitation temperature of $T\approx 2000\,\mathrm{K}$. As we see in Sect.~\ref{sec:excitation}, this is not fully the case here. With vibrational temperatures of $T_\mathrm{vib} \approx 3000\,-\, 4000\,\mathrm{K}$, we will overestimate the gas mass by a factor of up to two \citep{2017A&A...598A..55B}. Mentioning this, we will still keep Equation \ref{eq:h2mass} for better comparison with the literature. 

Near-infrared H$_2$ lines only trace the hot surface of the overall molecular gas reservoir. Empirical relations have been found between the NIR H$_2$ luminosities and the CO luminosities that trace the cold molecular gas. In the following, we use the conversion factor $M_{\mathrm{H}_2,\mathrm{cold}}/M_{\mathrm{H}_2,\mathrm{hot}} = (0.3-1.6)\times 10^6$ \citep{2013MNRAS.428.2389M}. Other consistent conversion factors have been published by \cite{2005AJ....129.2197D} and \cite{2006A&A...454..481M}.

The mass of ionized gas \ion{H}{ii} can be calculated as \citep[see][for details]{2015A&A...575A.128B}
\begin{equation}
\label{eq:ionized_gas}
M_{\ion{H}{ii}} = 2.9 \times 10^{22} \left( \frac{f_{\mathrm{Br}\gamma}}{\mathrm{W}\,\mathrm{m}^{-2}} \right) \left( \frac{D_L}{\mathrm{Mpc}} \right)^2 \left(\frac{n_e}{\mathrm{cm}^{-3}} \right)^{-1} \, M_\odot,
\end{equation}
where $f_{\mathrm{Br}\gamma}$ is the extinction-corrected Br$\gamma$ flux, $D_L$ the luminosity distance in megaparsecs, and $n_e$ the electron density for which we assume a typical value of $n_e = 10^2\,\mathrm{cm}^{-3}$.

We estimate the ionized and (hot) molecular gas masses over the 9$\arcsec \times 9\arcsec$ FOV. Here the emission line fluxes are corrected for an extinction with a typical value of $A_V \approx 2$ mag. The hot molecular gas mass is determined using the luminosity of the H$_2 \lambda 2.12$ emission line and is about $\sim 615 \,M_\odot$. This corresponds to a cold gas mass of $\sim (2-10) \times 10^8 \,M_\odot$ in the central $9\arcsec \times 9\arcsec$. Also using the Br$\gamma$ emission line luminosity, the ionized gas mass is determined to be $M_{\ion{H}{ii}} \sim 5.3 \times 10^6 \, M_\odot$. However, in order to exclude the AGN influence we also mask the spectra of a central aperture with a radius of $r \sim 0\farcs5$ , and derive the mass to be $M_{\ion{H}{ii}} \sim 4.3 \times 10^6 \, M_\odot$. This is consistent with typical masses found in nearby galaxies by the AGNIFS group \citep[][and references therein]{2016MNRAS.461.4192R} of $10< M_{H_2} <10^3\, M_\odot$ and $ 10^4 < M_{\ion{H}{ii}} < 10^7 \, M_\odot $. 
For NUGA sources, cold gas masses are estimated from the CO-emissions to be in ranges $10^8< M_{H_2} <10^{10} \, M_\odot$. 

Using Eqs. \ref{eq:h2mass} and \ref{eq:ionized_gas} we also estimate the gas masses in the apertures. The values are listed in Tables \ref{tab:emissionlines1}, \ref{tab:emissionlines2}, \ref{tab:h2lines1}, and \ref{tab:h2lines2}. Given the uncertainties in the assumed parameters that go into the equation (e.g., electron density), the model choice (assuming LTE), and the scatter of the scaling relation, we refrain from calculating formal errors but see the values as order-of-magnitude estimates.

\begin{figure}
\centering
\includegraphics[width=\columnwidth]{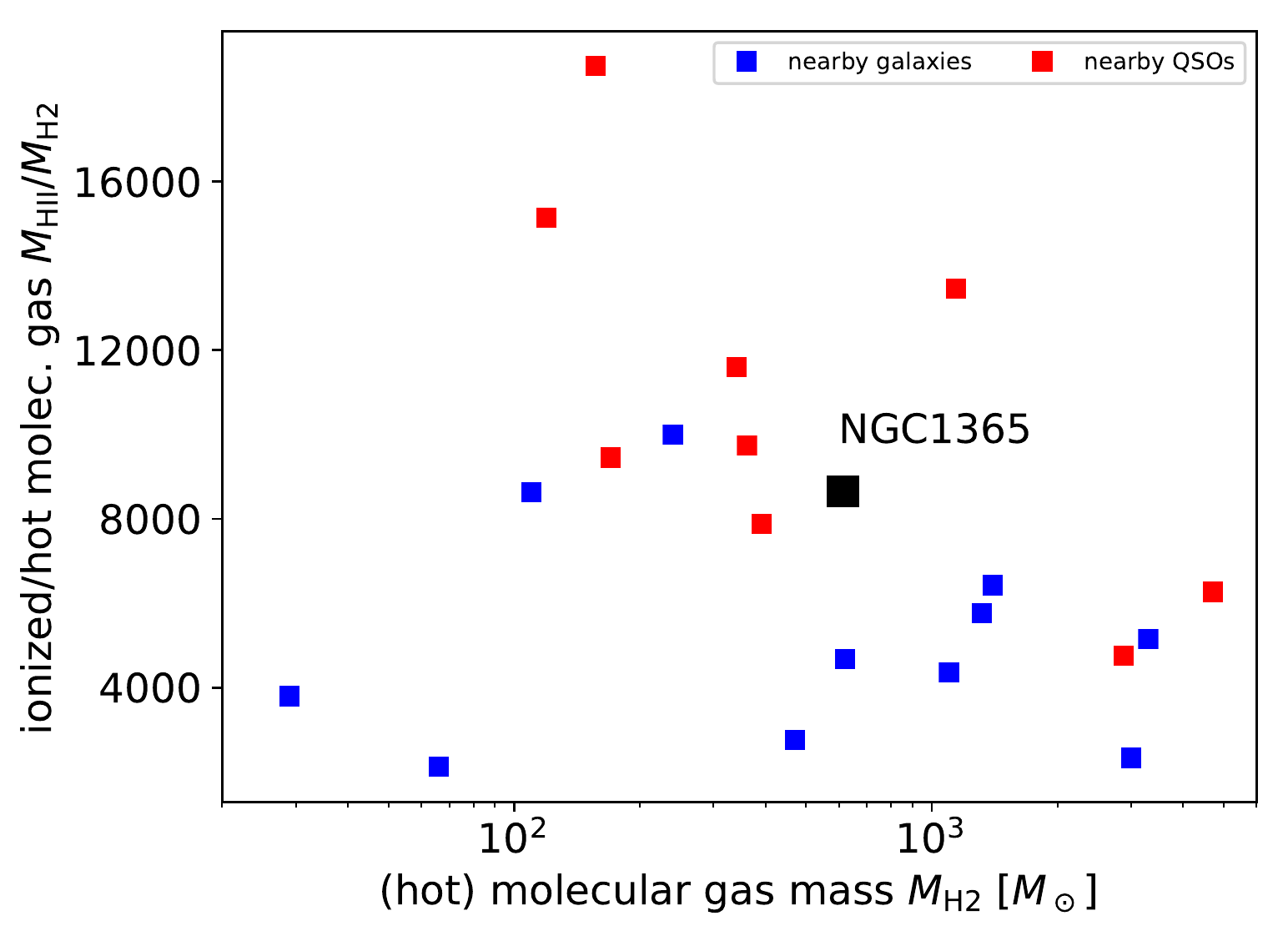}
\caption{Relation between (hot) molecular and ionized gas masses in the R (ring) and I (within ring) apertures of NGC 1365, compared to literature values for nearby galaxies (NGC 4051: \citet{2008MNRAS.385.1129R}, NGC 7582: \citet{2009MNRAS.393..783R}, NGC 4151: \citet{2009MNRAS.394.1148S}, Mrk 1066: \citet{2010MNRAS.404..166R}, Mrk 79: \citet{2013MNRAS.430.2249R}, NGC 1068: \citet{2014MNRAS.442..656R}, Mrk 766: \citet{2014MNRAS.445..414S}, NGC 5929: \citet{2015MNRAS.451.3587R}, NGC 2110: \citet{2015MNRAS.453.1727D}, NGC 4303: \citet{2016MNRAS.461.4192R}, and NGC 5548: \citet{2017MNRAS.464.1771S}) and nearby QSOs \citep{2015A&A...575A.128B,2016A&A...587A.138B}.}
\label{fig:gas_comp}
\end{figure}

The ratio between ionized and (hot) molecular gas masses integrated over the FOV in NGC 1365 is approximately $8000$. In order to compare our results with the literature, we collected molecular and ionized gas masses found for different categories of AGNs in Fig. \ref{fig:gas_comp}: 
nearby galaxies analyzed by the AGNIFS group and nearby QSOs \citep{2015A&A...575A.128B,2016A&A...587A.138B}\footnote{ In \citet{2016A&A...587A.138B} only line fluxes are stated. We convert these to gas masses using Equations \ref{eq:h2mass} and \ref{eq:ionized_gas}, assuming an electron density of $n_e\,=100\,\mathrm{cm}^{-3}$. Most ionized gas masses in the literature were calculated assuming this electron density as well. Exceptions are NGC1068 \citep{2014MNRAS.442..656R}, NGC2110 \citep{2015MNRAS.453.1727D}, and NGC4303 \citep{2016MNRAS.461.4192R} in which an electron density of $n_e\,=500\,\mathrm{cm}^{-3}$ was assumed. We adjusted these gas masses to have a common electron density of $n_e\,=100\,\mathrm{cm}^{-3}$ which results in five times higher ionized gas mass estimates. This underlines that the estimation of the electron density introduces an additional uncertainty.}. 
It becomes apparent that the ratio of ionized to molecular gas mass gets higher in nearby QSOs possibly due to a greater contribution of the narrow line region or the dissociation of molecular gas illuminated by the strong AGN. NGC 1365 has a higher ratio in comparison to nearby galaxies and a slightly lower ratio than nearby QSOs. This could be due to the AGN activity and the contribution of the strong circumnuclear star formation. 

\subsection{Gaseous kinematics and streaming motions}

In this section we study the gaseous kinematics of NGC 1365 and compare the results with the stellar kinematics. The line-of-sight velocity fields of stars and gas show the same direction and similar degrees of rotation. However, as shown in Fig.~\ref{fig:gas_losv} the gas shows more deviation from pure rotation compared to the stellar velocity field. The strongly twisted and irregular zero-velocity line, which could be due to noncircular motions induced by, for example, a secondary bar with an axis deviating from the large-scale velocity field or oval flows \citep[e.g.][]{2014ApJ...792..101D,2014MNRAS.438.2036M}, is especially prominent in the gas velocity field.
\subsubsection{The stellar ring}
\begin{figure}
\centering
\includegraphics[width=0.99\linewidth]{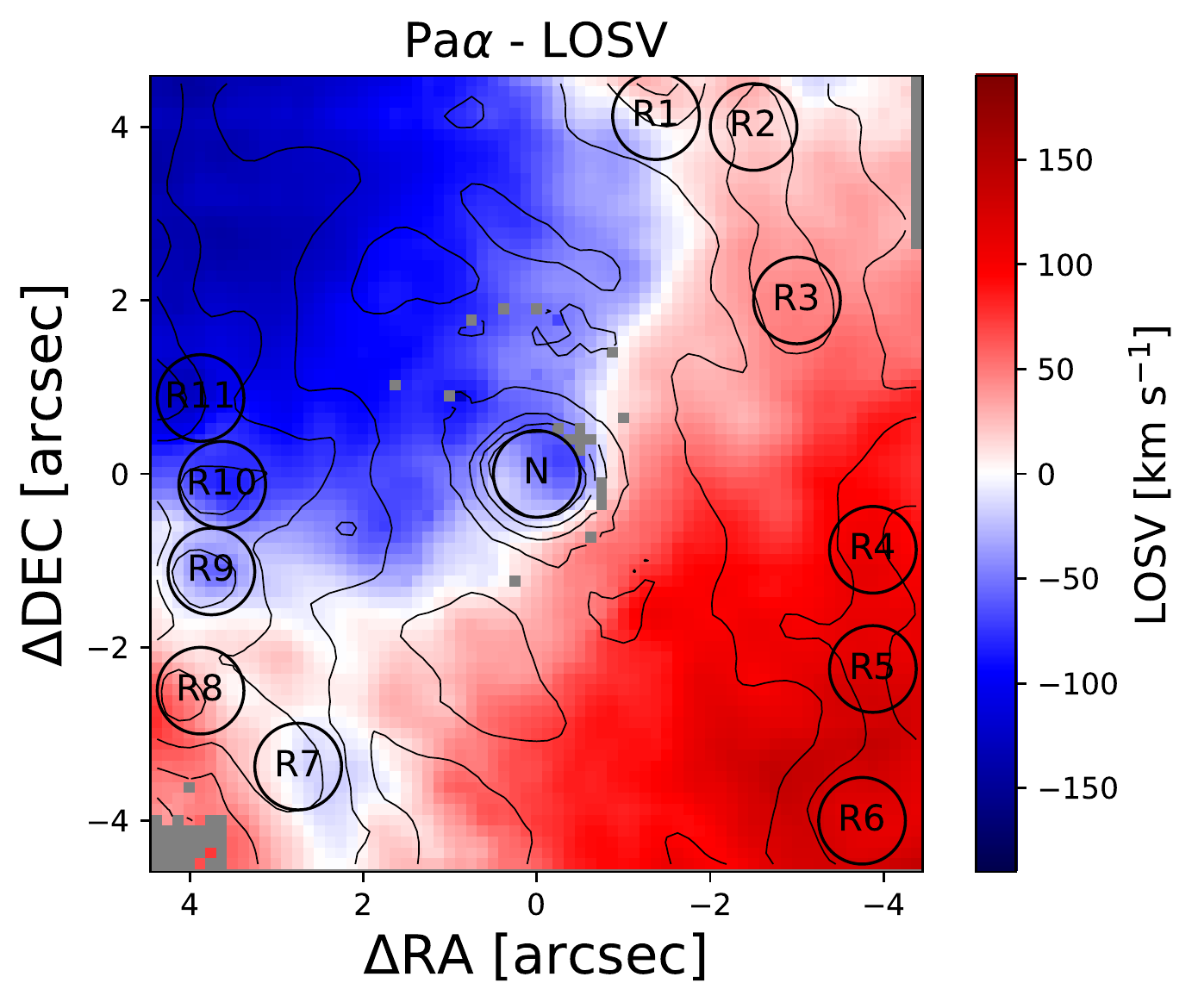}
\caption{Line-of-sight velocity map of ionized gas traced by Pa$\alpha$. Here we overlay the contours of the flux intensity map of the same line to show the alignment of the twist in the zero-velocity line and the star forming ring.}
\label{fig:pa_losv}
\end{figure}

We overlay the flux intensity contours of Pa$\alpha$ emission on its LOSV map in Fig.~\ref{fig:pa_losv} to show how the irregularity of the zero-velocity line is aligned with the location of the stellar ring. This could be due to a deviation between the kinematics of the gas in the stellar ring and in the disk. The noncircular motion of the ring is also observed in CO velocity fields which could be due to the ILR \citep{2007ApJ...654..782S}. 

\subsubsection{Streaming motions}
\begin{figure*}
\centering
\includegraphics[width=0.35\linewidth]{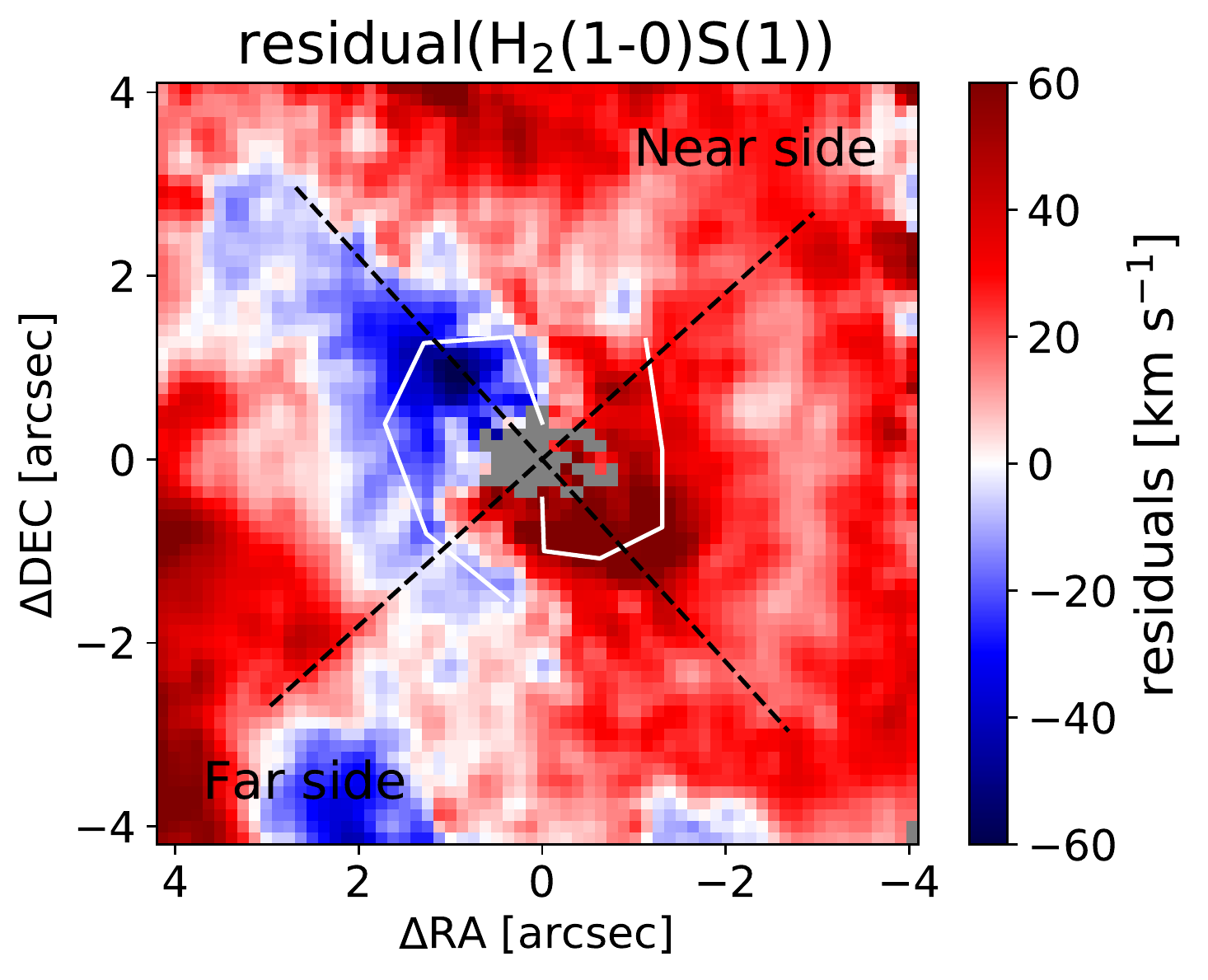}
\includegraphics[width=0.35\linewidth]{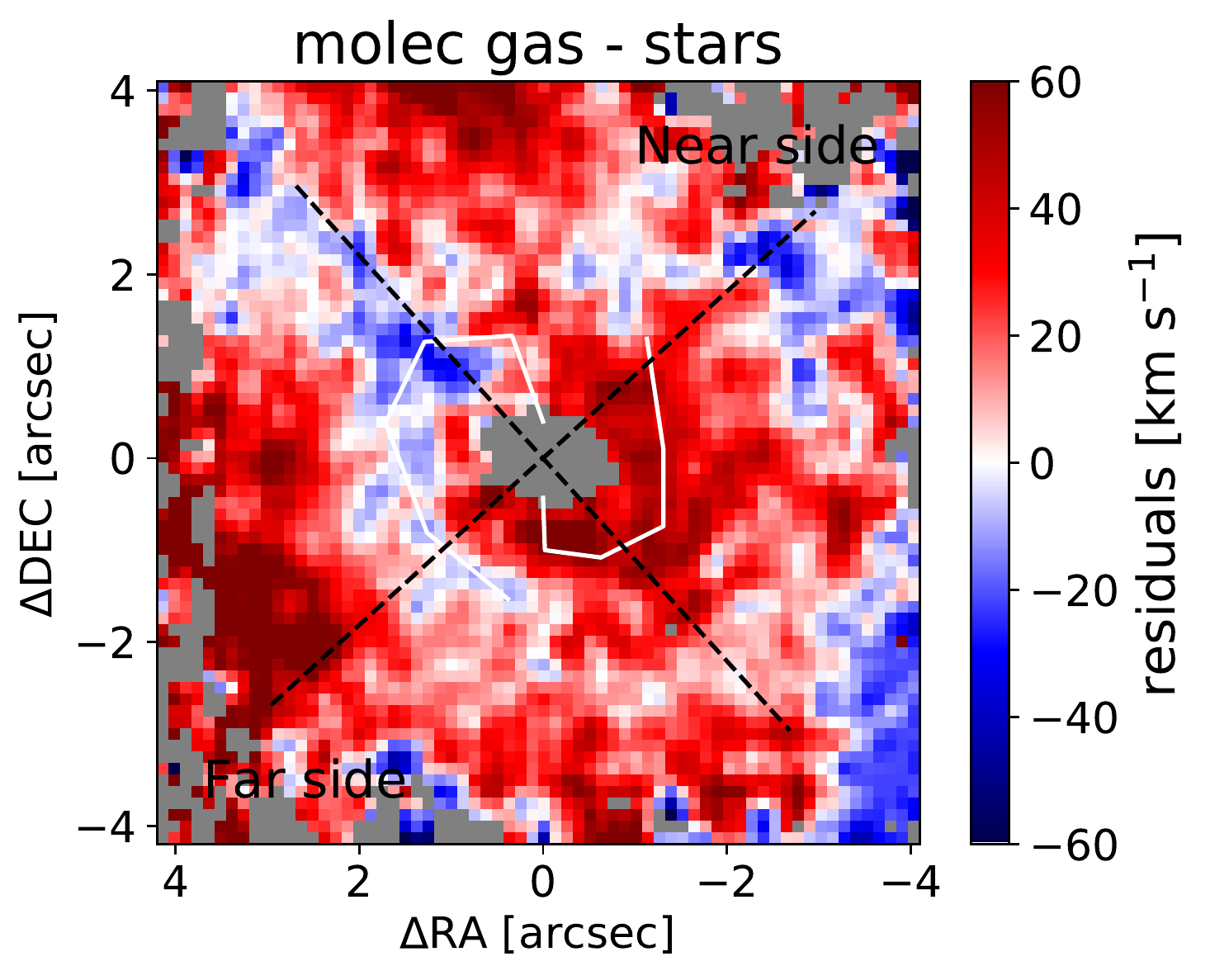}
\includegraphics[width=0.28\linewidth]{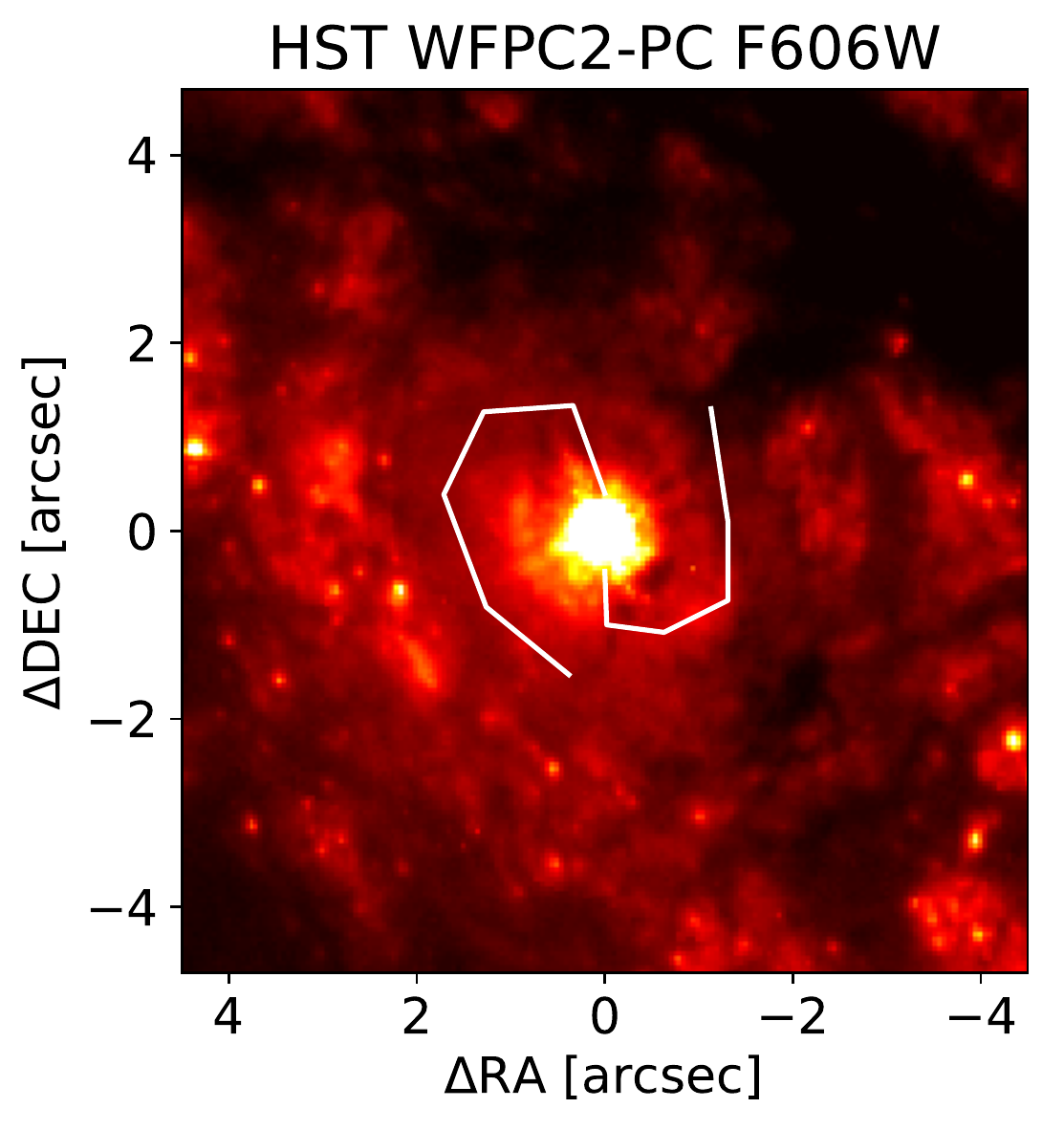}
\caption{Residuals after subtraction of a disk model (\emph{left}) and the stellar velocity field (\emph{middle}) from the molecular gas velocity field. In the \emph{right} panel we show the HST $F606W$ map to compare the residual velocities with dust lanes, with which inflowing motions are often associated.}
\label{fig:h2_residual}
\end{figure*}

We perform a fit of the rotating disk model as explained in Sect.~\ref{sec:stellarkinematics} to subtract the pure-rotation velocity field from the gas LOSVs. In the fit we fixed the central position and inclination similar to the stellar velocity fit. For all observed NIR lines, velocity fields show higher central redshifts compared to stars. From the fit results, we further find the position angle of the line of nodes for the gas. The molecular gas shows a slightly different position angle PA $\approx 42^{\circ} $ (compared to $51^\circ$ for the stellar field), which is in agreement with the large-scale kinematics position angle of \citep[PA$\approx 40^{\circ} $,][]{2008ApJ...674..797Z}. We aim to isolate the noncircular motions in the central regions of NGC 1365 by subtracting the rotation disk model, as shown in Fig.~\ref{fig:h2_residual} (left).
As a second approach to look for possible streaming motions, we also subtract the stellar velocity field from the gas velocity field (Fig.~\ref{fig:h2_residual}, middle). Both of the residual maps show consistent results. Specifically, in the central $200\,\mathrm{pc}$ ($\sim 2\arcsec $), the gas shows two-arm spiral features, which we indicate with solid white lines. Assuming trailing spiral arms, we can derive the orientation of the galaxy toward us. Considering the winding sense of the arms (Fig. \ref{fig:NGC1365}), the NW would be the near side and the SE the far side of the galaxy. 

Based on this orientation and assuming that they are located within the disk plane, we conclude that these spiral arms can be streaming motions towards the center.
In Fig.~\ref{fig:h2_residual} (right) we overlay the solid lines showing the spiral arms on the optical HST $F606W$ map and see that there are dust features coinciding with these possible inflowing streaming motions. \cite{2014ApJ...792..101D} suggest that there is a connection between the chaotic circumnuclear dust structures and external accretion for galaxies in moderately sized groups.
The amount of cold gas flowing in these streams and their associated gravitational torques can be quantified using high-resolution sub-millimeter observations with ALMA \citep[e.g.,][]{2005A&A...441.1011G}.

\subsubsection{Outflow}
\begin{figure}
\centering
\includegraphics[width=\columnwidth]{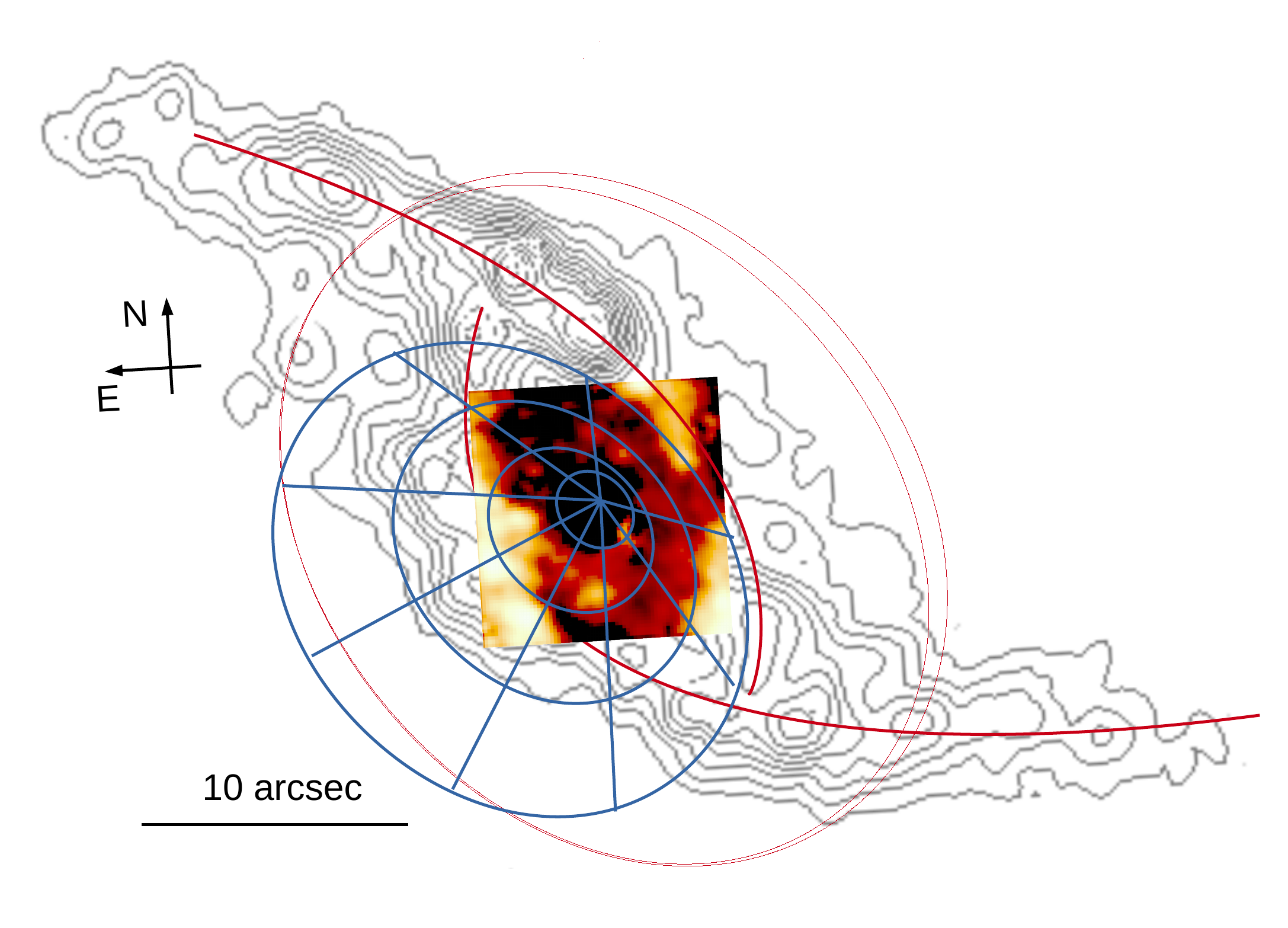}
\caption{Extinction corrected [\ion{Fe}{ii}] flux map and CO contours from \citet{2007ApJ...654..782S}, together with a sketch representing the galactic plane, the spirals (in red) and a cone that represents an outflow inferred from the [\ion{O}{III}] emission observations (in blue).}
\label{fig:sketch}
\end{figure}

NGC 1365 has a well-known outflow, which first was suggested by \cite{1960ApJ...132...30B}. \cite{1983MNRAS.203..759P} found that the forbidden [\ion{O}{III}] line shows extended emission with a two-component line split. A biconical outflow scenario is studied by several authors using [\ion{O}{III}] emission observations \citep{1984A&A...140..288J,1988MNRAS.234..155E,1991MNRAS.250..138S, 1999A&ARv...9..221L,2003AJ....126.2185V, 2010ApJ...711..818S}. Optical spectra show that the low-ionization lines like H$\alpha$ have kinematics dominated by the disk rotation and the high-ionization lines like [\ion{O}{III}] are more dominated by the outflow \citep{2017ApJS..232...11T, 2018arXiv180901206V}. However, \citet{2016MNRAS.459.4485L} claim that they observe a blueshifted outflow in the SE of their FOV (13\arcsec $\times $ 6\arcsec) from the [\ion{N}{ii}] emission line and in a few spaxels in this region line splitting of H$\alpha$ and [\ion{S}{ii}] emission lines. \cite{1996A&A...305..727H} suggest from the modeled outflow that it has a geometry extending from the nucleus out of the galactic plane towards the south and southeast, with an opening angle of $100^{\circ}$ and an axis aligned with the rotation axis of the galaxy. This geometry is illustrated in a sketch in Fig.~\ref{fig:sketch}. A projection of the outflow emission onto the galactic disk results in a double peaked [\ion{O}{III}] emission line. The bipolar outflow is mostly suggested to originate from the AGN, however \cite{1998A&A...339..345K} consider the possibility of a starburst-driven outflow.   

In our IFU NIR data we observe the outflow at its foot point. In the nuclear region we detect a strong blueshifted wing in the Br$\gamma $ line and a clear line split in Pa$\alpha$. The emission line maps extracted by fitting on a spaxel-by-spaxel basis show that the blueshifted line component is only extended in a region the size of the PSF (r $\sim 0\farcs5$). The Pa$\alpha$ line fit in an aperture with a radius corresponding to the FWHM of the PSF ($r=0\farcs 5$) is shown in Fig.~\ref{fig:nuclearPaafit}. The blueshifted components in both hydrogen recombination lines show a velocity offset of about $470 \pm 60\,\mathrm{km}\,\mathrm{s}^{-1}$. A blushifted wing component is also observed in the H$\alpha$ emission line. We show an integrated spectrum from a 4\arcsec central aperture, in the right panel of Fig.~\ref{fig:nuclearHafit}. This optical spectrum is taken from the Siding Spring Southern Seyfert Spectroscopic Snapshot Survey \citep[S7][]{2015ApJS..217...12D,2017ApJS..232...11T}. The line fits show slight differences in widths, for example, which could be, observationally, due to aperture size differences, or could be  intrinsic.  However it is remarkable that in both cases an extra blueshifted emission is present. In \citet[][Fig.~3 and 4 ]{2018arXiv180602839K} they show that in the center of NGC 1365 the low-ionization emission lines are doubly peaked but the [\ion{O}{III}] emission is symmetric and can be fit with one component. On the other hand a few arcseconds away from the center the high-ionization [\ion{O}{III}] line split is becoming prominent. This could be due to the fact that the outflow is getting radially accelerated, which can be induced by a pressure gradient on the path of the outflow. The fact that the hydrogen recombination lines have an asymmetric emission only in the central aperture could however also be due to a different phenomenon, for example a complex broad line region not related to the outflow.

Since [\ion{Fe}{II}] emission is a shock tracer (e.g., from the regions partially ionized by outflows or SN remnants) in the NIR \citep{1988ApJ...328L..41K, 2000ApJ...528..186M, 2014MNRAS.445.2353B}, we also examine this line to search for hints of the outflow in our FOV. The [\ion{Fe}{II}] flux map shows very strong emission from this line in the southeast and east side of our FOV (Fig. \ref{fig:sketch}). This coincides with the radio continuum emission at $3\,\mathrm{cm}$ \citep{1998MNRAS.300..757F} and is in the direction of the reaching cone of the outflow. However, these regions coincide with the clumps of the star forming ring and the strength of the [\ion{Fe}{II}] emission line might be due to starburst regions with supernovae remnants. We also observe a strong source of [\ion{Fe}{II}] emission towards the south of the nucleus (aperture I7). In this hot spot the velocity dispersion of the [\ion{Fe}{II}] emission line is also high. \cite{1997A&A...328..483K} finds clumps of enhanced [\ion{O}{III}] emission in this spot. This spot is also bright in 3-5-keV X-ray emission \citep[Fig. 12 in][]{2018arXiv180901206V}. All this could hint at an interaction of the outflow with the ISM but is not fully conclusive.

\subsection{Star formation in the ring}
\label{sect:starformation}
There are several starburst clusters in the inner regions of NGC 1365, observed as “hotspots” in an ILR \citep[first detected by][]{1958PASP...70..364M, 1965PASP...77..287S}. These clusters are bright in H$\alpha$ emission \citep{1981A&A...101..377A, 1997A&A...328..483K}, X-ray emission \citep{2009ApJ...694..718W}, nonthermal radio continuum \citep[e.g., from SN remnants,][]{1995A&A...295..585S, 1998MNRAS.300..757F}, and mid- and far-IR emission \citep[][but these authorsdid not identify all the bright compact sources]{2012MNRAS.425..311A}. The ILR has an oval shape and is located at the end of the prominent 3$\arcmin$ bar of the galaxy. \cite{2007ApJ...654..782S} detect large amounts of molecular gas associated with this ring.

\cite{2012MNRAS.425..311A} combine their AGN-subtracted $24\mm$ flux with the H$\alpha$ flux from \cite{2004A&A...419..501F} and estimate a star formation rate (SFR) of $7.3\,M_{\odot }\,\mathrm{yr}^{-1}$ of a $40\arcsec$ diameter aperture and claim that $85\%$ of the ongoing SFR inside the ILR region is taking place in dust-obscured regions. \cite{2007ApJ...654..782S} report a SFR of 9 $M_{\odot }\,\mathrm{yr}^{-1}$ in a radius of 1 kpc.
 
In our data we detect the starburst regions in the ILR (``R'' apertures). The morphology of the ILR parts we observe coincide very well with the oval ring of molecular gas from \cite{2007ApJ...654..782S} (Fig.\ref{fig:sketch}).
The high-resolution data provide an opportunity to detect further starburst regions located between this ring and the nucleus (Fig.~\ref{fig:line_fluxes}, ``I'' apertures). These regions show lower stellar velocity dispersion, indicative for younger stellar populations (Fig.~\ref{fig:sigmastarmap}) and are also visible in $K$-band continuum (Fig.~\ref{fig:kbandcont}). The regions are located in a ring-like shape and seem to be connected to the center with an elongated structure. However, no compelling evidence for a nuclear bar is found. It is therefore more likely that this structure is part of the nuclear disk \citep{2001A&A...368...52E}.
We also detect a few compact starburst regions in the NW where a dust lane is located which has not been indicated in the optical or the MIR data due to high extinction. 

\subsubsection{Star formation rates and efficiency}

Starburst events can be traced at several wavelengths, from X-ray to radio, through different physical mechanisms. Young, massive luminous O and B stars can be directly traced by their UV emission. This emission can be highly attenuated by reddening from the dust. On the other hand the UV luminous young stars heat up the dust and contribute to the thermal emission at wavelengths around $60\mm$. However, due to the contribution of older stellar populations in heating up the dust, estimating SFRs using dust emission can be complex. Therefore, hydrogen recombination lines, which are produced by the ionizing photons from massive and young stars that photoionize their surrounding gas, are one of the most commonly used starburst tracers. Since these lines are produced by stars with masses$\gtrsim  10  M_{\odot}$ and ages $\lesssim 20\,\mathrm{Myr}$ \citep{1998ARA&A..36..189K}, they can estimate the current SFR and are not dependent on star formation history.    

In the NIR, Pa$\alpha$ and Br$\gamma$, less affected by dust, are good tracers of starburst regions. We use the luminosity of the Br$\gamma$ emission line to estimate the SFRs in the starburst regions to compare their powers. Here the calibration from \citet{2003A&A...409...99P} is used to derive the SFRs:    

\begin{equation}
\mathrm{SFR}_{\mathrm{Br}\gamma} = \frac{L_{\mathrm{Br}\gamma}}{1.585\times 10^{32}\,\mathrm{W}}\, M_\odot\,\mathrm{yr}^{-1}.
\end{equation}

In the star-forming ring apertures, the SFR ranges from $0.013$ to $0.049\,M_{\odot}\,\mathrm{yr}^{-1}$ and in the apertures inside the inner ring the SFR ranges from $0.005$ to $0.023\,M_{\odot }\,\mathrm{yr}^{-1}$. All measurements are listed in Tables \ref{tab:SFR-R} and \ref{tab:SFR-I}. Since NGC 1365 has a strong AGN which dominates the emission in the center, we do not estimate the SFRs for the nuclear aperture. 

In order to better compare our SFR results with other measurements, we divide them by the surface area of our apertures ($r = 0\farcs5 \approx 45\,\mathrm{pc}$). The star formation surface densities then range from $2$ to $8\,M_{\odot }\,\mathrm{yr}^{-1}\,\mathrm{kpc}^{-2}$ in the ring and from $1$ to $4\,M_{\odot }\,\mathrm{yr}^{-1}\,\mathrm{kpc}^{-2}$ in the inner ring. Typical star formation surface density values are $(1-50)\,M_{\odot }\,\mathrm{yr}^{-1}\,\mathrm{kpc}^{-2}$ on scales of  hundreds of parsecs, and they become higher in the central regions, reaching about $(50-500)\,M_{\odot }\,\mathrm{yr}^{-1}\,\mathrm{kpc}^{-2}$ on scales of  tens
of parsecs, and can even reach up to approximately $1000\,M_{\odot }\,\mathrm{yr}^{-1}\,\mathrm{kpc}^{-2}$ on parsec scales \citep[][and references therein]{2012A&A...544A.129V}. We conclude that SFR surface densities in our apertures are of a typical order of magnitude.

\subsubsection{Star formation in the circumnuclear ring}

In this section our aim is to investigate the star formation history in the ILR. \cite{2008AJ....135..479B} introduce two models to explain the propagation of star formation in the rings. One is the ``popcorn'' model, which suggests that the gas accumulates in the ring, reaching a critical density. Subsequently gravitational instabilities \citep[induced by e.g., turbulences,][]{1994ApJ...425L..73E} force the gas to collapse into fragments and starbursts occur. Due to the random time and location of star formation in this model, hot spots do
not follow any age gradient but are ``popping up'' stochastically. The other model is ``pearls on a string'', suggesting that the star formation is only induced in particular regions of the ring, where the gas density can get sufficiently high. These regions are usually at the position where the gas enters the ring from the bars/spiral arms \citep{2003ApJ...582..723R}. Young clusters form at this position, from where they will rotate with the ring and in the meanwhile age, which results in an age gradient in the ring \citep[e.g.,][]{2014MNRAS.438..329F}. 

\begin{figure}
\centering
\includegraphics[width=0.98\linewidth]{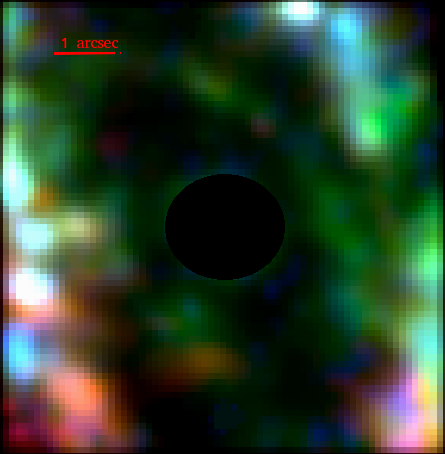}
\caption{Three-color plot of \ion{He}{i} in blue, Br$\gamma$ in green, and [\ion{Fe}{ii}] in red that suggests an age gradient in the western part of the ring. The AGN is masked.}
\label{fig:SF-plot}
\end{figure}

To estimate the relative cluster ages in the star forming ring, \cite{2008AJ....135..479B} suggest a method to use false-color (RGB) maps of the emission lines \ion{He}{I}, Br$\gamma$, and [\ion{Fe}{II}]. This method is based on the different ionization potentials at which each of these emission lines are produced. \ion{He}{I} has an ionization energy of $24.6\,\mathrm{eV}$ and emits stronger in regions with hotter and more massive (O and B) stars compared to Br$\gamma$ with an ionization energy of 13.6 eV. This line is therefore more prominent in early phases of starbursts, since the most massive stars, which are energetic enough to excite \ion{He}{i,} have the shortest lifetimes. On the other hand, [\ion{Fe}{II}] has stronger emission in regions dominated by stellar populations at ages of $\sim 10\,\mathrm{Myr}$ when massive stars explode in supernovae. Our false-color map is shown in Fig. \ref{fig:SF-plot}, where blue represents \ion{He}{I}, green Br$\gamma$, and red [\ion{Fe}{II}]. We see that in the west side of the map there seems to be an age gradient from NW (younger and blue) to SW (redder and older). However, on the east side there is no clear trend.  

Since we see only parts of the whole ring in our FOV, we try to assess the location of the gas downstream from the bars to the ring by comparing our image with the large-scale images. In \citet[][their Fig.~10]{2007ApJ...654..782S}, the $^{12}$CO(2-1) flux map is indicating the concentration of molecular gas in the bars and the ring. Figure \ref{fig:sketch} shows their CO map together with our FOV. \cite{2008A&A...492....3G, 2012A&A...545A..10G} report that the ring contains young (age$\sim 5.5-6.5\,\mathrm{Myr}$) massive clusters. \cite{2009ApJ...703.1297E} suggests that these clusters form in the bar and are on their way towards the ring.  

In order to get a better estimate of the absolute starburst ages, we use the models from \citet[][their Fig. 19]{2017A&A...598A..55B}. They simulate the Br$\gamma$ equivalent width and supernova rate normalized by the $K$-band continuum luminosity using the stellar synthesis code \textsc{STARBURST99} \citep[][and the references therein]{2014ApJS..212...14L}. These models can then be used to constrain the star formation history.

The [\ion{Fe}{II}] emission is produced in partially ionized regions excited by shocks of supernova remnants \citep{2000ApJ...528..186M} and is less observed in pure \ion{H}{II} regions. Therefore, under the assumption that supernova remnants are the only source of [\ion{Fe}{II}] emission, the luminosity of this line can be employed to crudely estimate the supernova rate in the apertures. We use the calibration from \cite{Alonso-Herrero2003a} to estimate the supernova rate:

\begin{equation}
\mathrm{SNR} = 8.08 \times \frac{L_{[\ion{Fe}{ii}]}}{10^{35}\,\mathrm{W}}\,\mathrm{yr}^{-1}.
\end{equation}

Results are shown in Tables \ref{tab:SFR-R} and \ref{tab:SFR-I}. 
\cite{1999MNRAS.306..479S} argue that the radio hot spots in NGC 1365 are made up of several supernova
rates. However, in the region towards the south and southeast (spot F in their Fig. 5) there might be shock contamination from a weak radio jet or the outflow. Therefore, we cannot completely rely on the age estimation of the apertures in this region. 
Both Br$\gamma$ equivalent width and [\ion{Fe}{II}] luminosity need to be normalized to the continuum luminosity, which contains the emission not only from young, but also from the underlying old stellar population \citep[e.g.,][]{2009MNRAS.393..783R}. We estimate this contribution using slits from the center through the star forming apertures to inspect the light profiles of the continuum emission. The results show old stellar population contributions of around $40\%$ and even in some regions up to $80\%$. This means that the equivalent widths and normalized supernova
rates should be multiplied by factors of 2 up to 9 in some apertures. On the other hand, many of the hot spots are located at the border of our FOV which makes it difficult to estimate continuum offsets and hence the starburst ages accurately. We find that the hot spots in the west side have ages of some 5-10 Myr which is in good agreement with \citet[][$\sim 6-8\,\mathrm{Myr}$]{2008A&A...492....3G}.

\begin{table*}
\caption{For all apertures in the main circumnuclear ring: SFR, star formation surface density, equivalent width of Br$\gamma$, supernova rate and supernova rate normalized by $K$-band continuum luminosity.}
\label{tab:SFR-R}
\centering
\begin{tabular}{cccccc}
\hline
\hline \\
 & $\mathrm{SFR}_{\mathrm{Br}\gamma}$ & $\log(\Sigma_\mathrm{SFR})$ & $W_{\mathrm{Br}\gamma}$ & SNR & $\mathrm{SNR}/L_K$ \\
Aperture & [$10^{-3}\,M_\odot\,\mathrm{yr}^{-1}$] & [$M_\odot\,\mathrm{yr}^{-1}\,\mathrm{kpc}^{-2}$] & [\AA] & [$10^{-5}\,\mathrm{yr}^{-1}$] & [$\mathrm{yr}^{-1}/10^{12}\,L_{\odot,K}$] \\
\hline

R1 & $28$ & $0.67$ & $11.1\pm0.4$ & $ 46$ & $4.6$ \\
R2 & $35$ & $0.76$ & $17.4\pm0.3$ & $ 41$ & $6.0$ \\
R3 & $13$ & $0.33$ & $5.7\pm0.2$ & $ 13$ & $0.8$ \\
R4 & $17$ & $0.45$ & $11.6\pm0.3$ & $ 16$ & $1.5$ \\
R5 & $38$ & $0.81$ & $11.0\pm0.3$ & $ 59$ & $3.9$ \\
R6 & $49$ & $0.92$ & $8.8\pm0.3$ & $179$ & $9.8$ \\
R7 & $30$ & $0.70$ & $11.1\pm0.2$ & $137$ & $12.7$ \\
R8 & $27$ & $0.66$ & $16.5\pm0.4$ & $ 38$ & $4.4$ \\
R9 & $37$ & $0.80$ & $15.4\pm0.3$ & $ 96$ & $9.0$ \\
R10 & $31$ & $0.72$ & $12.4\pm0.3$ & $ 60$ & $4.8$ \\
R11 & $22$ & $0.57$ & $13.6\pm0.4$ & $ 28$ & $2.3$ \\

\hline
\end{tabular}
\tablefoot{The apertures all have a radius of $0\farcs5$ which corresponds to an area of $6000\,\mathrm{pc}^2$.}
\end{table*}

\begin{table*}
\caption{Same as Table \ref{tab:SFR-R}, for all the apertures within the main circumnuclear ring.}
\label{tab:SFR-I}
\centering
\begin{tabular}{cccccc}
\hline
\hline \\
 & $\mathrm{SFR}_{\mathrm{Br}\gamma}$ & $\log(\Sigma_\mathrm{SFR})$ & $W_{\mathrm{Br}\gamma}$ & SNR & $\mathrm{SNR}/L_K$ \\
Aperture & [$10^{-3}\,M_\odot\,\mathrm{yr}^{-1}$] & [$M_\odot\,\mathrm{yr}^{-1}\,\mathrm{kpc}^{-2}$] & [\AA] & [$10^{-5}\,\mathrm{yr}^{-1}$] & [$\mathrm{yr}^{-1}/10^{12}\,L_{\odot,K}$] \\
\hline

I1 & $ 8$ & $0.13$ & $3.7\pm0.2$ & $ 12$ & $1.0$ \\
I2 & $12$ & $0.32$ & $3.3\pm0.2$ & $ 31$ & $2.1$ \\
I3 & $ 7$ & $0.09$ & $3.3\pm0.2$ & $  8$ & $0.5$ \\
I4 & $ 8$ & $0.11$ & $4.1\pm0.2$ & $  8$ & $0.6$ \\
I5 & $10$ & $0.21$ & $5.8\pm0.2$ & $ 11$ & $0.9$ \\
I6 & $ 5$ & $-0.09$ & $3.2\pm0.3$ & $  8$ & $0.7$ \\
I7 & $ 6$ & $0.03$ & $3.2\pm0.2$ & $ 16$ & $1.1$ \\
I8 & $ 8$ & $0.12$ & $3.3\pm0.2$ & $ 11$ & $0.7$ \\
I9 & $23$ & $0.58$ & $5.2\pm0.2$ & $ 63$ & $3.7$ \\
I10 & $ 7$ & $0.05$ & $4.0\pm0.2$ & $  7$ & $0.6$ \\

\hline
\end{tabular}
\tablefoot{The apertures all have a radius of $0\farcs5$ which corresponds to an area of $6000\,\mathrm{pc}^2$.}
\end{table*}
In order to estimate the travel time of a cluster along the ring, we calculate its physical rotational velocity in the ring, using a simplified version of a rotating disk (first order estimation):
\begin{equation}
v_{\mathrm{ring}} = \frac{v_{\mathrm{obs}}}{\sin(i) \cos(\theta)}
,\end{equation}
where $i$ is the inclination angle of $40^{\circ}$, $v_{\mathrm{obs}}$ is the observed velocity, and $\theta$ is the angle of the position in the plane of the ring (measured from the line of nodes) where we measure the velocity. We derive a velocity of $140\,\mathrm{km}\,\mathrm{s}^{-1}$ at $\theta= 0$ in the SW region of the Pa$\alpha$ LOSV map (Fig. \ref{fig:gas_losv}, aperture R6) and therefore estimate the rotation velocity to be $v_{\mathrm{ring}}\, \approx \, 215 \, \mathrm{km s}^{-1} $. This is in agreement with the orbital velocity reported by \citet{2009ApJ...703.1297E}. Looking at the CO maps \cite{2007ApJ...654..782S} we assess the semi-major axis of the ring orbit to be approximately $900\,\mathrm{pc}$. This means the that the time needed for a cluster to travel a quarter of the ring (NW to SW in our FOV) is        

\begin{equation}
t_{\mathrm{travel}} = \frac{\frac{\pi}{2} \times 900\, \mathrm{pc} }{215\,\mathrm{km\, s}^{-1}}\, \approx \,6.6 \, \mathrm{Myr}.
\end{equation}

It seems that the newly formed clusters in the NW of the ring, where it has more pure \ion{H}{II} region characteristics, need about 6 Myr to reach the SW. This is the lifetime of very massive stars until they explode in supernovae, which we observe in [\ion{Fe}{ii}]. We can therefore conclude that the estimated timescales support the pearls on a string model in the star forming ring of NGC1365, at least in the western part. In the eastern part, we do not find evidence for this scenario. However, this region might be affected by the outflow.

\cite{2013ApJ...769..100S} suggest that the SFR can influence the course of action in circumnuclear ring formation. They claim that the age gradient in the starburst clusters can mostly occur in rings with low SFR and for higher SFRs the age distribution would be more stochastic. They report a critical SFR of $1\sim \,M_{\odot }\,\mathrm{yr}^{-1}$ \citep[supported by observations,][]{2008ApJS..174..337M}. Summing up the SFRs in all the circumnuclear apertures we find a value of 0.4 $M_{\odot }\,\mathrm{yr}^{-1}$ for NGC 1365, which lies under the critical value for the pearls on a string scenario.

\section{Conclusions and Summary}
\label{sec:conclusions}
We have analyzed high-resolution NIR IFS observations of the central kiloparsec of the nearby Seyfert galaxy NGC 1365. Our main findings are summarized as follows:

\begin{itemize}
\item We detect parts of the starburst circumnuclear ring ($r\approx 1$ kpc) and resolve some weaker starburst regions within the inner ring and the nucleus ($r\approx 0.3$ kpc). The study of star formation history suggests ages to be $\lesssim $ 10 Myr. Inspecting the relative ages of the hot spots using the \ion{He}{I}, Br$\gamma,$ and [\ion{Fe}{II}] emission line ratios, we find an age gradient on the west side of the ring. We estimate the travel time for a sturburst cluster through the ring from the NW to the SW to be around 6 Myrs, which could explain the transition from a \ion{He}{I} dominated to an [\ion{Fe}{II}] dominated region.  

\item We find blackbody emission of hot dust close to its sublimation temperature ($T \approx 1300$ K) and detect broad components ($\mathrm{FWHM} \approx 2000\,\mathrm{km}\,\mathrm{s}^{-1}$) in the hydrogen recombination lines Pa$\alpha$ and Br$\gamma$, which is typical for type-1 AGNs. 

\item We calculate a black-hole mass of $(5-10)\times 10^6 \, M_{\odot}$ using both $M_{\mathrm{BH}}-\sigma_*$ and the broad components of hydrogen recombination lines.  

\item In our $9\arcsec \times 9\arcsec$ FOV ($9\arcsec$ corresponds to $\sim800\,\mathrm{pc}$), we estimate the molecular gas mass using the H$_2\lambda2.12$ emission line luminosity, which is observed in the entire FOV except for the nuclear region. We find the hot molecular gas to be $\sim 615 \, M_{\odot}$, which corresponds to a cold molecular gas mass of $\sim (2-10)\times 10^8 \,  M_{\odot}$. We also estimate the ionized gas mass using the Br$\gamma$ emission line to be $\sim  5.3\times 10^6 \, M_{\odot}$. We find that the ratio between the ionized and hot molecular gas is about $\sim 8000$, which is typical for nuclei of nearby galaxies as well as separate star forming regions. There is a molecular gas deficiency in the nuclear aperture. The estimated upper limit for the H$_2$ emission line suggests lower values than off-nuclear hot spots.   

\item We study the stellar kinematics using the CO band heads in $K$-band. The stellar velocity field shows an overall rotation. The velocity dispersion map shows a drop in the location of the starburst regions within the inner ring ($r\approx 0.3$ kpc) and the ring ($r\approx 1$ kpc).   

\item The ionized and molecular gas show an overall rotation in the general orientation with the stellar velocity. However, we detect a twisted zero-velocity line, which indicates deviations from disk rotation. In the ionized gas LOSV map the strong deviation from the disk coincides with the position of the starburst ring. After subtraction of the stellar velocity field from the molecular LOSV map, we find a spiral-shaped structure (scale $\lesssim 2\arcsec$) in the residual map, which could indicate inflowing streaming motions.

\item We detect a blueshifted line split of low-ionization Pa$\alpha$ emission in the nuclear region ($r\sim $ width of PSF). This could mean that we detect the outflow at its foot point. However, since the line split is only observed in the center and nowhere else in the FOV, an alternative explanation could be asymmetries in the geometry of the broad-line region.
The forbidden [\ion{Fe}{II}] emission is very strong in clumps towards the S and SE of the FOV. This may be a result of the outflow interacting with the disk in these regions, since the [\ion{O}{III}] observations suggest an approaching outflow cone and associated clumpy structures in the SE. 

\end{itemize}

\begin{acknowledgements}
The authors thank the anonymous referee for fruitful comments and suggestions. This research is carried out within the Collaborative Research Center 956, sub-project A2 (the studies of the conditions for star formation in nearby AGN and QSOs), funded by the Deutsche Forschungsgemeinschaft (DFG). N. Fazeli is a member of the Bonn-Cologne Graduate School of Physics and Astronomy (BCGS).
\end{acknowledgements}

\bibliographystyle{aa} % style aa.bst
\bibliography{references} % your references Yourfile.bib

\begin{appendix} %First online appendix
\section{All apertures}
\begin{figure*}
\centering
\includegraphics[width=\linewidth]{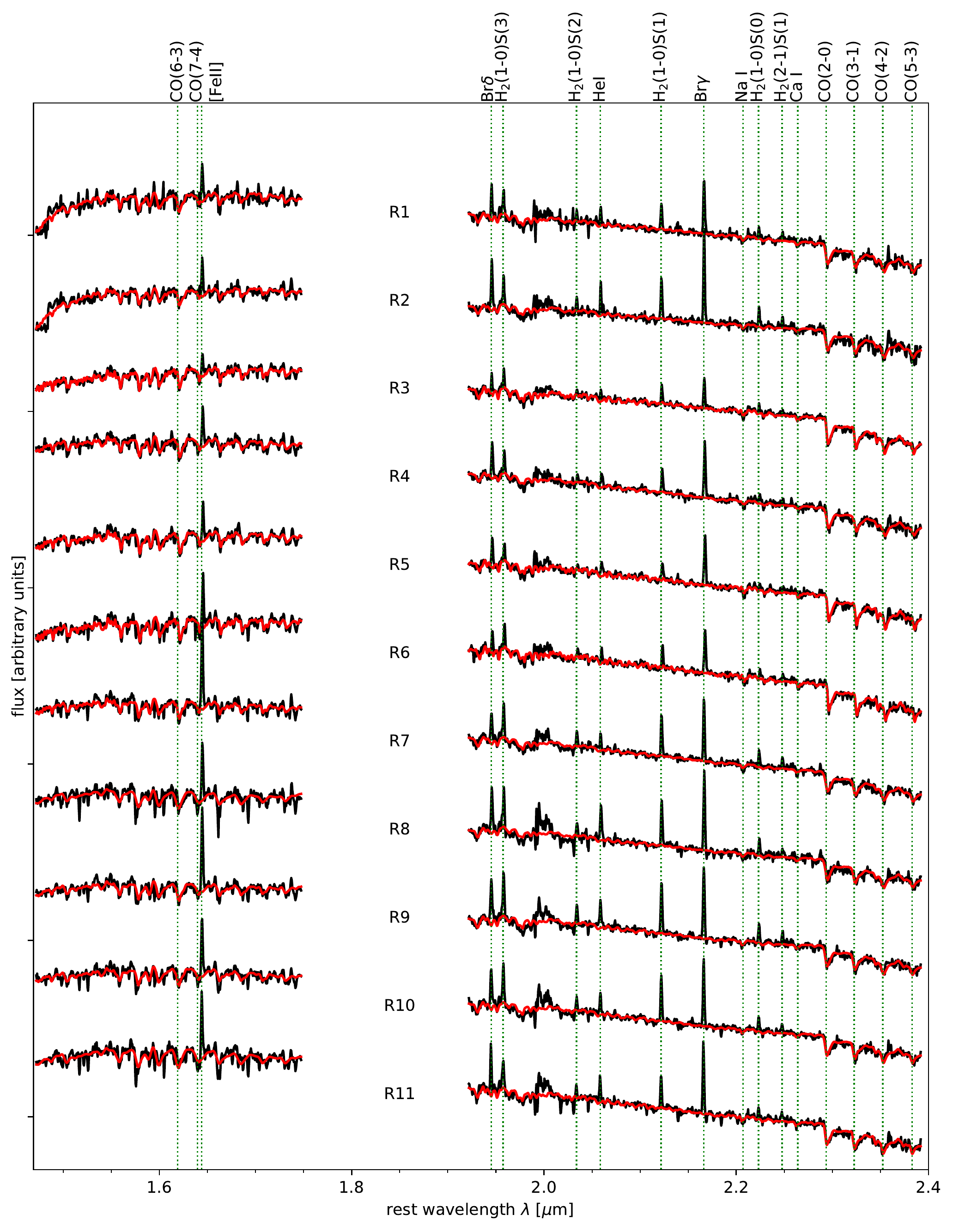}
\caption{As in Fig.~\ref{fig:regions_specs} for all R apertures.}
\label{fig:spectra_subt_R}
\end{figure*}

\begin{figure*}
\centering
\includegraphics[width=\linewidth]{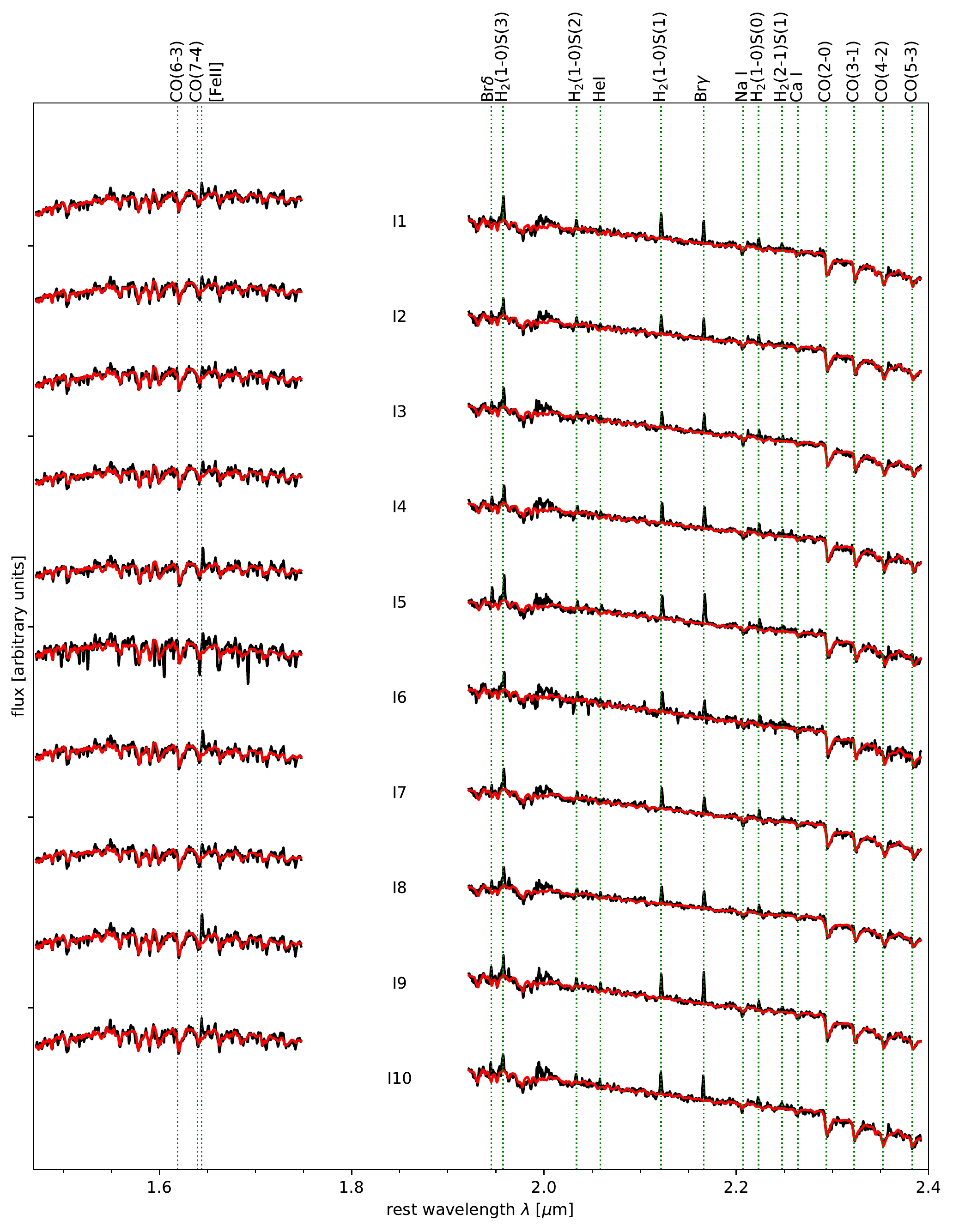}
\caption{As in Fig.~\ref{fig:regions_specs} for all I apertures.}
\label{fig:spectra_subt_I}
\end{figure*}
\end{appendix}
\end{document}